\documentclass[dvips,12pt]{thesis}

\usepackage{amsmath,amssymb,epsf,epsfig,amsthm,bm}

\theoremstyle{definition}
\newtheorem{theorem}{Theorem}[chapter]

\newtheorem{lemma}[theorem]{Lemma}
\newtheorem{proposition}[theorem]{Proposition}
\newtheorem{definition}[theorem]{Definition}
\newtheorem{conjecture}[theorem]{Conjecture}

\begin{document}

\begin{centering}
{\Large Ph.D. Thesis.}
\\
\vspace{1in} {\Huge {\bf Intersecting hypersurfaces and Lovelock
Gravity}}
\\\vspace{.3in}
{\Large Steven Willison\footnote{E-mail: steven.willison@kcl.ac.uk}
\\\vspace{.21in}
\it Department of Physics, Kings College, Strand, London WC2R 2LS,
U.K.
\\\vspace{1in}
{July 2004}}\\
\newpage

\vspace{1.91in} {\large {\bf Abstract}}
\\\vspace{.1in}
\end{centering}

A theory of gravity in higher dimensions is considered. The usual
Einstein-Hilbert action is supplemented with Lovelock terms, of
higher order in the curvature tensor. These terms are important for
the low energy action of string theories.

The intersection of hypersurfaces is studied in the Lovelock theory.
The study is restricted to hypersurfaces of co-dimension $1$,
$(d-1)$-dimensional submanifolds in a $d$-dimensional space-time.
It is found that exact thin shells of matter are admissible, with a
mild form of curvature singularity: the first derivative of the
metric is discontinuous across the surface. Also, with only this
mild kind of curvature singularity, there is a possibility of matter
localised {\it on the intersections}.
This gives a classical analogue of the intersecting brane-worlds in
high energy String phenomenology. Such a possibility does not arise
in the pure Einstein-Hilbert case.

\newpage
\begin{centering}
{\bf {\Large Acknowledgements}}\\\end{centering} \vspace{.21in}
I would like to thank  N. E. Mavromatos, my supervisor, for valuable
advice and encouragement; E. Gravanis for great fun working
together; my parents for constant support and understanding. I thank
God for all the opportunities I have had.

Thanks to CERN Theory department for their hospitality and to the
European Union for funding the trip through Network ref
HPRN-CT-2000-00152. The Ph.D. was
funded by EPSRC, to whom I am very grateful.\\\\

The material in Chapters \ref{inters}, \ref{tdimensions} and
\ref{example} represent the original work of Elias Gravanis and
myself as appearing in J. Math. Phys. {\bf 45}, 4223
(2004)~\cite{Gravanis-03} and gr-qc/0401062~\cite{Gravanis-04} with
emphasis on my own contributions. Some of the material in Chapter
\ref{junction} is my own original previously unpublished work. The
rest is a survey of the literature and is fully cited.

\tableofcontents

\chapter{Introduction}

The General Theory of Relativity was a great conceptual
advance and a great piece of twentieth century physics.
The rigorous pursuit of the principle of the invariance
of the laws of physics led to the idea that space-time
was a dynamical manifold.

This brought into physics the abstract study of non-Euclidean
geometry by mathematicians such as Riemann. It also elevated space
and time to the status of a dynamical structure. The application of
geometry to physics also plays an important role in gauge theories.
The possibility to represent gauge theories in terms of connections
on fiber bundles has been of great use.

The latter kind of geometrical theory has proved to
be very much compatible with the principles of quantum
mechanics. The discovery that non-abelian gauge theories
are re-normalisable allowed the description of the
electro-magnetic, weak and strong nuclear forces
in terms of a re-normalisable quantum theory, the Standard
Model of particle physics.

Gravity has proved much more difficult to quantise.
A major goal of theoretical physics is to find a consistent
quantum theory of gravity. The canonical
approach of quantisation to General Relativity leads
to a non-renormalisable theory.

Various approaches to the problem of quantum gravity have been
attempted. String theory conceives of extended particles in a flat
background. The modes of the strings give rise to gravitons as well
as various other fields. Loop quantum gravity is a background
independent theory which takes very seriously the principles of GR.
The Ashtekkar variables have allowed for the treatment of GR as a
gauge theory. There are also approaches, motivated by work in
condensed matter physics, where the properties like general
coordinate invariance are emergent properties. In this case the
underlying theory is not important - only the quasi-particle
spectrum. Other approaches are simplicial quantum gravity,
Chern-Simons theories, non-commutative geometry. The list of
proposed quantum theories of gravity seems to be endless. Also, it
has been proposed that quantum gravity may actually be a
deterministic theory~\cite{'tHooft-99}.

\section{Extended objects}

The idea of fundamental particles as point particles is conceptually
very tidy. To date, experiment has not revealed any substructure to
certain particles such as the electron. The idea of matter as
continuous and infinitely sub-divisible has also enjoyed the favour
of scientists at various times in history. Einstein thought that a
fundamental particle would be a stable, smooth, matter
configuration. There are no-go theorems prohibiting such solitons in
GR and Einstein-Maxwell theory. Some smooth solitonic solutions have
been found in more general Einstein-Yang-Mills
theories~\cite{Bartnick-88}.

Another possibility is particles not as points but membranes. In the
last thirty years there have been a number of advances in theories
of extended particles featuring extra dimensions. Type I strings,
Type IIa and IIb strings, Heterotic $E8\times E8$ and Heterotic
$SO(32)$ strings all live in ten dimensions. These theories are
related to each other and also to the eleven dimensional
super-gravity by dualities and are believed to be part of an
underlying theory called M-Theory~\cite{Polchinski}.
\begin{figure}
\begin{center}\mbox{\epsfig{file=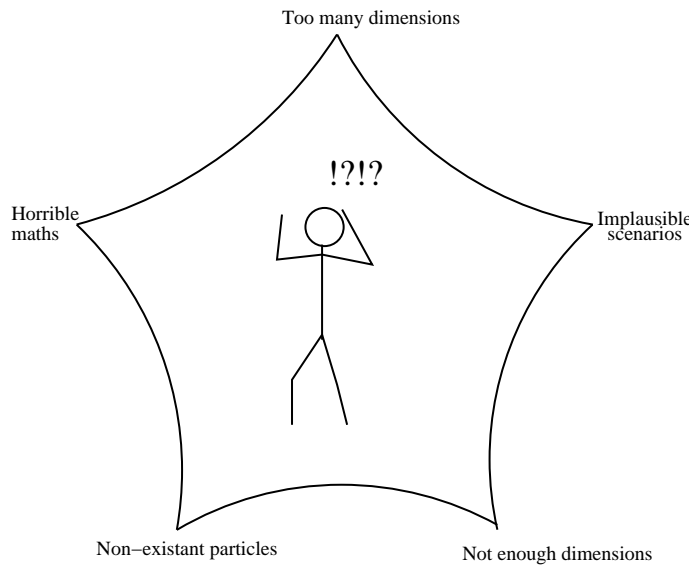, width=9cm}}
\caption{Many roads to quantum gravity}
\label{TOE}
\end{center}
\end{figure}

A feature of strings and eleven dimensional super-gravity is the
existence of membrane solutions. They are solitons of the
super-gravity theories. In open string theories they are
$p$-dimensional surfaces on which open strings can have their
end-points.

\section{Extra dimensions}

Another common feature, as already mentioned, is extra dimensions.
The interest in extra dimensions long predates string theory. As
early as the 1920's people looked into extra dimensions to generate
a gauge field. The Kaluza-Klein compactification of a fifth
dimension on a circle gives a photon type field. It was noticed
later that there is also a scalar field which can not be ignored.

We are familiar with three dimensions of space and one of time,
but the string/M theory tells us we should have another
six or seven dimensions lurking about somewhere.
It was thought that if such a theory is to be taken seriously,
these extra dimensions must be compactified very tightly
else they would already have been ruled out by experiment.

Questions as to why the theory should compactify down to four dimensions
arise. This is part of a more general problem of fine tuning.
Many physical parameters seem to be finely tuned to be amenable to life.
e.g. the stability of atoms in four dimensions.
Anthropic arguments can be invoked, whereby
the seeming fine tuning of various physical quantities for
life is not viewed as fundamental but just a part of the
complicated geography of the universe. We just live in a
part that happens to be hospitable for life. But an explanation
in terms of a unique solution to a fundamental theory is desirable to
many people. Consequently, a dynamical unique
compactification to four dimensions has been sought after.

Lately, the idea that we live confined to a membrane in higher
dimensions has been suggested as in the much hyped Randall-Sundrum
case~\cite{Randall-99a,Randall-99b} and a similar earlier idea by
Rubakov and Shaposhnikov~\cite{Rubakov-83}. It is interesting that
certain branes in string theory necessarily have gauge fields
localised on them. Gravity, as a closed string mode would be free to
propagate in the higher dimensions, but if they are compactified
tightly enough, the deviations from the inverse-square law are small
enough to evade current experimental falsification. As such, they
constitute some kind of prediction for future gravity experiments.
Another possibility is that gravity can be effectively localised
even with large extra dimensions as demonstrated by Randall and
Sundrum.

Another interesting phenomenological feature of strings is the
Standard-Model-like effective field theories of intersecting D
$p$-Branes. The open strings stretching between the intersecting
stacks of branes can be localised around the intersection due to the
tension of stretching them. In particular, there are Chiral matter
fields living on the intersection~\cite{Berkooz-96}. Thus, around a
3+1 dimensional intersection there is potentially a realistic
standard model low energy physics.

\section{Thin walls and gravity}

I will consider the role of membranes in higher dimensions from a classical
perspective. The initial motivation for such a study was phenomenological-
the modeling of Braneworlds of the Randall-Sundrum type,
in general gravitational backgrounds, including the branes own gravity.

Brane cosmology is very popular at the moment. Presumably, this is
because it offers possibilities of solving some perceived problems
in cosmology and phenomenology.
\\\\
$*$ Brane worlds may resolve the Gauge Hierarchy problem.
\\\\
$*$ Problems of fine tuning such as the cosmological constant
problem and the cosmic coincidence problem could be explained away.
\\\\
$*$ The initial singularity could be avoided and replaced by a ``big
bounce" or some other kind of eternal universe. The problem of
initial conditions is avoided.
\\\\
$*$ Contact is made between String/M-theory and four-dimensional
physics.
\\


The need to reduce the number of free parameters in any theory of
physics is an understandable one. It would be nice to have a
solution to the gauge hierarchy problem and have one energy scale
rather than two. This is in accordance with a basic intuition on
which science is based: that diverse phenomena can be explained in
terms of unifying principles. The desire to avoid special initial
conditions is a different matter though. This seems to represent an
atheistic approach to a question which goes beyond the normal scope
of Physics. This is an incorrect approach in my opinion.
\\

In general, if a manifold is filled with membranes, they will
intersect in all sorts of ways. I have already mentioned
intersecting branes. People have also studied colliding
brane-worlds, the ``Ekpyrotic" scenario~\cite{Khoury-01}, as a
cosmological model.

As well as String inspired cosmology, there are several other
motivations for studying thin walls of matter. For example, they
have been used to make simple models of quantum black holes. The
principle of holography is also important in quantum gravity. The
realisation that the degrees of freedom of a black hole live on the
surface area has profound implications. The study of hypersurfaces
may be relevant to this.

Another situation where hypersurfaces and their intersections is
relevant is in the various inflationary models. There has been much
study on the behaviour of bubbles of false vacuum, in the context of
phase transitions in the early Universe. There has been speculation
that a bubble of false vacuum (de Sitter space) in some primordial
space-time foam could have inflated to give birth to a `universe'.
Possibly this could have happened many times, with many universes
being created out of the quantum foam. The simplest approximation is
to model the interface between the false vacuum and surrounding
space-time as a singular wall of matter. Such a wall is known as a
shell in this context. The work has been concentrated on isolated
spherical bubbles, but a more general foam of bubbles would be
interesting to study.

There are situations in astrophysical GR where a thin wall
approximation is useful, such as the ejection of matter from
supernovae or the motion of globular clusters~\cite{Barkov-01}.

The above are concerned with speculations and approximations. But I
will attempt to show that the study is profitable in its own right
and will give some insights into classical gravity as a geometrical
theory.

In this study, I will restrict myself to surfaces of co-dimension
one, which I shall call hypersurfaces or walls, and their
intersections. For example, a wall in 4+1 dimensions of space and
time will be a $3+1$ or $4$ dimensional surface. Clearly, the study
of membranes of other co-dimensionalities is also relevant and I
will comment on this in the concluding section.

\section{Higher curvature gravity}

The modification of the Einstein tensor has been suggested in many
contexts: counter-terms in general relativity to regulate
singularities; scalar-tensor theories in inflationary contexts;
terms appearing in super-gravity; low energy actions from strings.

GR is not a renormalisable theory. The Lagrangian has a negative
dimension coupling $\propto M_\text{Planck}^{-2}$. Each graviton
self interaction term $i$ contributes $- N_i \Delta$ to the
superficial degree of divergence, $D$, where $N_i$ is the number of
vertices and $\Delta = [\text{coupling}] = -2$. For Feynman diagrams
with internal closed loops, one has integrals over momenta of the
form $\int_0^{\infty} dk k^{D-1}$ which diverge when $D > 0$.
General relativity is renormalisable at one loop but, since $- N_i
\Delta$ is positive, at each order in the loop expansion, more
divergences arise. This requires absorbing divergences by
re-normalising an infinite number of parameters. Consequently none
of its interactions are re-normalisable. Couplings to matter (and
gravity couples to everything) make the divergences even more
marked. Also, in higher than four dimensions $\Delta = -(d-2)$ and
the same problems arise

 A non-renormalisable theory is
not considered to be a reasonable candidate for a final theory. A
naive approach to quantum gravity means one has to add every term,
including all the non-gravitational terms, allowed by the
symmetries, as counterterms. This makes the theory fairly useless at
high energies from the point of view of predictive power- there are
an infinite number of phenomenological parameters. This is why
non-renormalisable theories tend to be regarded as effective field
theories. Integrate out a massive particle $\phi$ , of mass $M$, and
consider a low energy theory with partition function
\begin{gather*}
\exp\left(i\smallint {\cal L}_\text{eff} \right) = \int [d\phi] \exp
\left(i\smallint {\cal L}\right)
\end{gather*}
at energies, $E$, much lower than this mass scale. The result is
that ${\cal L}_\text{eff}$ is composed of non-renormalisable terms
suppressed by $E/M$ to some power. For example, starting with QED,
integrate out the electron/positron and you get non-renormalisable
theory for photon-photon scattering~\cite{Weinberg-95}. Similarly,
GR can be viewed as the low energy effective theory of some unknown
fundamental quantum theory which may be very different at the Planck
energy and length scale.

Certain theories of gravity with curvature tensor squared terms have
been suggested to render gravity renormalisable in four
dimensions~\cite{Stelle-77}. Consider a general action quadratic in
curvature.
\begin{gather}\label{Curv_squared}
\int_M(\alpha_1 \sqrt{-g}R + \alpha_2 L_2)d^dx, \\\nonumber L_2 =
\sqrt{g}(
R_{\mu\nu\kappa\lambda}R^{\mu\nu\kappa\lambda}+aR_{\mu\nu}R^{\mu\nu}
+b R^2).
\end{gather}
The quadratic term has a dimensionless coupling constant,
$[\alpha_2] = 4-d$. Consequently, its graviton self interaction
terms are expected to be re-normalisable in $4$ dimensions. This has
motivated the study of such quadratic theories. However, this
approach to the re-normalisation problem proved to be a dead-end due
to a loss of unitarity of the S-Matrix. Adopting the covariant
perturbative approach to quantisation, one expands about a flat
metric\footnote{The expansion has only a finite number of terms, so
$h$ need not be small. The only assumption is that space-time has
topology $\mathbb{R}^d$ and is asymptotically flat.} $g^{\mu\nu} =
\eta^{\mu\nu}+h^{\mu\nu}$. Then the field $h^{\mu\nu}$ is treated
like a spin $2$ field in Minkowski space. The Einstein-Hilbert term
contains only two derivatives, the quadratic term contains four. The
higher derivatives dominate at high frequency and potentially render
the theory re-normalisable. The higher derivatives are also
responsible for the appearance of ghosts. To see the problem, one
need look at the propagator. Let us do the calculation.
\begin{align*}
R & = \Gamma^{\mu\nu}_{\ \ \nu,\mu} - \Gamma^{\mu\nu}_{\ \ \mu,\nu}+
\Gamma^{\mu}_{\lambda\mu} \Gamma^{\lambda\nu}_{\ \
\nu}-\Gamma^{\mu}_{\lambda\nu} \Gamma^{\lambda\nu}_{\ \ \mu},
\\\nonumber
\Gamma^{\alpha}_{\beta \gamma} & =
\frac{1}{2}(\eta+h)^{\alpha\lambda}(h_{\beta\lambda,\gamma}+h_{\lambda\gamma,\beta}
-h_{\beta\gamma,\lambda}).
\end{align*}
The comma denotes partial differentiation. The diffeomorphism
invariance allows one to impose the harmonic gauge fixing condition
$h^{\mu\nu}_{\ \ ,\,\nu} = \frac{1}{2}h^{,\,\mu}$. Let $\doteq$ mean
equality in harmonic gauge.
\begin{gather*}
\sqrt{-g}R \doteq -\frac{1}{4}h^{\mu\nu}\Box
h_{\mu\nu}-\frac{3}{8}h\Box h +\partial(\cdots)+ O(h^3).
\end{gather*}

There will be additional quadratic terms in $h^{\mu\nu}$ coming from
the curvature-squared terms in
(\ref{Curv_squared})~\cite{Zwiebach-85}:
\begin{gather*}
L_2 \doteq (a+4)h^{\mu\nu}\Box^2 h_{\mu\nu} + (b-1) h \Box^2
h+\partial(\cdots) + O(h^3).
\end{gather*}
The higher order terms in the field $h^{\mu\nu}$ do not affect the
propagator. They will contribute to the self interaction of gravity
in perturbation theory.

The propagator is:
\begin{align*}
\left[-\alpha_1\left\{\frac{1}{4}\eta_{\mu\alpha}\eta_{\nu\beta}+\frac{3}{8}\eta_{\mu\nu}
\eta_{\alpha\beta}\right\} \Box + \alpha_2\Big\{
(a+4)\eta_{\mu\alpha}\eta_{\nu\beta}+(b-1)\eta_{\mu\nu}\eta_{\alpha\beta}\Big\}\Box^2\right]^{-1}.
\end{align*}

It is important to note that the fourth derivative contribution to
the propagator vanishes if and only if $a=-4$, $b=1$. In four
dimensions, there is the Gauss-bonnet identity: The term
\begin{gather}\label{GBhello}
R^2-4R_{\mu\nu}R^{\mu\nu}+R_{\mu\nu\alpha\beta}R^{\mu\nu\alpha\beta}
\end{gather}
is actually a total derivative and does not contribute to the local
degrees of freedom of the theory at all. In higher dimensions, this
term is not a total derivative. It contributes to the graviton
self-interactions but not the propagator.

In momentum space, the components of the propagator are of the form
\begin{gather}
\frac{-1}{k^2 + \lambda k^4}
\end{gather}
with $\lambda \sim \alpha_2/\alpha_1$. If we expand this as:
\begin{gather}
\frac{-1}{k^2} +\frac{\lambda}{1+\lambda k^2},
\end{gather}
then whatever the sign of $\lambda$, the $k^2$ in the second term
comes with the wrong sign. This wrong sign indicates that there are
propagating ghosts in the quantum theory.

The appearance of ghosts is a quite generic feature of higher
derivative theories. To get more insight into this, consider for
simplicity a scalar field instead  of a spin 2 field. A theory with
higher derivatives $(\phi,\Box \phi, \Box^2 \phi)$ can always be
re-cast as second order by field redefinitions, say $\psi_i :=
\psi_i(\phi,\Box \phi)$, so that the fields are $(\psi_i, \Box
\psi_i)$. The higher derivatives introduce extra particles. For the
higher derivative gravity, it turns out that there is always a
massive graviton with the wrong sign in the propagator, which spoils
the physical interpretation of the theory. Only the Gauss-Bonnet
combination introduces no extra particles. There is just the
massless graviton of the normal theory. Consider the (scalar) field
theory described by the higher derivative Lagrangian:
\begin{gather*}
 {\cal L}= \alpha_1 \phi \Box \phi + \alpha_2 \phi \Box^2 \phi
 + (\text{interactions}).
\end{gather*}
The ``kinetic" term is:
\begin{gather*}
\alpha_2 \phi \Box(\Box + \alpha_1/\alpha_2) \phi.
\end{gather*}
To bring this into a second derivative form, make the field
redefinitions:
\begin{gather*}
\psi_1 = \frac{\alpha_2}{\alpha_1}\Box \phi, \qquad \psi_2 =
\frac{\alpha_2}{\alpha_1}(\Box + \alpha_1/\alpha_2)\phi.
\end{gather*}
Then, up to total derivatives:
\begin{align*}
\alpha_2 \phi \Box(\Box + \alpha_1/\alpha_2) \phi & =
\alpha_1(\psi_2 - \psi_1) (\Box + \alpha_1/\alpha_2)\psi_1
\\ & = \alpha_1(\psi_2 - \psi_1) \Box \psi_2
\\ & = \frac{\alpha_1^2}{\alpha_2}\psi_1\psi_2.
\end{align*}
Taking a linear combination of these, we get, up to total
derivatives, for the kinetic terms:
\begin{gather*}
\alpha_1\psi_2 \Box \psi_2 - \alpha_1 \psi_1 (\Box +
\alpha_1/\alpha_2)\psi_1.
\end{gather*}
We have a massless particle plus another, massive particle. This
massive particle inevitably comes with a minus sign in the kinetic
term relative to the massless particle.

The interaction terms will be products of the linearly independent
combinations
\begin{align*}
\phi= \psi_2 -  \psi_1, \qquad \Box\phi =
\frac{\alpha_1}{\alpha_2}\psi_1.
\end{align*}
The two particle species will interact with each other. Classically,
there are two fields, one with positive energy and one with
negative. Although the choice of sign is arbitrary, the relative
minus sign is significant, the interactions causing both fields to
grow exponentially. The same argument applies for the spin 2 field.
From the point of view of our perturbative expansion of the metric
about flat space-time, it means that Minkowski space is unstable.

If $\psi_1,$ $\psi_2$ are quantum fields, the negative energy
interpretation can be avoided by choosing $\psi_2$ to be a ghost
field, a boson which has Poisson bracket replaced by anti-commutator
instead of commutator. Then, the energy is positive but a state
representing an odd number of particles has negative norm. The
negative probabilities make the physical interpretation very
problematic. The violation of the spin-statistics law means that
either the  S-Matrix is non-unitary or the principle of local
causality is violated. These are two very important physical
principles underlying quantum field theories.

We are left with two cases: \\1) Gauss-Bonnet combination:
non-renormalisable but makes sense as a low energy effective theory,
Minkowski space is stable.
\\ 2) Any other curvature-squared combination:
In principle re-normalisable in four dimensions\footnote{This is for
pure gravity. Couplings to matter fields would not be expected to be
re-normalisable.}, but inevitable presence of ghosts makes the
theory unphysical.
\\

In four dimensions, the Gauss-Bonnet combination is locally a total
derivative. It does not affect the local degrees of freedom.
Thinking now of the path integral approach to quantisation, it
merely contributes a topology-dependent weight to the functional
integral.

In higher dimensions, although the ``kinetic term" is a total
derivative, the Gauss-Bonnet term does contribute to the
self-interactions of the graviton. In higher dimensions then, the
Gauss-Bonnet is the only curvature squared term, containing the
metric and its derivatives, which does not suffer from ghosts, and
furthermore, the particle content is the same as GR: a massless spin
two particle.
\\

It is worthwhile to look at modifications to GR. New gravitational
physics may have something to say on: the cosmological constant
problem; the dark matter problem; black holes and the problems
associated with them such as unitarity, entropy~\cite{Myers-98}.

A particularly interesting higher curvature theory is the Lovelock
gravity. This is a generalisation of General relativity in a special
sense. The Einstein-Hilbert action is the dimensional continuation
of the two-dimensional Euler Characteristic. The Gauss-Bonnet term
(\ref{GBhello}) is similarly related to the 4-dimensional Euler
Characteristic. The general Lovelock action is a sum of terms which
are similarly the dimensional continuation of Euler Characteristics
of each even dimension. Aside from a cosmological constant, the
other Lovelock terms are only non-zero in higher than four
dimensions.

If we are looking for terms which are some low energy expression of
a fundamental quantum theory of gravity in higher dimensions, such
as String Theory, which is unitary and which has flat space as a
vacuum, the Gauss-Bonnet term is a natural one to consider. Thinking
of purely gravitational terms, other curvature-squared combinations
would not be expected to arise~\cite{Zwiebach-85,Aragone-87}.
However, there would be other fields coupling to gravity in various
ways. Also, generically, curvature cubed terms and higher do not
contribute to the propagator. The cubic Lovelock term is not special
in this sense.

There are many nice properties related to the geometrical origins
and quasi-linearity property which will be discussed in chapter
\ref{Einstein_Lovelock}, for example, the Cauchy problem is very
similar to GR.\\

The Lovelock theories have been studied extensively. Higher
dimensional black holes have been
studied~\cite{Boulware-85,Banados-94,Cai-02, Clunan-04}. This has
shed some interesting light on questions of black hole entropy.

Some cosmological metrics have been studied~\cite{Madore-86,
Deruelle-90}. In particular, the Lovelock contributions, motivated
by string theory, have played a big role in studying brane-world
cosmology\footnote{some of the many works are
~\cite{Mavromatos-00,Deruelle-00,Charmousis-02,Germani-02,Mavromatos-02,Davis-02,Gregory-03}}.
Such higher curvature terms, as well as other stringy fields like
dilaton and moduli, have led to all sorts of possibilities. It has
been suggested that the Gauss-bonnet gravity can account for
cosmological expansion without the problem of missing dark
matter~\cite{Dehghani-04}. Localisation of gravity in large extra
dimensions has also been realised in this context.

The discovery by Chamseddine~\cite{Chamseddine-89} that certain
Lovelock theories are equivalent to Chern-Simons gauge theories of
the de Sitter or anti-de Sitter gauge group has also generated
interest~\cite{Zanelli-00}.

Apart from anything else, as a mathematical theory with close connections to
interesting geometry, the Lovelock theory deserves some attention.
It should be stressed that the motivations for study are all theoretical.
There is no experimental evidence as yet for such modifications to gravity.
\\

The special relevance of Lovelock terms to intersecting braneworlds
was realised by Kim, Kyae and Lee~\cite{Kim-01a,Kim-01b}. Very
recently after the work presented here was first announced,
intersecting brane-worlds in pure Lovelock gravity have been studied
~\cite{Lee-04,Navarro-04} by others and an example has been found of
a 4-dimensional intersection in 10 bulk dimensions. I hope that this
work, being a general treatment of intersections, will be useful in
such studies.

\section{Summary of contents}

In the second chapter I will discuss the Lovelock Theories of
Gravity in higher dimensions.

In the third chapter the singular hypersurfaces method of Israel is
introduced. The notion of distributional curvature is discussed
along with the problems with extending these concepts to the higher
curvature theories. The study of boundary terms in gravitational
actions is also introduced. The possibility of using boundary terms
in the Lagrangian to describe membranes is discussed and the method
is compared to integrating the field equations over distributional
sources.

In the fourth chapter I present the results of [hep-th/0306220]
co-authored with E. Gravanis. First, I deal with the Euler density
of manifolds containing a certain kind of discontinuity in the
geometry. We found a chain of terms, starting with the Euler density
in $2n$ dimensions, obeying a simple relation: A sum of terms gives
the total derivative of another term. By a simple inductive
argument, it is shown that these terms `live' on the intersection of
hypersurfaces of discontinuity in the geometry.

I then generalise to the dimensionally continued Euler density. The
reason for studying the discontinuities is made clear. It
corresponds to the Israel type junction and distributional sources.
We show that it is possible to construct an action which gives all
the junction conditions for the intersections, as well as the
gravitational field equations.

In the fifth chapter I present the results of [gr-qc/0401066], again
co-authored with E. Gravanis. The problem is re-formulated in terms
of the simplices dual to the intersections. Again turning to the
dimensionally continued Euler density, the action constructed in
chapter four is further justified by Lemma \ref{Finalaction}.


Chapter six has examples illustrating the possible physical
applications.

In the final chapter, the possibilities for future research and
possible interpretations of the geometrical features are discussed.

Mathematical details, definitions of simplices and a glossary
are in the appendices.




\chapter{Einstein and Lovelock Gravity}\label{Einstein_Lovelock}

\section{General Relativity}

Gravity, via the \emph{equivalence principle}, leads to relative
acceleration between local Lorentz frames at different places. In
other words, space-time is curved. There is no global inertial
frame. We have to do away with rigid Minkowski space in favour of a
dynamical manifold. In particular, space-time is no longer a vector
space. There is no meaningful way to define or add displacement
vectors. Instead vectors live on the tangent bundle, corresponding
to infinitesimal directional derivatives.

Matter and (non-gravitational) energy, in the form of the Energy-momentum
tensor is the source of the curvature via the famous Einstein Equation. This
is a form of
{\it Mach's principle}, the definition of inertial frames being dependent on
the matter-energy content of the universe.
The Einstein Equations are
\begin{gather}
T_{\mu\nu}=G_{\mu\nu}
\end{gather}
or, including the cosmological constant,
\begin{gather}
T_{\mu\nu}=G_{\mu\nu} + \Lambda g_{\mu\nu},
\end{gather}
where $G^{\mu\nu}$ is the Einstein tensor constructed solely from
the metric tensor and it's derivatives:
\begin{gather}\label{EinsteinTensor}
G_{\mu\nu} = R_{\mu\nu}-\frac{1}{2}Rg_{\mu\nu}.
\end{gather}
$g_{\mu\nu}$ is the metric which gives us a measure of distance.
$R_{\mu\nu}$ is the Ricci Tensor. It is derived from
the Riemann curvature tensor $R^\alpha_{\mu\beta\nu}$
\footnote{I use the convention $R^{\alpha}_{\mu\beta\nu}
\equiv \partial_\beta \Gamma^\alpha_{\mu\nu} + \Gamma^\alpha_{\lambda\beta}
\Gamma^\lambda_{\mu\nu} - (\beta \leftrightarrow \nu)$} by contraction
of first and third indices.
The Riemann tensor is constructed from the metric compatible,
torsion-free connection coefficients $\Gamma^\alpha_{\beta\gamma}$.
$\Gamma^\alpha_{\beta\gamma}$ are known as the Levi-Civita connection
coefficients or Christoffel Symbols. The metric compatible connection
defines the parallel transport of vectors such that their metric
products (norm and angle) are preserved.
The Riemann tensor measures the non-commutativity
of the associated covariant derivative.
$R = R^\mu_\mu$ is the
Ricci scalar.
\\

Einstein's equations (with cosmological constant) obey three very
important principles~\cite{Zanelli-02}:
\\\\
1) they are independent of reference frame determined by
choice of co-ordinates,
\\\\
2) there is well defined Cauchy problem for the evolution of the
metric tensor,\footnote{for globally hyperbolic space-time (see Wald
Ch 8,10~\cite{Wald-84})}
\\\\
3) they reduce to Newtonian Gravity in the weak field,
non-relativistic case.
\\

Condition 1 is satisfied because Einstein's equations constitute
a tensor relation.

Condition 2 means that the Cauchy conditions are necessary and sufficient
to integrate the vacuum equations.
Specifying the field and its first derivative on an initial Cauchy surface
will fully determine the time evolution of the field.
This means that there exists a Hamiltonian formulation. If higher derivatives
need to be specified, one can always redefine them as new variables and the
theory is again second order. But we then
have new observables and the theory is qualitatively different from the
Newtonian physics. The form of the Einstein tensor ensures that the metric has
a well defined Cauchy problem. Whether the whole Einstein equations have such
a well-defined initial value problem will depend on the matter content.

Condition 3 is in accord with measurement from everyday physics to
celestial mechanics that all non-relativistic, classical, weak
gravity systems obey Newtonian physics. As an example consider a
spherically symmetric Schwarzschild solution around a spherical mass
distribution. The weak field condition means $ r/r_0 >> 1$ where $r$
is the radial co-ordinate and $r_0$ is the Schwarzschild radius. In
this limit, the Newtonian physics is indeed
recovered~\cite{Wald-84}.

If we demand in addition that Minkowski space be a vacuum solution
then the Cosmological term is eliminated (or very small).

The Einstein tensor (\ref{EinsteinTensor}) is the only tensor
that is:
\\\\
A) Symmetric,\\\\
B) Covariantly conserved: $\nabla_\mu G^\mu_\nu =0$,\\\\
C) depending only on the metric, its first and second derivatives,
\\\\
D) linear in second derivatives of the metric.\\\\
As such it is a suitable ingredient for the field equations and
obeys the three physical conditions above.

The importance of conditions (A) and (B) is apparent since an energy-momentum
tensor, coming from the variation of a matter lagrangian
with respect to the metric, is both symmetric and divergence free.
These conditions are automatically satisfied by the existence of the
Lagrangian for gravity. In other words, if the whole matter-gravity
system comes from a lagrangian, then conditions (A) and (B) are
satisfied for both sides of the Einstein equation. The action, which
yields the Einstein field equations, is known as the Einstein-Hilbert
action:
\begin{gather}
{\cal S} = \int_M d^4x \sqrt{g}\, R,
\end{gather}
$g \equiv \det{g_{\mu\nu}}$.
Condition (B), the contracted Bianchi Identity, gives the local
conservation of energy and momentum.

Conditions (C) and (D) are important for the physical conditions (2)
and (3). This second order condition is not absolutely essential if
we are describing an effective theory for the low energy of some
more fundamental theory. However, the second order field equations
are important for the classical stability of solutions and are the
most natural choice by analogy with classical mechanics.

\section{Lovelock gravity}

In higher dimensions, there are other tensors admissible if
condition D is relaxed to quasi-linearity~\cite{Deruelle-03}. The
definition of quasi-linearity is basically that there are no squared
or higher order terms in second derivatives of the metric with
respect to a given direction.

So for our higher dimensional theory we have the following
reasonable conditions:
\\\\
i) The classical equations of motion are equivalent to a principle
of extremal action;
\\\\
ii) The Euler variation with respect to the metric produces a tensor
depending only on the metric, its first and second derivatives;
\\\\
iii) The tensor is quasi-linear in second derivative terms.
\\

The tensor quadratic in the Riemann tensor was found by Lanczos. All
such tensors were found by Lovelock~\cite{Lovelock-71} along with
the corresponding action.

\begin{gather}\label{Lovelocktensor}
H^{\mu}_{\nu} = -\sum_{n=0}^{[d/2]}\frac{1}{2^{n+1}}
\beta_n\delta^{\mu\mu_1... \mu_{2n}}_{\nu\nu_1...\nu_{2n}}
R^{\nu_1\nu_2}_{\mu_1\mu_2}\cdot\cdot\cdot
R^{\nu_{2n-1}\nu_{2n}}_{\mu_{2n-1}\mu_{2n}},
\\
{\cal S} = \sum_{n=0}^{[d/2]}\int_M \frac{1}{2^n}\beta_n
\delta^{\mu_1... \mu_{2n}}_{\nu_1...\nu_{2n}}
R^{\nu_1\nu_2}_{\mu_1\mu_2}\cdot\cdot\cdot
R^{\nu_{2n-1}\nu_{2n}}_{\mu_{2n-1}\mu_{2n}}
\sqrt{g}\,d^dx.\label{Lovelockscalar}
\end{gather}
The delta is the generalised totally anti-symmetrised Kronecker delta.
It is the determinant of a matrix with elements $\delta^M_N$,
\begin{gather*}
\delta^{\mu_1... \mu_{p}}_{\nu_1...\nu_{p}}
=det(\delta^M_N)=p!\delta^{\mu_1}_{[\nu_1}
\delta^{\mu_{p}}_{\nu_{p}]}
\end{gather*}
with the index $M$ running from $\mu_1,...,\mu_p$ and likewise
$N=\nu_1,...,\nu_p$.

The $n=0$ term in (\ref{Lovelockscalar}) is the Cosmological constant.
The $n=1$ term is the Einstein-Hilbert term. The $n=2$ term is proportional to:
\begin{gather*}
R^2 -4R_{\mu\nu}R^{\mu\nu}+R_{\mu\nu\alpha\beta}R^{\mu\nu\alpha\beta}
\end{gather*}
and is variously known a Gauss-Bonnet term, Lanczos term or Lovelock term.

The quasi-linearity can be seen from the totally anti-symmetric form
of the Kronecker delta in (\ref{Lovelocktensor}). The terms
containing products of second derivatives of the metric are of the
form:
\begin{gather*}
\delta^{\mu\nu\kappa\lambda \dots}_{\alpha\beta\gamma\delta\dots}
\partial_\mu \Gamma^{\alpha\beta}_{\ \ \nu}
\partial_\kappa \Gamma^{\gamma\delta}_{\ \ \lambda} \cdots
\end{gather*}
The anti-symmetry means that the derivatives are in orthogonal
directions. One of the consequences of quasi-linearity w.r.t. the
second time derivatives is that the Lovelock theory is ghost free in
perturbation theory about a flat
background~\cite{Zwiebach-85,Zumino-86}. On the down side, there are
some problems with the evolution of the classical solutions due to
the multiple solutions of polynomial equations~\cite{Deruelle-03}.

In $d=2n$, the integrand
\begin{gather}\label{GBintegrand}
\sqrt{g}\delta^{\mu_1... \mu_{2n}}_{\nu_1...\nu_{2n}}
R^{\nu_1\nu_2}_{\mu_1\mu_2}\cdot\cdot\cdot
R^{\nu_{2n-1}\nu_{2n}}_{\mu_{2n-1}\mu_{2n}}
\end{gather}
goes by the name of the Lipschitz-Killing curvature. This tensor
appears in the Gauss-Bonnet formula for 2n dimensions (Appendix
\ref{Eulerappend}). This similarity with the Gauss-Bonnet formula
explains many of the interesting properties of the
theories~\cite{Patterson-81,Zumino-86}. It will be especially
important for our study of intersecting hypersurfaces.

\section{Orthonormal frames}

Rather than working with tensors, we will use the method of exterior
differential forms. The method replaces metric and Christoffel
symbol degrees of freedom with orthonormal frame (or vielbein) and
spin connection (or connection 1-form). This would be essential when
dealing with spinors on curved space-time. We will not be interested
in spinors but the formalism will be useful. It leads to quite an
abstract approach. We shall see returns on this investment in
mathematics in the end when the formulae will become very simple.
Another reason for using frames is that we can choose to restrict
ourselves to frames {\it adapted} to an embedded submanifold.

The frame field over a $d$-dimensional manifold is the map from an
orthonormal basis of sections of the tangent bundle to a basis of
$\mathbb R^d$. In English, the vielbein is a non-co-ordinate basis
which provides an orthonormal basis for the tangent space at each
point on the manifold.
\begin{gather*}
g_{\mu\nu}dx^\mu \otimes dx^\nu
= \eta_{ab} E^a\otimes E^b
\end{gather*}
Strictly speaking $E^a$, being one-forms, are the {\it co}-frames,
but I will not distinguish.

In the vicinity of an embedded submanifold of co-dimension $d$, we
can adapt the frame field to the submanifold. Split the Minkowski
space into $\mathbb{R}^{d-p} \otimes\mathbb{R}^p$. $\mathbb{R}^p$
gets mapped to the normal bundle by the frame field. So we can
choose $E^i$ to be tangential directions and $E^\lambda$ to be
normal say, where $i = 1,...,d-p$ and $\lambda = d-p+1,...,d$.

We also introduce the metric compatible Lorentz connection and
associated connection one-form $\omega^a_b$ (or spin connection).

We shall make use of exterior differential calculus.
Exterior differential forms and exterior calculus are reviewed in
the Appendix. The action will now be built out of the vielbein and
connection and their exterior derivatives.

The curvature is, according
to the first of Cartan's structure equations:
\begin{gather}
\Omega^a_b = d\omega^a_b +\omega^a_c\wedge\omega^{cb}.
\end{gather}
The second of Cartan's structure equations defines the Torsion:
\begin{gather}
T^a = dE^a + \omega^a_b\wedge E^b = DE^a.
\end{gather}
In GR the torsion tensor is constrained to vanish. When this
constraint is not imposed, we have the Einstein-Cartan theories.
When the torsion is non-vanishing, parallel transport
and distance are independent of each other.

The exterior derivative of the structure equations gives the Bianchi
identities.
\begin{align}
D \Omega^a_b & = d\Omega^a_b +\omega^a_c \wedge \Omega^c_b -
\Omega^a_c\wedge \omega^c_b = 0,
\\
DD E^a & = dT^a + \omega^a_b\wedge T^b = \Omega^a_b \wedge E^b.
\end{align}

The Einstein-Hilbert action can be written compactly as
\begin{gather*}
{\cal L} = \alpha_1\,\Omega^{ab}\wedge E^c
\wedge E^d\epsilon_{abcd}
\end{gather*}
$\epsilon_{abcd}$ is the totally anti-symmetric Levi-Civita tensor
density such that $\epsilon_{0123} = +1$.

When finding the equations of motion, we make use of:
\begin{gather*}
\delta \Omega^{ab} = D(\delta \omega)^{ab}
\end{gather*}
The explicit Euler variation with respect to the connection is
\begin{align*}
\delta_{\omega} {\cal L} & = \alpha_1\, D(\delta \omega)^{ab}\wedge
E^c
\wedge E^d\epsilon_{abcd}\nonumber\\
& = d \left(\alpha_1\, \delta \omega^{ab}\wedge E^c \wedge
E^d\epsilon_{abcd}\right) - 2\alpha_1\, \delta \omega^{ab}\wedge T^c
\wedge E^d\epsilon_{abcd}
\end{align*}
The first term is a total derivative and does not contribute to the
equations of motion. The second term will vanish if the torsion is
zero. Thus, unless there are spinors in the matter side of the
action coupling to the spin connection, or the $\det$(vielbein)
vanishes, the vanishing of the torsion is actually an equation of
motion. The Einstein field equation comes just from the explicit
variation with respect to the vielbein.
\begin{gather*}
\delta_E {\cal L} = \alpha_1\, \delta E^c{\cal E}^{(1)}_c,
\\
{\cal E}^{(1)}_c = 2\Omega^{ab}\wedge E^d\epsilon_{abcd}
\end{gather*}
In the familiar tensor language, the Einstein Tensor comes only from
the variation of the metric, where $g^{\mu\nu}$ and
$\Gamma^{\alpha}_{\beta\gamma}$ are varied independently. This is the
Palatini formulation.

The Lovelock action can also be written very compactly in the
differential form language. The action is:
\begin{gather}\label{Lovelockforms}
{\cal L}(\omega,E) = \sum_n \beta_n\, \Omega^{a_1...a_{2n}}
\wedge e_{a_1...a_{2n}},
\end{gather}
where $\Omega^{a_1...a_{2n}}\equiv \Omega^{a_1a_2}\wedge\cdot\cdot
\cdot\Omega^{a_{2n-1}a_{2n}}$ is the $n$-fold wedge product of
curvature two-forms and
\begin{gather}
e_{a_1...a_r}\equiv \frac{1}{(d-r)!}\epsilon_{a_1...a_d}
E^{a_{r+1}}\wedge\cdot\cdot\cdot E^{a_d}.
\end{gather}

The Euler variation with respect to the spin connection will again
vanish (up to a total derivative) if the torsion vanishes:
\begin{align*}
\delta_{\omega} {\cal L} & = \sum_n \beta_n D(\delta
\omega)^{a_1a_2}\Omega^{a_3...a_{2n}} \wedge e_{a_1...a_{2n}}
\\\nonumber & = \sum_n \beta_n \left\{
d\left(\delta \omega^{a_1a_2}\Omega^{a_3...a_{2n}}
\wedge e_{a_1...a_{2n}}\right) - \delta \omega^{a_1a_2}
\Omega^{a_3...a_{2n}}\wedge T^{a_{2n+1}} e_{a_1...a_{2n+1}}\right\}
\label{varytot}
\end{align*}
but now there are also other solutions for $det(e) \neq 0$. If we
require zero torsion it must be imposed as a constraint. Assuming
this constraint, again the gravitational field equation comes from
the explicit variation with respect to the vielbein:
\begin{gather}
\delta_E {\cal L} = \sum_n \beta_n\, \delta E^{b}\wedge
\Omega^{a_1...a_{2n}}\wedge e_{a_1...a_{2n}b}.
\end{gather}
\\

We can once again reformulate our guiding principles
~\cite{Zumino-86,Regge-86,Mardones-91}
\\\\
a) The Lagrangian is invariant under Local Lorentz transformations,
constructed from vielbein, spin-connection and their exterior
derivatives, without reference to Hodge duality.
\\\\
b) The torsion is zero.
\\\\
One can also generalise to a Lovelock-Cartan theory with torsion
degrees of freedom~\cite{Mardones-91}.

\section{Some notation}

Sometimes it will be more convenient to write
\begin{gather*}
{\cal L}(\omega,E) = \sum_n \alpha_n\, \Omega^{a_1...a_{2n}}
\wedge E^{a_{2n+1}...a_d} \epsilon_{a_1...a_{2n}}
\end{gather*}
where $E^{a_{2n+1}...a_d}\equiv E^{a_{2n}}\wedge\cdot\cdot\cdot
E^{a_d}$ is the $(d-2n)$-fold wedge product of vielbein frames. It
should be remembered that the coefficients differ by a factor:
\begin{gather*}
\beta_n = (d-2n)!\alpha_n.
\end{gather*}

In much of the following chapters I will suppress the indices
by writing
\begin{gather}\label{notation}
f(\psi) \equiv \psi^{a_1\dots a_d}\epsilon_{a_1\dots a_d}
\end{gather}
for $\psi$ a d-form. For example:
\begin{gather*}
f(\Omega^n E^{d-2n}) = \Omega^{a_1...a_{2n}}
\wedge E^{a_{2n+1}...a_d} \epsilon_{a_1...a_{2n}}.
\end{gather*}
The wedge notation has also been suppressed.

\section{Dimensionally continued Euler densities}\label{topapproach}

As mentioned earlier, the Lovelock action is closely related to the
Euler Characteristic. The Euler Characteristic Class is reviewed in
the appendix. The important relation for us is the following: Let
$g$ be a metric on a $2n$ dimensional manifold, $M$, of a certain
topology. Let $g'$ be another metric over $M$. Let $\omega$,
$\omega'$ be the torsion free, metric compatible, connection 1-forms
associated with $g$ and $g'$ respectively. Then
\begin{gather}\label{Transgress}
f(\Omega(\omega)^n)-f(\Omega(\omega')^n) = n d \int_0^1 dt \,
f(\theta\,\Omega(t)^n).
\end{gather}
$t$ is some parameter interpolating between the $1$-form fields
\begin{gather*}
\omega(t) \equiv \omega + t \theta,
\qquad \theta \equiv \omega-\omega',\\
\Omega(t)= d\omega(t) +\omega(t)\wedge\omega(t).
\end{gather*}
The difference is just a total derivative. The proof involves the
introduction of the homotopy parameter $t$. We shall make use of
this and a natural generalisation to study intersections in later
chapters.

The Lovelock action (\ref{Lovelockforms}) is a sum of terms
very similar to the Euler density. There is just that extra
factor of $E^{(d-2n)}$. There are many nice features inherited from the
Euler density. In particular, the variation of the connection,
when restricted to infinitesimal variation, is still a total derivative,
as we saw in equation (\ref{varytot}).

\chapter{The junction conditions}\label{junction}

The action principle for General Relativity on a space-time manifold
can be generalised to a manifold with boundary. The inclusion of a
certain boundary term (Gibbons-Hawking) makes the action principle
well defined on the boundary. Also, singular hyper-surfaces of
matter~\cite{Israel-66} can be incorporated into a suitably smooth
space-time. We will see that these are part of the general
properties of actions built out of dimensionally continued
topological invariants.

\section{Hypersurfaces}\label{Hypersection}

I will use the following definitions throughout:
\begin{definition}$M$ is a $d$-dimensional manifold (usually assumed to
have a Lorentzian metric). A {\it  hypersurface}, $\Sigma$, is an
embedded $d-1$ dimensional submanifold. The {\it bulk} is the
compliment of the hypersurface in $M$, $M- \Sigma$. Hypersurfaces
typically divide $M$ into disconnected bulk regions.\end{definition}

Let us think of a simple situation. We have an
oriented hypersurface
$\Sigma$ in Euclidean (or Minkowski) space. The position
of the hypersurface is defined by the constraint:
$f({\bf x}) = 0$. There is a normal vector at each point
on $\Sigma$.
The normal vector can be thought of as a vector in
the Euclidean space. In other words it is assumed to live
in the same vector space as the position vectors.

In General Relativity space-time is intrinsically curved, with the
curvature determined by matter content. as mentioned in the previous
chapter, this means we have to do away with rigid Minkowski space in
favour of a manifold. Not only do the positions now live in a more
exotic object- a manifold, but the tangent vectors now live on the
tangent bundle. In particular, the normal vector of the hypersurface
lives on the tangent bundle.

The non-Euclidean nature of the geometry makes the study of
hypersurfaces more complicated. One useful tool is the Riemann
normal co-ordinate system. On a local co-ordinate neighbourhood, we
can always choose co-ordinates such that a smooth hypersurface is
located at $x^d=0$~\cite{Kobayashi,Misner}. It is not always
convenient to work with this Riemann normal co-ordinate system.
Sometimes the bulk solution will have a much simpler form in another
co-ordinate system. Alternatively, the adapted frames introduced in
the previous chapter may be useful.

We have a hypersurface $\Sigma$, which can be defined by $\phi({\bm
x})=0$. In a surrounding region D we define a vector field ${\frak
n}$ which coincide with normals on $\Sigma$. ${\frak n}_\mu \propto
\partial_{\mu} \phi$ Using the Riemann normal co-ordinates mentioned
above, the integral curves will be parameterised by some normal
co-ordinate $w (= \phi)$.

We define the induced metric (for a non-null hypersurface):
\begin{gather}
h_{ab} = g_{ab} - \frac{{\frak n}_a{\frak n}_b}{{\frak n}
\!\cdot\!{\frak n}}
\end{gather}
and we will assume ${\frak n}$ is normalised to
${\frak n}\!\cdot\!{\frak n} = \pm 1$ according to whether the
hypersurface is time-like or space-like.
In terms of ${\frak n}$ and $h$, the extrinsic curvature is:
\begin{gather}
 K_{ab} = -h_a^c\nabla_c{\frak n}_b
\end{gather}
The details are in Appendix \ref{hyperappendix}.
\begin{figure}
\begin{center}\mbox{\epsfig{file=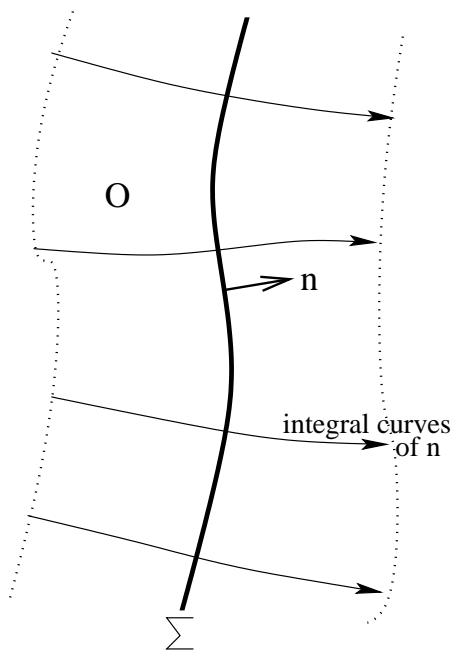, width=6cm}}
\caption{{\small D is a region of space-time around the hypersurface,
$\Sigma$.}}
\label{hyper}
\end{center}
\end{figure}

The extrinsic curvature tells us how the hypersurface is embedded
into the bulk manifold. In flat space-time, this is purely determined by the
intrinsic geometry of $\Sigma$. In non-Euclidean geometry there is
the interesting possibility that the extrinsic curvature can actually be
different on each side.

The theory of singular hypersurfaces was introduced by
Israel~\cite{Israel-66}. The Energy momentum tensor is assumed to be
of the form:
\begin{gather*}
T^{ab}_{{\rm Bulk}} + \tilde{T}^{ab} \delta(x \in \Sigma)
\end{gather*}
so that $\tilde{T}^{ab}$ is singular with respect to the normal
direction.

The Israel Junction condition relates the energy-momentum of the wall
to the discontinuity or {\it jump} of the extrinsic curvature:
\begin{gather}
\tilde{T}^{ab} = -[K^{ab}-K h^{ab}]\equiv -(K^{ab}-K h^{ab})|^{\Sigma_+}_{\Sigma_-},
\end{gather}
$\Sigma_\pm$ being the right or left hand side respectively.
This is the singular part of the tangential-tangential component of the
Einstein equation. The normal-normal and normal-tangential components
of $S^{ab}$ must vanish.

As discussed in chapter \ref{Einstein_Lovelock}, the matter content
is related to the geometry. In the thin shell formalism, with
$T^{ab}_{{\rm Bulk}}=0$, this is realised in a simple way. Since the
Einstein Tensor is only non-zero at the location of the
hypersurface, it merely determines the way in which the geometry of
the two sides are to be matched together. It is not, however, as
simple as matching flat regions. One has to match vacuum regions
where the Weyl Tensor (the part of the curvature not appearing in
Einstein's equation) may not be zero. This part of the curvature
depends, through the Bianchi Identity upon the global details of the
matter-energy distribution such as the shape of the hypersurface
itself.

This formalism gives a well defined mathematical treatment of
singular walls in GR. As an aside, I note that the initial value
problem for hypersurface singularities was studied by
Clarke~\cite{Clarke-98} and by Vickers and Wilson~\cite{Vickers-01}.

There are some strange features of bubbles enclosed by singular
shells in non-Euclidean space-time. For example, a bubble of de
Sitter space inside a region of Schwarzschild space-time could be
observed from just inside to be expanding while to an observer just
outside it would be contracting~\cite{Blau-87}. The reason is that a
wall of a given intrinsic shape can be embedded in a very different
way into the half-manifolds on each side.


We know that the effect of gravity is that, as we move along in space,
the local inertial frames are relatively accelerated. If there is a smooth
matter distribution
this  will be a continuous acceleration. If you step through a wall of
singular mass distribution, there is a sudden jump in the acceleration.
Imagine a man geodesically moving across such a wall.
The geodesic is described by:
\begin{gather*}
\ddot{x}^\mu +\Gamma^\mu_{\alpha\beta}\dot{x}^\alpha\dot{x}^\beta =
0.
\end{gather*}
The 4-velocity $\dot{x}$, should be assumed continuous.
The acceleration $\ddot{x}$
will be suddenly changed because the connection $\Gamma$ is discontinuous.
There would be a tidal force between the man's
front leg on one side of the wall and rear leg on the other side.

The above are curious properties of {\it co-dimension 1} surfaces.
When there is matter of co-dimension greater than one, a different
geometrical feature becomes possible. If we parallel transport a
vector in $T(M)$ around some infinitesimal closed loop about our
matter source, the final vector can be different from the initial.
This type of curvature singularity with the localised holonomy is
not possible with a co-dimension one brane since such a loop will
always cut the brane once in each direction.

However, such possibilities seem to lead to ambiguities in the
non-linear theory of GR, as was shown by  Geroch and
Traschen~\cite{Geroch-87}. The hypersurface is the only kind of
singular source possessing a curvature tensor that is well defined
as a unambiguous limit of a smooth concentration of matter.

\section{Distributional curvature}\label{distributional}

The curvature tensor involves second derivatives of the metric.
Normally one would consider a metric that is at least $C^2$
(continuous, twice differentiable). This requires that the manifold
is $C^3$, i.e. the co-ordinate transformations between overlapping
charts is at least $C^3$. Then the curvature is a $C^0$ tensor. By
Einstein's Equations, this implies that the Energy-momentum Tensor
is also $C^0$. As mentioned in the introduction, Einstein himself
felt that space-time should be smooth and conceived of fundamental
particles as solitonic solutions of the field equations. We want to
consider a less smooth case. We require only that the metric is
piecewise $C^2$. At the hypersurface it is only $C^0$. The Riemann
curvature tensor may then be singular. This requires that the
differentiable structure of the manifold itself is piece-wise $C^3$
and $C^1$ at the hypersurface~\cite{Clarke-87}.

Taub~\cite{Taub-80} defined the distribution-valued curvature
tensors in such a way that the Bianchi identity was still satisfied
as an equation relating distributions.

In GR, the curvature of a singular hypersurface is well defined as a
distribution, as shown by Geroch and Traschen~\cite{Geroch-87}. The
singular solution, viewed as the limit of some non-singular matter
is independent of the way that the limit is taken. This is in
contrast to lower dimensional singular matter such as cosmic strings
and point particles\footnote{It was shown by
Garfinkle~\cite{Garfinkle-99} that co-dimension 2 sources can be
well defined in terms of distributional curvature, but they are not
the unique limit of smooth matter.}.

However, the tensor product of distributions may not generally be a
distribution. The field equations of higher order Lovelock theories
involve the product of curvature tensors. It is not obvious whether this
is well defined.

Instead of addressing the problem of distributional field equations,
for the Lovelock theory we shall just work with the Lagrangian. The
boundary terms in the Lagrangian will {\it define} the equations of
motion.

\section{Gravity actions and boundary terms}

General relativity can be generalised to a manifold with boundary.
The variational principle will however be a problem. The total
derivative terms in the Euler-Lagrange variation will give a
boundary term on $\partial M$. We would like to have the following
variational principle:
\begin{quote}
{\it For fixed $\delta h^{\mu\nu}$ on the boundary\footnote{Because
of diffeomorphism invariance, this is the same as fixing $\delta
g^{\mu\nu}$ on the boundary.} , the action is to be minimised.}
\end{quote}

Let us calculate the Euler-Lagrange variation
of a general gravity action, keeping track of the total derivative
terms.
We have an action functional built by contracting
products of the Riemann tensor with the metric tensor.
\begin{gather*}
S[g^{ab},\partial_cg^{ab},\partial_c\partial_dg^{ab}]
\end{gather*}
The Euler variation of the action with respect to the metric tensor
is (see Appendix \ref{Boundvar}):
\begin{gather}
\delta S=H^{ab}\delta g^{ab} + \partial_c V^c,
\end{gather}
\begin{gather}
H^{ab} \equiv \frac{\partial {\cal L}}{\partial g^{ab}}
-\partial_c
\frac{\partial {\cal L}}{\partial g^{ab}_{\ \ ,c}}
+\partial_d\partial_c\left(
\frac{\partial {\cal L}}{\partial g^{ab}_{\ \ ,cd}}\right),
\end{gather}
\begin{gather}\label{Vc}
V^c \equiv \delta g^{ab}\frac{\partial {\cal L}}{\partial g^{ab}_{\ \ ,c}}
-\delta g^{ab}\partial_d\left(
\frac{\partial {\cal L}}{\partial g^{ab}_{\ \ ,cd}}\right)
+\partial_d \delta g^{ab}
\frac{\partial {\cal L}}{\partial g^{ab}_{\ \ ,cd}}.
\end{gather}
$V^c$ is the term which will appear on the boundary.
If there were only terms
proportional to $\delta g^{\mu\nu}$ and its derivative {\it along} the
boundary, there would be no problem. The problem arises when there are
normal derivatives of $\delta g^{\mu\nu}$ appearing on the boundary,
coming from the last term in (\ref{Vc}).
These will not be fixed just by fixing $\delta h^{\mu\nu}$ on
$\partial M$.

In GR the inclusion of a certain boundary action (Gibbons-Hawking)
cancels the normal derivatives of the metric variation. Thus the
action can be extremised whilst keeping only the surface geometry
fixed. This action was originally due to York~\cite{York-72} and was
proposed within the context of the Hamiltonian formalism.
It is also important for the path integral approach to quantum
gravity~\cite{Hawking-79}.

In all of these cases, the boundary term which fixes the action
is:
\begin{gather*}
2\, {\frak n}\!\cdot\!{\frak n} \int_{\partial M} d^{d-1}x  \sqrt{|h|}\, K
\end{gather*}

The generalisation of this boundary term to Lovelock gravity was
suggested by Myers~\cite{Myers-87}. It is simply related to the
dimensional continuation of the Transgression associated with the
Euler density (see appendix \ref{Eulerappend}). In terms of the
moving frames, the general boundary term for the nth order Lovelock
term is:\footnote{The expansion of this for $n=2$ in terms of
extrinsic and intrinsic curvature tensors is in \cite{Myers-87} but
there is a typo. The correct expression is in \cite{Davis-02}.}
\begin{gather}\label{Myersterm}
{\cal L}(\omega,\omega_0,e)\equiv n\, \beta_n\int_{\partial M}\int_0^1 dt\ \theta^{a_1a_2}\wedge\Omega(t)^{a_3a_4}\wedge
\cdot\cdot\cdot \wedge \Omega(t)^{a_{2n-1}a_{2n}}\wedge
e_{a_1...a_{2n}},
\end{gather}
where, as in Section \ref{topapproach},
\begin{gather*}
\omega(t) \equiv \omega_0 + t \theta,
\qquad \theta \equiv \omega-\omega_0,\\
\Omega(t) = {\bm d}\omega(t)+\omega(t)\wedge\omega(t).
\end{gather*}
It was argued by Myers~\cite{Myers-87} that the transgression
formula (\ref{Transgress}) means that variation w.r.t. $\omega$ with
$\omega'$ fixed just gives a closed form:
\begin{gather}
\delta {\cal L}(\omega) = {\bm d} \delta {\cal L}(\omega,\omega_0)
\end{gather}
in the vicinity of the boundary
and that this relationship carries over to the dimensionally
continued case. Hence, the combined action including the boundary
term is one-and-a half order in the connection (for a fixed geometry on
the boundary). Since the explicit variation of the action with respect to the
vielbein can not produce any derivatives of a variation, we are
guaranteed a well defined variational principle.

This argument has been elucidated by Muller-Hoissen
~\cite{Muller-Hoissen-90a} and Verwimp~\cite{Verwimp-91}. Our
approach will be somewhat different. The question that motivates the
whole of what follows is:
\begin{center}
{\it Can these boundary terms be used, and generalised,\\
to describe membranes and intersections?}\end{center}

\section{Gluing manifolds together}\label{Glue}

A manifold with boundary is defined in the appendix.
The boundary inherits its differential structure from the manifold. If
$f$ is some smooth tensor in the interior and can be extended smoothly across
the boundary, then the derivative on the boundary is defined
accordingly.
 We want to describe situations where matter is singular on hypersurfaces
by an equivalent theory involving fields in the bulk and separate
fields defined on hypersurfaces.

\begin{figure}
\begin{center}\mbox{\epsfig{file=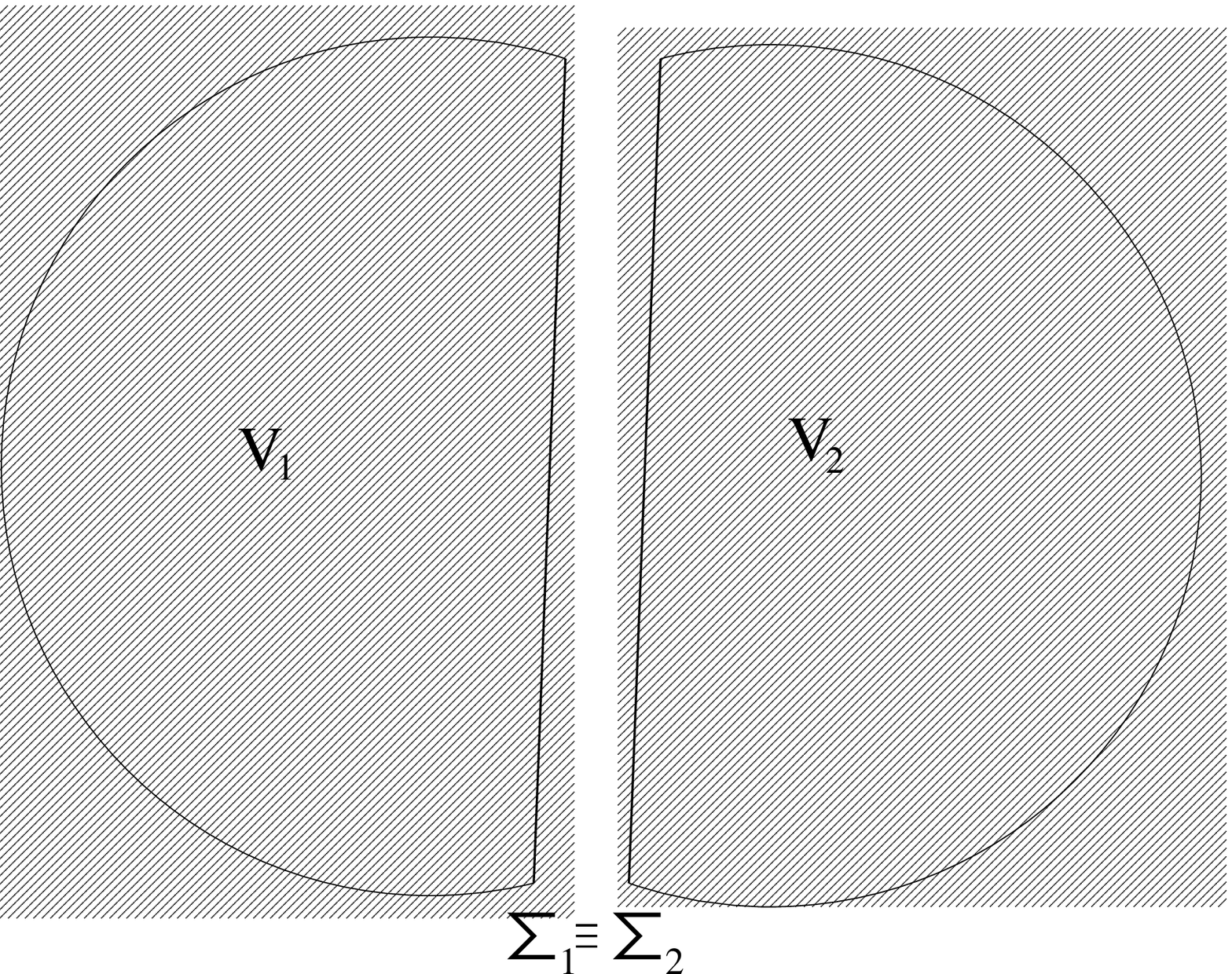, height=7cm}}
\caption{{\small The regions are glued together by identifying $\Sigma_1$
with $\Sigma_2$}}
\label{Gluepic}
\end{center}
\end{figure}

I will treat the hypersurface as the shared boundary of two
disconnected (open) regions $V_1$ and $V_2$. In other words, the manifold is
the union of the two regions and the hypersurface, $M=V_1\cup\Sigma\cup V_2$
(fig. \ref{Gluepic}).

For a well-defined embedding, this gluing process must also match up
the tangent spaces in the correct way. It was shown by Clarke and
Dray~\cite{Clarke-87} that this gluing is well defined if the
naturally induced metric on the hypersurface from each side are the
same. The intrinsic geometry of $\Sigma$ is independent of the
embedding. There exists a unique $C^1$ ($C^3$ away from the
hypersurface) co-ordinate atlas over $M$ (giving it a differentiable
structure) which allows for all the components of the metric to be
continuous across $\Sigma$.


Define some (d-1)-form ${\bm\sigma}$ which is discontinuous at
$\Sigma$ and $C^1$ elsewhere. We want to integrate the exterior
derivative of ${\bm\sigma}$ over $M$. In region $V_1$, ${\bm
\sigma}$ coincides with some smooth $(d-1)$-form field ${\bm
\sigma}_1$. Similarly, in $V_2$ we have a smooth field ${\bm
\sigma}_2$. Because of the discontinuity at $\Sigma$ :
\begin{gather*}
\int_M \bm{d\sigma} \neq \int_{V_1}\bm{d\sigma}_1
+ \int_{V_2} \bm{d\sigma}_2.
\end{gather*}
The inequality is due to the fact that $d\sigma$, evaluated on $\Sigma$,
is not that induced by say $\sigma_1$ continued across $\Sigma$.
It is some kind of delta function.
We can define
\begin{gather*}
\int_M \bm{d\sigma} = \int_{V_1}{\bm d}{\bm \sigma}_1
+ \int_{V_2} {\bm d}{\bm \sigma}_2\
+I_{\Sigma},
\end{gather*}
in such a way that Stokes' Theorem is still valid on $M$. If $M$ is
without boundary:
\begin{gather*}
0 = \int_{\Sigma}i^*({\bm \sigma}_1 - {\bm \sigma}_2)\
+I_\Sigma.
\end{gather*}
we have used Stokes' Theorem and $\partial V_1 = -\partial V_2
= \Sigma$.
Here $i^* \sigma_1$ means the pullback of $\sigma_1$ with respect to the
embedding $i:\Sigma\to M$. I will usually neglect the $i^*$ notation
and write, for example
\begin{gather*}
I_\Sigma = \int_{\Sigma}(\bm{\sigma}_1 - {\bm \sigma}_2).
\end{gather*}
So, integrating without this boundary term sees $\sigma$ each side of the
boundary as if it were to be smoothly continued across the boundary. In order
to account properly for the derivative across the boundary, a boundary
term must be added.

In the Lovelock gravity action, when there is hyper-surface matter, we will
encounter the exterior derivative of a discontinuous $(d-1)$-form.
We would like to represent it by an equivalent action involving fields in the
bulk and separate fields with their support on hypersurfaces.
The above calculations suggest that, if the
divergent terms only appear in a total derivative, this is possible.
\\

It can be shown that Israel's method for singular hyper-surfaces is
equivalent to an action principle with boundary terms. In this case
the variation of the metric evaluated on the hypersurface should not
be fixed- it gives the junction conditions. The joining together of
two boundaries with identical intrinsic geometry gives an action
which yields the junction conditions. This has been done explicitly
for Einstein's theory by Hayward and Luoko~\cite{Hayward-90} and the
doubly covariant form of the action was shown by
Mukohyama~\cite{Mukohyama-01}. The doubly covariant action is where
the intrinsic geometry is treated independently of the geometry of
the bulk regions on either side. The equations of motion must, of
course, generate the continuity of the geometry.

I will now generalise to The Lovelock theory, giving a more simple
proof where the bulk metric is varied and the intrinsic geometry is
just treated as that induced from the bulk.   I will show that the
Lovelock {\it action} in the presence of singular hypersurfaces can
be written in terms of smooth bulk integrals plus a boundary term.
The question of the equivalence of the {\it equations of motion} is
a more difficult one which I will go on to consider. This action for
the $n=2$ Gauss-Bonnet theory was first written down by
Davis~\cite{Davis-02} and also independently by E. Gravanis and  the
author~\cite{Gravanis-02}. A general treatment of hypersurfaces in
this theory has been done by Maeda and Torii~\cite{Maeda-03}.

Let $\omega$ be the Levi-Civita connection for the physical metric
$g$. The manifold $M$ admits also the metric
\begin{gather}
\widehat{ds}^2 = h_{ij}(x^i, z=0)\,dx^idx^j
+ ({\frak n}\!\cdot\!{\frak n})(dz)^2
\end{gather}
at least in some local co-ordinate neighbourhood, $D$, of $\Sigma$.
Let $\omega_0$ be the connection associated with this metric. We are
using explicitly Gaussian normal co-ordinates but $\theta \equiv
\omega-\omega_0$ will be co-ordinate invariant. The one-form
$\theta$ is the {\it Second fundamental form} of $\Sigma$.
\begin{figure}
\begin{center}\mbox{\epsfig{file=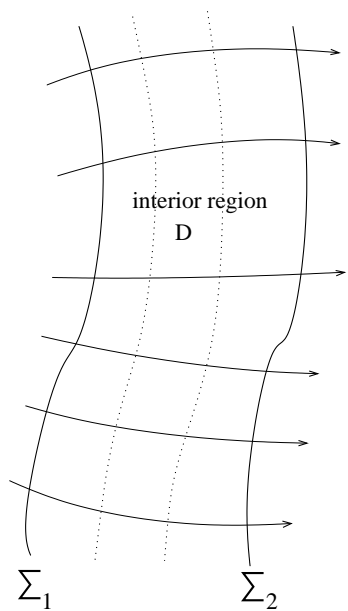, height=8cm}}
\caption{{\small region D enclosed by two hypersurfaces.}}
\label{foliate}
\end{center}
\end{figure}
Following Mukohyama~\cite{Mukohyama-01}, we foliate the region $D$
with hypersurfaces according to some congruence of differentiable
curves (fig. \ref{foliate}). We will eventually take the limit $D
\to \Sigma$.

We rewrite the action in terms of boundary terms
which identically cancel\footnote{for notation see (\ref{notation})}:
\begin{gather}
{\cal S} = I_{1} + I_D + I_2
\end{gather}
\begin{align*}
I_{1} & = \alpha_n\int_{V_1}f(\Omega_{\{1\}}^n E^{d-2n}) +
I_{\Sigma_i},
\\
I_{2} & = \alpha_n\int_{V_2}f(\Omega_{\{2\}}^n E^{d-2n}) -
I_{\Sigma_2},
\\
I_{D} & = \alpha_n\int_{D}f(\Omega^n E^{d-2n}) +
I_{\Sigma_2}-I_{\Sigma_1}.
\end{align*}
There is a relative minus sign because the regions $V_1$ and $V_2$ induce
opposite orientations on the wall $\Sigma$. That is to say the boundary of
region $V_1$ is $\partial V_1 = \Sigma +...$ while the boundary of $V_2$ is
$\partial V_2 = -\Sigma +...$. The boundary terms are chosen to be those
already discussed (eqn. \ref{Myersterm}):
\begin{gather}
I_{\Sigma_i} \equiv n\, \alpha_n\int_{\Sigma_i}\int_0^1 dt\ f(\theta_{\{i\}}
\Omega(t)^{n-1}E^{d-2n}).
\end{gather}


Let us show that there are no second normal derivatives appearing in
$I_D$. Our approach is somewhat simpler than that of Muller-Hoissen
~\cite{Muller-Hoissen-90a}. It will serve as an introduction to the
formalism we shall use in later chapters to find the intersection
terms.

We interpolate between the connection $\omega$
and the connection $\omega_0$ associated with the induced
metric. \footnote{Note, we will not interpolate
between E and $E_0$. We will do this in chapter (\ref{tdimensions})}
\begin{gather*}
\omega(t) \equiv \omega_0 + t \theta,
\qquad \theta \equiv \omega-\omega_0
\end{gather*}

Making use of several relevant identities in the Appendix
\ref{interpappendix}:
\begin{align}\label{fbreakup}
f(\Omega^n E^{d-2n}) - f(\Omega_0^n E^{d-2n}) & = \int_0^1 dt
\frac{\partial}{\partial t} f(\Omega(t)^n E^{d-2n})
\\
& = n\int_0^1 dt\, f(D(t)\theta\, \Omega(t)^{n-1} E^{d-2n})
\nonumber
\\\nonumber & \hspace{-1.45in}= n \,  {\bm d}\! \int_0^1 dt\,
f(\theta\, \Omega(t)^{n-1} E^{d-2n}) + n(d-2n)\int_0^1 dt\,
f(\theta\, \Omega(t)^{n-1} D(t)E\, E^{d-2n-1}).
\end{align}
The formula (\ref{Transgress}) for the Euler density
becomes (\ref{fbreakup}) for the dimensionally continued
theory. The difference is the appearance of the last term:
\begin{gather}\label{Zterm}
{\cal Z} \equiv  n(d-2n)\int_0^1 dt\, f(\theta\, \Omega(t)^{n-1}
D(t)E\, E^{d-2n-1}).
\end{gather}

The total derivative term gives a contribution to the
boundary, which is precisely minus our boundary term.
So we can rewrite $I_D$:
\begin{gather}\label{ID}
I_D = \alpha_n\int_D \left\{f(\Omega_0^n E^{d-2n}) + n(d-2n)\int_0^1
dt\, f\left(\theta\, \Omega^{n-1} D(t)\!E\, E^{d-2n-1}\right)
\right\}.
\end{gather}
This last term, ${\cal Z}$, contains no second normal derivatives
of the metric. This is because $\theta$ has a normal
component in the local frame. Contraction with the
anti-symmetric Levi-Civita symbol means that only
components of $\Omega(t)$ with tangential indices
contribute.
It is shown in the Appendix \ref{Zappend}
that these curvature components contain no second normal derivatives.

We require that even in the limit $D \to \Sigma$
the metric is continuous. So the absence of second normal
derivatives of the metric guarantee that $I_D$ is finite.
We now take the limit $D \to \Sigma$ and $I_D$ must vanish,
justifying the original definition.
This leaves us with the bulk integrals
on $V_1$ and $V_2$ and the boundary terms
which become integrals on $\Sigma^{\pm}$.
\begin{gather}\label{singlag}
{\cal S} = \int_1 {\cal L}(\omega_1,E)
+\int_2 {\cal L}(\omega_2,E) + \int_\Sigma {\cal L}(\omega_2,\omega_0,E)
- {\cal L}(\omega_1,\omega_0,E),
\end{gather}
where
\begin{align*}
{\cal L}(\omega,E) & \equiv \sum_n \alpha_n f(\Omega^n E^{d-2n}),\\
\nonumber {\cal L}(\omega_i,\omega_0,E) & \equiv \sum_n \alpha_n
\int_0^1 dt\ f(\theta_{\{i\}}\Omega(t)^{n-1}E^{d-2n}).
\end{align*}

As an explicit case we give the argument due to Mukohyama for the
Einstein theory~\cite{Mukohyama-01}.
\begin{gather*}
I_D = \int_{D} d^dx \sqrt{-g}\ (R-2\Lambda)
+2\,({\frak n}\!\cdot\!{\frak n})\!\int_{\partial D_+} d^{d-1}x \sqrt{|h|} K
-2\,({\frak n}\!\cdot\!{\frak n})\!\int_{\partial D_-} d^{d-1}x \sqrt{|h|} K
\end{gather*}
By a Gauss-Codacci decomposition~\cite{Wald-84}, equivalent to
equation (\ref{Gauss-Codacci}), and partial integration, the
boundary term is cancelled:
\begin{gather}\label{Einstein ID}
I_D =  \int_{D} d^dx \sqrt{-g} \big({}^{d-1}R +
({\frak n}\!\cdot\!{\frak n})K^2
-({\frak n}\!\cdot\!{\frak n})Tr(K\!\cdot\!K)
-2\Lambda\big)
\end{gather}
As anticipated, there are no derivatives of $K$ appearing
in (\ref{Einstein ID}), so $I_D$ vanishes as $D\to\Sigma$.

This leaves us with the action:
\begin{gather*}
{\cal S} = \int_1 d^dx \sqrt{-g}\ (R-2\Lambda)
+\int_2 d^dx \sqrt{-g}\ (R-2\Lambda)
+ ({\frak n}\!\cdot\!{\frak n})\!\int_\Sigma d^{d-1}x \sqrt{|h|} (K_1 - K_2)
\end{gather*}

So we see that the {\it Lagrangian} (\ref{singlag})
is well-defined as the singular
limit of an integral over a smooth geometry.
What does this mean for the field equations?

I will illustrate the Einstein case, showing that the boundary method does
indeed reproduce the Israel Junction conditions.
The second fundamental form is, in terms of the extrinsic
curvature tensor (\ref{thetainK}):
\begin{gather*}
\theta^{ab} = 2({\frak n} \cdot {\frak n}) {\frak n}^{[a} K^{b]}_{\ c}E^c.
\end{gather*}
The boundary term is:
\begin{gather*}
\alpha_1\int_{\Sigma} (\omega_2-\omega_1)^{ab}{\bm e}_{ab}.
\end{gather*}
\begin{gather*}
(\omega_2 - \omega_1)^{ab} =
(\theta_2 - \theta_1)^{ab} =
2({\frak n} \cdot {\frak n}) {\frak n}^{[a} (K_{\{2\}} -K_{\{1\}})^{b]}_c E^c.
\end{gather*}
We make use of the formula (\ref{Etimese}).
\begin{gather*}
{\cal L}(\omega_1, \omega_2,E) =2 \alpha_1
 ({\frak n} \cdot {\frak n}) {\frak n}^{a} (K_{\{2\}} - K_{\{1\}})^{b}_c
(\delta^c_b {\bm e}_a -\delta^c_a {\bm e}_b)
\\\nonumber = 2 \alpha_1(K_{\{2\}} - K_{\{1\}})
({\frak n} \cdot {\frak n}) {\frak n}^a {\bm e}_a,
\end{gather*}
where in the last line $K \equiv K^a_a$. Also
$n^aK^b_a = 0$ was used.

So the hypersurface term is just the difference between the
trace of the intrinsic curvature on each side, multiplied
by the (d-1)-dimensional volume element (\ref{bndryelement}):
\begin{gather*}
({\frak n} \cdot {\frak n}) {\frak n}^a {\bm e}_a
 = ({\frak n} \cdot {\frak n})
\sqrt{|h|}\,d^{d-1}\zeta = \tilde{\bm e}
\end{gather*}
in terms of the local co-ordinates on $\Sigma$.
The surface term is the difference between the two
Gibbons-Hawking boundary terms of regions 1 and 2.

As shown in the next chapter, the infinitesimal variation of ${\cal
L}(\omega_{1},\omega_2,E)$ with respect to $\omega_i$ gives two
terms. One is a total derivative. The other cancels with the total
derivative term coming from the bulk. The surface contribution to
the equations of motion is just given by the variation with respect
to the frame:
\begin{gather*}
\delta E^C\wedge(\omega_2 -\omega_1)^{ab} \wedge {\bm e}_{abc}.
\end{gather*}
The term multiplying $\delta E$ is:
\begin{align*}
2({\frak n} \cdot {\frak n}) {\frak n}^a(K_{\{2\}} -K_{\{1\}})^b_{\
d} E^d \wedge {\bm e}_{abc} & = 2({\frak n} \cdot {\frak n}) {\frak
n}^a (K_2 -K_1)^b_{\ d}(\delta^d_a {\bm e}_{bc} -\delta^d_b {\bm
e}_{ac}+\delta^d_c {\bm e}_{ab})
\\\nonumber & =
2 \tilde{{\bm e}}_{ab}\big((K_{\{2\}} -K_{\{1\}})^b_{\ c}
- (K_{\{2\}} -K_{\{1\}})
h^b_{\ c}\big)
\end{align*}
As found in (\ref{juncminushalf}) this equals
$-2T^b_{\ c} \tilde{{\bm e}}_{ab}$.
Equating the two terms,
this is the Israel Junction condition
\begin{gather*}
\tilde{T}^{ab} = -\beta_1 \bigl((K_{\{2\}} -K_{\{1\}})^{ab}
- (K_{\{2\}} -K_{\{1\}})
h^{ab} \bigr).
\end{gather*}
\\

There are two approaches one can take to singular hypersurfaces:
\\\\
1) Including boundary terms to give a well defined action principle.
\\\\
2) Solving the equations of motion in the limit that sources become
singular.
\\
\\
Here, we have adopted the former approach. For Einstein's theory, we
have shown that the boundary approach reproduces Israel's method,
which is the unique singular limit of a smooth source. Methods 1 and
2 are equivalent.

For the higher order Lovelock Theory, we also have a well defined
action principle with boundary terms. Consider some thick matter
distribution where the width is taken to zero.  Is integrating the
field equations in the singular limit equivalent to our variational
principle with a singular Lagrangian? This has been discussed by
Barcelo et al~\cite{Barcelo-03} and by Deruelle and
Germani~\cite{Deruelle-03b}. It has been shown that the singular
Lagrangian is equivalent to \emph{a} solution of the field equations
with distributional sources. What is not clear is whether this is a
\emph{unique} limit of smooth sources.

If we want to approximate a dense but finite thickness shell of
matter, such as domain walls or in simple models of black hole
formation, method 2 would be appropriate. If we think of a membrane
as some kind of fundamental particle, with matter strictly
localised, then the singular Lagrangian approach, being a preferred
way of describing the distributional field equations, seems to be
more appropriate.
\begin{conjecture}
For Lovelock gravity, methods 1 and 2 are equivalent.
\end{conjecture}
A proof of this would mean that our boundary approach is always a
good approximation on scales larger than the thickness of the
hypersurface. A general proof is not known to me.


\chapter{Intersections}\label{inters}

As mentioned in the introduction, much work has been done on the study
of single and two brane worlds. Also there has been much work on the
dynamics of a single false vacuum bubble in the context of inflation.

We want to look at what happens in a more general
network of hypersurfaces. It is well known from the obvious
example of soap bubbles that when spherical bubbles come together
they do not tend to remain as spheres.
The surface energy of the bubbles tends to  make them form into
something like a honeycomb. There are shared boundaries and
corners and edges.

There have been several studies of collisions of shells of matter in
General Relativity~\cite{Dray-85,Dray-86,Langlois-02,Berezin-02}.
Langlois et al~\cite{Langlois-02} found an interesting geometrical
treatment of colliding shells in AdS black hole backgrounds.

We address the problem of intersecting/colliding walls in a more
generalised way, motivated by the topological invariant approach
presented in section \ref{topapproach}. Our treatment will be for all
Lovelock theories of gravity.

We consider a smooth\footnote{$C^1$: see section \ref{distributional}.
This will amount to excluding any deficit angle at intersections.}
manifold with embedded arbitrarily intersecting
hypersurfaces of singular matter. As in section \ref{Glue}, we can
view a hypersurface as the shared boundary of two adjacent regions.
The gravity of localised matter can be described by a boundary
term in the action. We allow for the possibility of matter being
localised on the surfaces of intersection also.

More generally, the space-time is divided up into polygonal regions
bounded by piecewise smooth hypersurfaces-like a matrix of bubbles
or a honeycomb. We exploit the topological nature of the theory to
write the action in terms of different connections in different
regions. This generates our surface actions remarkably simply. We can
derive the Junction conditions and the junction conditions for
any intersections purely from the explicit Euler variation w.r.t the
vielbein.

To illustrate the problem we can think of a 3-way junction
on the plane. The intrinsic geometry is well defined
but the extrinsic curvature is discontinuous.
At the 2-way junctions, although the metric is not differentiable,
we can approximate differentiation in an unambiguous way.
The three way junction is more problematic. There is no second
derivative of the metric and no obvious way in which it can be
well defined under integration.

Our approach will be effectively to treat the exterior derivative
as a certain kind of co-boundary operator in co-homology theory.
In this way, the derivatives in the bulk contribute only to the walls
and not the intersection. Similarly, the total derivatives on the
walls contribute only to the co-dimension 2 intersections.

The crucial fact is that the Lovelock Lagrangian is the dimensional
continuation of the Euler density. The integral of the Euler density
is a topological number and is independent of the discontinuities.
We can dimensionally continue certain results over to the Lovelock
theory to get a well defined unambiguous way of dealing with the
lack of a second derivative.

\section{Euler densities}

As mentioned in chapter \ref{Einstein_Lovelock}, the Lovelock gravity
is closely related to the Euler density. We shall first of all consider a
`Lagrangian' which is proportional to the Euler density.
Such a theory is topological in the sense that it has no local
degrees of freedom. The equations of motion are identically zero.

We need to treat manifolds where the metric is continuous but not
differentiable ($C^0$), the manifold being only once differentiable
($C^1$)~\cite{Clarke-87}. The integral of the Euler density over a
compact Riemannian manifold is equal to the Euler number. The Euler
number is even defined for non-smooth objects such as polyhedra.
This gives meaning to the integral over a manifold even when it is
not smooth. This gives hope that we might be able to do something
similar with a Lorentzian manifold and a dimensionally continued
Euler density.

The Euler density obeys the Transgression formula (\ref{Transgress}).
\begin{gather*}
f(\Omega_1^n)-f(\Omega_2^n) =
n\, {\bm d}\! \int_0^1 dt \, f\big((\omega_2 -\omega_1)\, \Omega(t)^n \big).
\end{gather*}

Let $\omega$ be a metric connection which has a discontinuity at some
hyper-surface $\{12\}$. In region 1, the connection coincides with the smooth
field $\omega_1$ and likewise region 2.
Following the argument of section \ref{Glue}, we can write:
\begin{gather}
\int_M {\cal L}(\omega) = \int_1 {\cal L}(\omega_1)
+ \int_2 {\cal L}(\omega_2)
+ \int_{\{12\}} {\cal L}(\omega_1,\omega_2),\nonumber\\\nonumber\\
{\cal L}(\omega) = f(\Omega) ,\nonumber\\\label{the2lag}
{\cal L}(\omega_1,\omega_2)
=  n\int_0^1 dt \, f\big((\omega_2 -\omega_1)\, \Omega(t)\big).
\end{gather}
If $M$ is Riemannian and compact, the above integral, with the
hypersurface term, does indeed give the Euler number of $M$.

When there exists an everywhere vanishing connection, the
term ${\cal L}(\omega,0)$ is a Chern-Simons term.
In that case ${\cal L}(\omega)$ is an exact form:
$ {\cal L}(\omega)= {\bm d} {\cal L}(\omega,0) $.
Sometimes the more general term ${\cal L}(\omega_1,\omega_2)$ is
known also as a Chern-Simons term.

If there are many regions separated by walls of discontinuity
in the connection, the walls will intersect.
We exploit the topological nature of the theory to write the action in
terms of different connections in different regions.
Integrating ${\cal L}(\omega)$ over the manifold, when $\omega$ is the
discontinuous connection form, one has to add the surface terms
(\ref{the2lag}) integrated over the walls for the final result to have
well defined variations with respect to $\omega$ (and to be diffeomorphism
invariant). One should also add appropriate generalizations of the
surface terms integrated over the intersection. In the next subsection
we find these terms.

\subsection{From boundary to intersection action terms}\label{biat}

Given an Euler density the Transgression formula can be found by
interpolating smoothly between the given connection $\omega$ and
another one $\omega'$. We can continue by interpolating between the
latter and a new connection.

In general, let us define the $p$-parameter family of connections,
interpolating between $p+1$ connections, $\omega_1,..,\omega_{p+1}$,
\begin{gather}\label{omega(p)}
\omega_{(p)}\equiv\omega(t^1..t^p) \equiv\omega_{1}-(1-t^1)
\Upsilon_{1}-...-(1-t^1)..(1-t^p) \Upsilon_{p}
\end{gather}
where
\begin{gather*}
\Upsilon_{r}\equiv\omega_{r}- \omega_{r+1}, \quad r=1,..,p
\end{gather*}
Define
\begin{gather*}
\frac{\partial}{\partial t^q}\omega_{(p)}= \frac{\partial}{\partial
t^q}\omega(t^1..t^p)= \sum^p_{r \geq q} (1-t^1)..
\widehat{(1-t^q)}...(1-t^r) \Upsilon_{r}=
\Upsilon_{q}(t^1...\widehat{t^q}...t^p)\equiv\Upsilon_{q(p)}
\end{gather*}
where the hat over a term or index means it is omitted. Let
$\Omega_{(p)}$ be the $p$-parameter curvature 2-form associated with
$\omega_{(p)}$. Then
\begin{align} \label{id1}
\frac{\partial}{\partial t^q}\Omega_{(p)}= & d\Upsilon_{q(p)} +
\Upsilon_{q(p)}\wedge \omega_{(p)}+\omega_{(p)}\wedge\Upsilon_{q(p)}
\nonumber\\
= &  D_{(p)}\Upsilon_{q(p)}
\end{align}
There is a generalised Bianchi identity for  $\Omega_{(p)}$:
\begin{gather}\label{genBian}
D_{(p)}\Omega_{(p)}  =0
\end{gather}
One way to prove this is to note that $\omega_{(p)} = \sum_i C^i
\omega_i$ where $\sum_i C^i =1$ , see (\ref{Bianchi}).
\\\\

\begin{lemma}\label{Gravanisterms} The generalisation of the term
(\ref{the2lag}) to $p+1$
curvature entries is:
\begin{equation} \label{genl}
{\cal L}(\omega_1,..,\omega_{p+1})= \zeta_p
\int_0^1 dt^1..dt^p \
f \big(\Upsilon_{1(p)} \Upsilon_{2(p)}\cdots \Upsilon_{p(p)}
\,(\Omega_{(p)})^{n-p}\big)
\end{equation}
where
\begin{gather}
\zeta_p=(-1)^{p(p+1)/2}\frac{n!}{(n-p)!}
\end{gather}
\\
{\bf Proof:}
The proof is by induction.
If we define
\begin{equation*}
\omega_{p+1}(t^{p+1})\equiv t^{p+1} \omega_{p+1} + (1-t^{p+1})
\omega_{p+2}
\end{equation*}
then:
\begin{gather*}
\omega_{p+1}\to \omega_{p+1}(t_{p+1})
\Rightarrow \Upsilon_{q(p)} \to \Upsilon_{q(p+1)}, \quad \forall q \leq p
\end{gather*}

\begin{gather*}
{\cal L}\big(\omega_0,..,\omega_{p}(t^{p+1})\big)=
\zeta_p \int_0^1 dt^1..dt^p \
f(\Upsilon_{1(p+1)}\cdots\Upsilon_{p(p+1)}
\,(\Omega_{(p+1)})^{n-p})
\end{gather*}

We can write the difference of two terms by using this interpolation:
\begin{align}\label{original}
{\cal L}(\omega_1,..,\omega_p,\omega_{p+1})- {\cal L}
(\omega_1,..,\omega_p,\omega_{p+2}) & = \int_0^1 dt^{p+1}
\frac{\partial}{\partial t^{p+1}} {\cal L}
\big(\omega_1,..,\omega_p,\omega_{p+1}(t^{p+1})\big) \nonumber\\ &
\hspace{-2.15in} =  \zeta_p \int_0^1 dt^1..dt^pdt^{p+1} \
\frac{\partial}{\partial t^{p+1}} \ f
\big(\Upsilon_{1(p+1)}\Upsilon_{2(p+1)}\cdots \Upsilon_{p(p+1)}
(\Omega_{(p+1)})^{n-p}\big)
\end{align}
From the multi-linearity of the invariant polynomial $f$ we have
\begin{align}\label{parta}
\frac{\partial}{\partial t^{p+1}}
f\big(\Upsilon_{1(p+1)}\Upsilon_{2(p+1)}\cdots \Upsilon_{p(p+1)}\,
(\Omega_{(p+1)})^{n-p}\big)
= &
\\\nonumber & \hspace{-2.4in} =
\sum_{r=1}^p f \big(\Upsilon_{1(p+1)}\cdots
\frac{\partial}{\partial t^{p+1}}
\Upsilon_{r(p+1)} \cdots \Upsilon_{p(p+1)}\,
(\Omega_{(p+1)})^{n-p}\big)
\\\nonumber &\hspace{-2.3in} +(n-p)
f \big(\Upsilon_{1(p+1)}\Upsilon_{2(p+1)}\cdots\Upsilon_{p(p+1)}
\frac{\partial \Omega_{(p+1)}}{\partial t^{p+1}}
(\Omega_{(p+1)})^{n-p-1}\big)
\end{align}
Using (\ref{id1}) we can write the last term as
\begin{gather}\label{partb}
(n-p)
f \big(\Upsilon_{1(p+1)}\cdots\Upsilon_{p(p+1)}
D_{(p+1)}\Upsilon_{p+1(p+1)}(\Omega_{(p+1)})^{n-p-1}\big)
\end{gather}
Now the covariant derivative $D_{(p+1)}\Upsilon_{p+1(p+1)}$ does not
just give a total derivative term.
There will be other terms of the form $D_{(p+1)}\Upsilon_{q(p+1)}$.
Making use of the generalised Bianchi Identity (\ref{genBian}) and
the invariance of the polynomial:
\begin{align}
&\sum_{r=1}^{p+1} (-1)^{r-1} f \big(\Upsilon_{1(p+1)}\cdots
D_{(p+1)}\Upsilon_{r(p+1)}\cdots\Upsilon_{p+1(p+1)}
(\Omega_{(p+1)})^{n-p-1}\big)\nonumber\\
  &\hspace{.6in}= D_{(p+1)} f \big(\Upsilon_{1(p+1)}
\cdots\Upsilon_{p+1(p+1)}(\Omega_{(p+1)})^{n-p-1}\big)
\nonumber\\ &\hspace{.65in}=
d f \big(\Upsilon_{1(p+1)} \cdots\Upsilon_{p+1(p+1)}
(\Omega_{(p+1)})^{n-p-1}\big)\label{partc}
\end{align}
\\

There is a total derivative term plus some other terms.
It is natural to guess that they come from other ${\cal L}$ terms.
To prove that this is the case, first note that
\begin{align} \label{omission}
\omega(t^1..t^p)|_{t^r=0}= & \omega_1 -(1-t^1)\Upsilon_1 -
\cdots-(1-t^1)\cdots(1-t^{r-1})(\Upsilon_r-\Upsilon_{r-1})-\cdots
\nonumber\\
= & \omega_1 -(1-t^1)\Upsilon_1 -\cdots
-(1-t^1)\cdots(1-t^{r-1})(\omega_{r-1}-\omega_{r+1})-\cdots
\nonumber\\
=& \omega(t^1,..,t^{r-1},t^{r+1},..,t^p)
\end{align}
where $\omega(t^1,...,t^{r-1},t^{r+1},...,t^p)$ is the connection
interpolating between $\omega_1,...,\omega_{r-1}$, $
\omega_{r+1},...,\omega_{p+1}$ just as
in (\ref{omega(p)}).
It follows that other ${\cal L}$'s can be written:
\begin{align}
& \hspace{-.1in}\zeta_p \int_0^1 dt^1\cdots dt^{p+1}
\frac{\partial}{\partial t^{r}}
f \big(\Upsilon_{1(p+1)} \cdots
\widehat{\Upsilon_{r(p+1)}}\cdots \Upsilon_{p+1(p+1)}
(\Omega_{(p+1)})^{n-p}\big)\nonumber\\\nonumber
&= \zeta_p \int_0^1 dt^1\cdots\widehat{dt^r}\cdots dt^{p+1}
\big. f \big(\Upsilon_{1(p+1)} \cdots
\widehat{\Upsilon_{r(p+1)}}\cdots \Upsilon_{p+1(p+1)}
(\Omega_{(p+1)})^{n-p}\big)\big|_{t^r =0}^{t^r=1}\\
&= 0 - {\cal L}(\omega_1,...,\widehat{\omega_r},...,\omega_{p+2})
\label{equalsL}\end{align} Evaluating the $t$-derivatives we do
indeed get $D_{(p+1)}\Upsilon_{r(p+1)}$ terms as well as others:
\begin{multline}\label{tderivterms}
\frac{\partial}{\partial t^r} f \big(\Upsilon_{1(p+1)} \cdots
\widehat{\Upsilon_{r(p+1)}}\cdots \Upsilon_{p+1(p+1)}
(\Omega_{(p+1)})^{n-p}\big) = \\
\sum_{q<r} f \big(\cdots \frac{\partial \Upsilon_{q(p+1)}}
{\partial t^r}
\cdots\widehat{\Upsilon_{r(p+1)}}\cdots
(\Omega_{(p+1)})^{n-p}\big)
+\\\sum_{q>r} f \big(\cdots \widehat{\Upsilon_{r(p+1)}}
\cdots\frac{\partial \Upsilon_{q(p+1)}}{\partial t^r}\cdots
(\Omega_{(p+1)})^{n-p}\big)
+\\
f \big(\Upsilon_{1(p+1)}\cdots
D_{(p+1)}\Upsilon_{r(p+1)} \cdots
\Upsilon_{p+1(p+1)} (\Omega_{(p+1)})^{n-p-1}\big)
\end{multline}
\\

Bringing all the terms together by combining (\ref{parta}),
(\ref{partb}),
(\ref{partc}) and (\ref{tderivterms}):
\begin{multline}
\sum_{r=1}^{p+1}(-1)^{r-1}\frac{\partial}{\partial t^{r}}
f \big(\Upsilon_{1(p+1)} \cdots
\widehat{\Upsilon_{r(p+1)}}\cdots \Upsilon_{p+1(p+1)}
(\Omega_{(p+1)})^{n-p}\big) = \\\hspace{-1.15in}
=(n-p) d f \big(\Upsilon_{1(p+1)} \cdots \Upsilon_{p+1(p+1)}
(\Omega_{(p+1)})^{n-p-1}\big) + \\
+\sum_{r=1}^{p+1}(-1)^{r-1}\left\{
\sum_{q<r} f \big(\cdots \frac{\partial \Upsilon_{q(p+1)}}{\partial t^r}
\cdots\widehat{\Upsilon_{r(p+1)}}\cdots
(\Omega_{(p+1)})^{n-p}\big)\right.\\\left.
+\sum_{q>r} f \big(\cdots \widehat{\Upsilon_{r(p+1)}}
\cdots\frac{\partial \Upsilon_{q(p+1)}}{\partial t^r}\cdots
(\Omega_{(p+1)})^{n-p}\big)\right\}\label{alltogether}
\end{multline}
In the last two terms, if we change variables $ r \leftrightarrow s$
in the latter and using this identity
\begin{equation}
\frac{\partial}{\partial t^s} \Upsilon^r_{p+1}=
\frac{\partial}{\partial t^s} \frac{\partial}{\partial t^r}
\omega_{p+1}= \frac{\partial}{\partial t^r} \Upsilon^s_{p+1}
\end{equation}
we see that the terms cancel.

Now if we substitute (\ref{original})(\ref{equalsL}) for the LHS
in (\ref{alltogether}) we get the result:
\begin{gather*}
\sum_{r=1}^{p+2} (-1)^{r} {\cal L}
(\omega_1,...,\widehat{\omega}_r,...,\omega_{p+2})
= (n-p)\,\zeta_p\, d f \big(\Upsilon_{1(p+1)} \cdots \Upsilon_{p+1(p+1)}
(\Omega_{(p+1)})^{n-p-1}\big)
\end{gather*}
We will multiply through by a factor of $(-1)^{p+1}$ for later convenience.
By comparison with (\ref{genl}), with $p\to p+1$ we get
\begin{gather}\label{thecomprule}
\sum_{r=1}^{p+2} (-1)^{p+r+1} {\cal L}
(\omega_1,...,\widehat{\omega}_r,...,\omega_{p+2})
= d {\cal L}(\omega_1,...,\omega_{p+2})
\end{gather}
provided
\begin{gather*}
\zeta_{p+1} = (n-p)\zeta_p(-1)^{p+1}\nonumber
\\ \Rightarrow \zeta_p = (-1)^{p(p+1)/2} \frac{n!}{(n-p)!}
\end{gather*}
We have proved (\ref{genl}) by induction.$\Box$
\end{lemma}

It is not hard to show that ${\cal L}$ is fully anti-symmetric in
its entries, so we can write (\ref{thecomprule}) in the form
\begin{equation} \label{ncomprule}
\sum_{s=1}^{p+1} {\cal L}(\omega^1,..,\omega^{s-1},\omega',\omega^{s+1},..,\omega^{p+1})=
{\cal L}(\omega^1,..,\omega^{p+1})+d{\cal L}
(\omega^1,..,\omega^{p+1},\omega')
\end{equation}
where $\omega'$ is arbitrary. \footnote{ In different contexts the
boundary variation of Chern-Simons terms are important (e.g.
Gravitational and gauge anomalies~\cite{Alvarez-85} and black hole
thermodynamics~\cite{Gegenberg-00}). Here ${\cal L}_3$ is the
boundary action of three ``intersecting" Chern-Simons theories and
as such it is gauge invariant under local Lorentz transformations.}

The polynomial $f$ is anti-symmetric with respect to interchanging two
$\Upsilon$'s so we have $f(..,\Upsilon^r,..,\Upsilon^r,..,\Omega_{p},..\Omega_{p})=0$ and we
can write (\ref{genl}) explicitly in terms of $\Upsilon^r=\omega^r-\omega^{r+1}$, $r=1..p\ $
in the
form \begin{eqnarray} \label{genl2}
&& {\cal L}(\omega^1,..,\omega^{p+1})=  \int_0^1 dt_1..dt_p \
\hat{\zeta}_p \  f(\Upsilon^1,\Upsilon^2,..\Upsilon^p,\Omega_p,..,\Omega_p),
\\ &&  \label{zeta}
\hat{\zeta}_p=(-1)^{\frac{p(p+1)}{2}} \frac{n!}{(n-p)!}
\ \
\prod_{r=1}^{p-1}(1-t_r)^{p-r}.
\end{eqnarray}
\\\\

Let us finally show that ${\cal L}_p$'s, constructed from Characteristic
Classes, are invariant under
local Lorentz transformations. The
connections transform:
\begin{equation*}
\omega^r_{(g)}=g^{-1} \omega^r g+ g^{-1} dg
\end{equation*}
for all $r=1,..,p+1$, where $g$ belongs to the
adjoint representation of SO(d-1,1). Then, $\Upsilon^r_{(g)}=g^{-1} \Upsilon g$
and $\Omega(t)_{(g)}=g^{-1} \Omega(t) g$, so
\begin{equation}
\label{ginv}
{\cal L}(\omega^1_{(g)},..,\omega^{p+1}_{(g)})={\cal
L}(\omega^1,..,\omega^{p+1}).
\end{equation}
\\\\

In fact, one can derive (\ref{ncomprule}) without reference to the
invariant polynomial , by use of the Poincare lemma and the
following observation (inspired by the form of (\ref{ncomprule})).
If $f(x_1,..x_n)$ is an anti-symmetric function  of n variables and
\begin{equation*}
Af(x_1,..x_p,x_{p+1})=f(x_1,..x_p)-\sum_{i=1}^p
f(x_1,..,x_{i-1},x_{p+1},x_{i+1},..x_p)
\end{equation*}
(antisymmetrising over n+1 variables) then $AAf(x_1,..,x_{p+2})=0$.
The proof is trivial.

We can now show (\ref{ncomprule}) by induction assuming only that it
is true for the p=0 case. Assume (\ref{ncomprule}) for p=k-1 (let us
use the symbol ${\cal L}_k (\omega^1..\omega^k)$ for the
intersection forms in this proof)
\begin{equation*}
A{\cal L}_k (\omega^1..\omega^{k+1})=
-d{\cal L}_{k+1}(\omega^1..\omega^{k+1})
\end{equation*}
then $A d{\cal L}_{k+1}(\omega^1..\omega^{k+2}) =0$. Since $d$ is a
linear operator, $Ad-dA=0$ and so we get $dA{\cal L}_{k+1}
(\omega^1..\omega^{k+2})=0$. By Poincare's lemma, we have that there
exists an invariant form, ${\cal L}_{k+2}$, locally, such that
\begin{gather*}
A{\cal L}_{k+1}(\omega^1..\omega^{k+2})
=-d{\cal L}_{k+2}(\omega^1..\omega^{k+2}).
\end{gather*}
Since d is linear, the right hand side must be anti-symmetric w.r.t.
the $k+2$ entries. This completes the induction.

There is similarity between our composition rule and Stora-Zumino
descent equations~\cite{Weinberg-96}, occurring in the study of
anomalies in non-abelian gauge theories. Anomalies arise in a
quantum field theory when the quantum effective action is not
invariant under a classical gauge symmetry. To be well defined, an
anomaly must satisfy the Wess-Zumino consistency conditions. It is
useful to introduce ghost fields and the BRST operator which mixes
between particles and ghosts. One can define a generating functional
such that the consistency conditions are equivalent to the BRST
invariance of this functional. The BRST operator is nilpotent:
$ss=0$. It anti-commutes with the exterior derivative: $ds+sd=0$.
Taking a functional $\int Tr F^2$, $F$ being the field strength, one
can use the above procedure to establish the Stora-Zumino descent
equations. This allows one to find BRST invariant integrals of
non-BRST invariant terms (Chern-Simons type terms). The similarity
with our derivation of the composition rule is the existence in both
cases of a nilpotent operator, $A$ in our case and the fermionic
BRST operator there, which commutes and anti-commutes respectively
with the derivative operator $d$.
\\\\

\subsection{Manifolds with discontinuous connection 1-form}

We now construct the action functional of gravity on a manifold
containing intersecting surfaces. It will also enable us to draw
conclusions for the general dimensionally continued topological density.

If the functional $\int_M {\cal L}$ is independent of the local form
of the metric of the manifold $M$, then it can be evaluated using a
continuous connection as well as a connection that is discontinuous
at some hypersurfaces. Either way the result will be the same. We
use this formal equivalence to give a meaning to $\int_M {\cal
L}(\omega)$ when $\omega$ is discontinuous.

Let us start with the case of a topological density ${\cal
L}(\omega_0)$ ($\omega_0$ continuous) integrated over $M$ which
contains a single hypersurface. Label 1 and 2 the regions of $M$
separated by the hypersurface $\{12\}$. Introduce two connections,
$\omega^1$ and $\omega^2$ which are smooth in the regions 1 and 2
respectively. We now write
\begin{gather} \label{1}
\int_M {\cal L}(\omega_0) = \int_{1}{\cal L}(\omega_1)+d{\cal
L}(\omega_1,\omega_0) +\int_{2}{\cal L}(\omega_2)+d{\cal
L}(\omega_2,\omega_0)
\end{gather}
Label the surface, oriented with respect to region $1$, with $12$.
(Formally $\int_{12} = -\int_{21}$).
\begin{align}\nonumber  \label{2}
\int_M {\cal L}(\omega_0) =& \int_{1}{\cal L}(\omega_1)
+\int_{2}{\cal L}(\omega_2)+ \int_{12}{\cal
L}(\omega_1,\omega_0)-{\cal L}(\omega_2,\omega_0)
\\=&\int_{1}{\cal L}(\omega_1)
+\int_{2}{\cal L}(\omega_2)+ \int_{12}{\cal
L}(\omega_1,\omega_2)+d{\cal L}(\omega_1,\omega_2,\omega_0)
\end{align}
That is, for a smooth surface the r.h.s. is independent of
$\omega_0$.

Consider now a sequence of co-dimension $p=1,2,3..h$ hyper-surfaces which
are intersections of $p+1=2,3..h+1$ bulk regions respectively.
A
co-dimension p hyper-surface is labelled by $i_0..i_p$ where $i_0,..,i_p$
are the labels of the bulk regions which intersect there. We call this
configuration a {\it simplicial intersection}.

\begin{figure}
\begin{center}\mbox{\epsfig{file=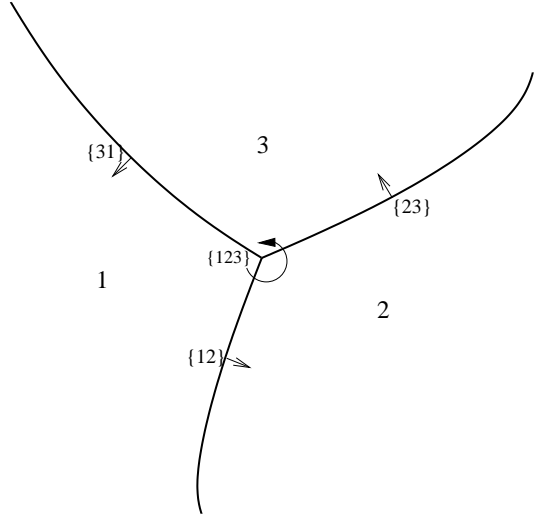, width=7cm}}
\caption{{\small The simplicial intersection of co-dimension 2 (h=2).
The totally
antisymmetric symbol \{123\} specifies the intersection including the
orientation}}\label{123}
\end{center}
\end{figure}

\begin{figure}
\begin{center}\mbox{\epsfig{file=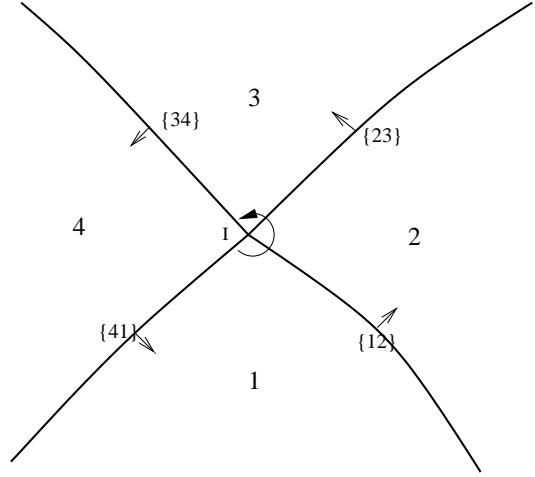, width=7cm}}
\caption{{\small A non-simplicial intersection of co-dimension 2 (h=2). The
intersection, including the orientation, would not be properly represented
by the totally anti-symmetric symbol \{1234\}}}\label{1234}
\end{center}
\end{figure}

We take the example $h=2$ (fig.\ref{123}), where the intersections are
$\{12\}$, $\{13\}$, $\{23\}$, $\{123\}$. An exact form integrated over
$\{12\}$ will contribute at $\{123\}$
the opposite that when integrated over $\{21\}$, that is, for the latter
integration the intersection can be labelled by $-123=213$, if
we assume anti-symmetry of the label. The arrows of positive
orientations in fig.\ref{123} tell us that a fully anti-symmetric symbol
$\{123\}$ will adequately describe the orientations of the intersection
$123$. 
This is in contrast to the non-simplicial intersection (fig. \ref{1234}).
\begin{definition}[for a simplicial intersection]
$\{i_0...i_p\}$ is the set $\overline{i_0}\cap
\cdots \cap \overline{i_p}$
where $\overline{i_r}$ is the closure of the open set $i_r$ (a bulk region).
$\overline{i_r}$ overlap such that $\partial i_r = \sum_{s=0,\neq r}^h
 \overline{i_r}\cap\overline{i_s}$
and $i_r \cap i_s = \emptyset$ for all $s\neq r$.
By $\partial (\overline{A}\cap \overline{B}) = (\partial \overline{A}\cap
\overline{B})\cup  (\overline{A}\cap \partial\overline{B})$, for A, B open sets,
we can write:
\begin{gather}\label{defthesets}
\partial \{i_0...i_p\} = \sum_{i_{p+1}} \{i_0...i_{p+1}\}.
\end{gather}
Full anti-symmetry of the symbol $\{i_0...i_p\}$ keeps track of the orientations
properly in (\ref{defthesets}). As a check:
\begin{gather*}
\partial^2 \{i_0...i_p\} = \sum_{i_{p+1},i_{p+2}} \{i_0...i_{p+1}i_{p+2}\}=0.
\end{gather*}
\end{definition}

\begin{definition}{\it A simplicial lattice}
is a lattice with all intersections being simplicial
intersections (fig. \ref{simplat}).\end{definition}

\begin{definition}\label{localisedcurvdef}{\it localised curvature}
on a surface of co-dimension
$>1$ is a singularity in the Riemann tensor such that the parallel transport
of a vector around an infinitesimal closed curve produces a finite change.
This means that there will not be a well defined ortho-normal frame
at the intersection
(see the Outlook section for a further discussion of this).
\end{definition}

\begin{lemma}\label{Gravanisinter}~\cite{Gravanis-03}
For a simplicially valent lattice, with no
localised curvature on intersections of co-dimension $>1$ ,
the contribution from each intersection
$\{i_1...i_k\}$ is
\begin{gather}
\int_{\{i_1...i_k\}} {\cal L}(\omega_{i_1},...,\omega_{i_k})
\end{gather}
up to a boundary term on $\partial M$.
\begin{figure}
\begin{center}\mbox{\epsfig{file=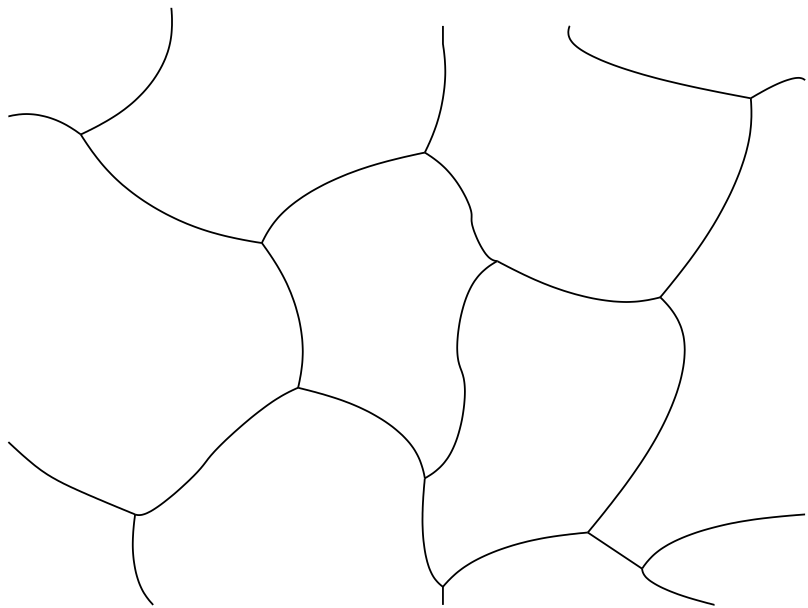, height=8cm}}
\caption{{\small A simplicially valent lattice in two dimensions.
Each intersection is the meeting of three bulk regions.}}
\label{simplat}
\end{center}
\end{figure}
\\\\
{\bf Proof:}
Assume, for  $l < h$,
\begin{gather*}
\int_M {\cal L}(\omega_0)= \sum^{l-1}_{k=1}
\frac{1}{k!} \sum_{i_1..i_k} \int_{\{i_1...i_k\}}
{\cal L}(\omega_{i_1}..\omega_{i_k}) +
\frac{1}{l!} \int_{\{i_1...i_l\}}
{\cal L}(\omega_{i_1}..\omega_{i_l}) +
d{\cal L}(\omega_{i_1}..\omega_{i_l},\omega_0).
\end{gather*}
We have already seen that this is true for $l=1$ and $l=2$.
The exact form gives
\begin{equation*}
\frac{1}{l!} \sum_{i_1..i_li_{l+1}} \int_{\{i_1...i_li_{l+1}\}}
{\cal L}(\omega_{i_1}..\omega_{i_l}\omega_0)
\ + (\text{a term on}\ \partial M).
\end{equation*}
From the anti-symmetry of $\{i_1..i_li_{l+1}\}$
and of ${\cal L}$ we have
\begin{equation*}
\frac{1}{l!} \sum_{i_1..i_{l+1}} \int_{\{i_1...i_li_{l+1}\}}
\frac{1}{l+1} \sum_{r=1}^{l+1} {\cal L}
(\omega_{i_1}..\omega_{i_{r-1}}\omega'\omega_{i_{r+1}}..
\omega_{i_{l+1}})
\end{equation*}
Applying the composition rule (\ref{ncomprule}) we get:
\begin{align*}
\int_M {\cal L}(\omega_0)= & \sum^{l}_{k=1} \frac{1}{k!}
\sum_{i_1..i_k} \int_{\{i_1...i_k\}}
{\cal L}(\omega_{i_1}..\omega_{i_k}) \nonumber\\ & +
\frac{1}{(l+1)!}\sum_{i_1..i_{l+1}} \int_{\{i_1...i_{l+1}\}}
{\cal L}(\omega_{i_1}..\omega_{i_{l+1}})
+d{\cal L}(\omega_{i_1}..\omega_{i_{l+1}},\omega_0)
\end{align*}
The total derivative term on the highest co-dimension intersections
(order $h$), can only contribute to $\partial M$.

Note that apart from our composition formula we have used only
Stokes' theorem, which is valid on a topologically non-trivial
manifold $M$ assuming a partition of unity ${f_i}$ subordinated to a
chosen covering\footnote{This is reviewed in appendix
\ref{manifoldint} as well as in the
textbooks~\cite{Choquet-Bruhat-82,von-Westenholz-78}.}. By
(\ref{ginv}) each of the terms appearing will be invariant w.r.t.
the structure group. So the last formula is valid over $M$
understanding each ${\cal L}$ as $\sum_i f_i {\cal L}$. By induction
we have proved the Lemma.$\Box$
\end{lemma}

We began with a smooth manifold with an Euler density action which
is completely independent of the choice of $\omega_0$. This gives
only a topological invariant of the manifold and is entirely
independent of any embedded hypersurfaces. The $\omega_i$'s, as well
as their number, are arbitrary also. So we see that we have
constructed a theory for arbitrarily intersecting hypersurfaces
which is a topological invariant. It is a trivial theory in that the
gravity does not see the hypersurfaces even if they contain matter.

\section{Dimensionally continued action}\label{Dimext}

Now we consider the dimensionally continued Euler density for
intersecting hypersurfaces separating bulk regions counted by $i$.
Inspired by Lemma \ref{Gravanisinter}, we postulate the action:
\begin{gather}\label{fundaction}
S_g=\sum_i \int_{i} {\cal L}_g(\omega_i,e) + \sum^h_{k=2}
\frac{1}{k!} \sum_{i_1..i_k} \int_{\{i_1...i_k\}} {\cal L}_g
(\omega_{i_1},..,\omega_{i_k},e)
\end{gather}

We will show that this action is `one and a half order' in the
connection.  We will need to revisit our derivation of the
composition rule in section \ref{biat}, this time interpolating
between the different vielbein fields $E^i(x)$, where the index
represents the region (the local Lorentz index being suppressed).
Physically, we require the metric to be continuous at a surface
$\Sigma_{1...p+1}$: $i^*E^i=E$ which implies $i^*(e^i)=e$. Here
$i^*$ is the pullback of the embedding of $\Sigma_{1...p+1}$ into
$M$. Define the Lagrangian on the surface $\Sigma_{1...p+1}$ to be:
\begin{align}\label{palatini}
{\cal L}(\omega^1,\dots,\omega^{p+1},e)
= & \int_0^1 dt_1\cdots dt^{p} \ \widehat{\zeta_{p}}
f(\Upsilon_1\cdots\Upsilon_{p}\Omega_{(p)}^{n-p}e_{(p)}),
\\\nonumber
(e_{(p)})_{a_1...a_{2n}} = & \frac{1}{(d-2n)!}
(E_{(p)})^{a_{2n+1}}\wedge..\wedge(E_{(p)})^{a_{d}}
\epsilon_{a_1...a_d}.
\end{align}
where $E_{(p)}=E_1-(1-t^1)(E_1-E_2)-...
-(1-t^1)...(1-t^{p})(E_{(p)}-E_{(p+1)})$ and $\zeta_p$ is given by
(\ref{zeta}).

Following through the calculation of section (\ref{biat}) , we pick
up extra terms, involving derivatives of $E_{(p+1)}$, from using the
Leibnitz Rule on $f$.
\begin{eqnarray*}
&& \frac{\partial}{\partial t_{p+1}}
f(\Upsilon_{1(p+1)}\cdots\Upsilon_{p(p+1)}
\Omega_{(p+1)}^{n-p}e_{(p+1)})=
\\ &&
\sum_{s=1}^{p+1} f(\Upsilon_{1(p+1)}\cdots\widehat{\Upsilon_{s(p+1)}}
\cdots\Upsilon_{p+1(p+1)}\Omega_{(p+1)}^{n-p} \frac{\partial
e_{(p+1)}}{\partial t^{s}}) +\\&& (n-p)
f(\Upsilon_{1(p+1)}\cdots\Upsilon_{p+1(p+1)}
\Omega_{(p+1)}^{n-p-1}D_{(p+1)}e_{(p+1)})+(...) \nonumber
\end{eqnarray*}
The $(...)$ denote terms which appear just as in section
(\ref{biat}). We will verify our assertion that the action is
one-and-a-half order by infinitesimally varying the metric and
connection in one region. We vary them as independent fields. Using
$t^{p+1}$ to interpolate between $E_{p+1}$ and $E_{p+1}+\delta
E_{p+1}$ and the corresponding variation of $\omega_{p+1}$:
\begin{gather}
\delta{\cal L} (\omega_1,\dots,{\omega_{p+1}},e) =  \int_0^1
dt^1\cdots dt^{p+1} \ \widehat{\zeta_p}\ \Xi\ +(...),
\end{gather}
\begin{align}\label{Xi}
\Xi = & \prod_{i=1}^p (1-t^i) \sum_{s=1}^{p}
f(\Upsilon_{1(p+1)}\cdots\widehat{\Upsilon_{s(p+1)}}
\cdots\delta\omega_{p+1}\Omega_{(p+1)}^{n-p} \frac{\partial
e_{(p+1)}}{\partial t^{s}}) \\\nonumber & -
f(\Upsilon_{1(p+1)}\cdots\Upsilon_{p(p+1)}\Omega_{(p+1)}^{n-p}
\frac{\partial e_{(p+1)}}{\partial t^{p+1}})\\\nonumber
& +\prod_{i=1}^p (1-t^i) (n-p+1)
f(\Upsilon_{1(p+1)}\cdots\Upsilon_{p(p+1)}\delta\omega_{p+1}
\Omega_{(p+1)}^{n-p-1}D_{(p+1)}e_{(p+1)}).
\end{align}
The $(...)$ denote terms which will cancel when intersections are
taken into account, just as in the topological theory (provided that
the metric is continuous). Above, we have made use of
$\Upsilon_{p+1(p+1)} =-(1-t^1)\cdots(1-t^P)\delta\omega_{p+1}$.

We want to check that the terms in (\ref{Xi}) involving
$\delta \omega_{p+1}$ vanish. Now
$E_{(p+1)}=E_1-(1-t^1)(E_1-E_2)-\cdots +(1-t^1)\cdots(1-t^{p+1})
\delta E_{p+1}$. Making use of formula (\ref{varvol})
\begin{align}
\frac{\partial }{\partial t_s}(e_{(p+1)})_{a_1...a_{2n}} = &
\frac{\partial}{\partial t_{s}}(E_{(p+1)})^{b} \wedge
(e_{(p+1)})_{a_1...a_{2n}b}\\\nonumber &\hspace{-.75in} =\sum_{i=1}^p
(1-t^1)\cdots\widehat{(1-t^s)}\cdots(1-t^i) (E_i-E_{i+1})^b \wedge
(e_{(p)})_{a_1...a_{2n}b}+{\cal O}(\delta E_{p+1}).
\end{align}
So we see the first term in (\ref{Xi}) vanishes if
$i^*(E_{i+1})=i^*(E_i)$ for all $i=1,\dots,p+1$ i.e. the metric is
continuous. Given this, we see that:
\begin{align}
i^*(D_{(p+1)} E_{(p+1)}) =& \,i^*\left(dE_{(p+1)} +\omega_{(p+1)}
\wedge E_{(p+1)}\right)
\\\nonumber
=&\,i^*\Big(d\big\{E_1+t^1(E_2-E_1)+\cdots+t^1\cdots t^{p+1}\delta
E_{p+1}\big\} \Big.\\\nonumber &\quad+\Big. \big\{\omega_1
+t^1\Upsilon_1+\cdots +t^1\cdots t^p \delta
\omega_{p+1}\big\} \wedge (E +
t^1\cdots t^{p+1}\delta E_{p+1})\Big)\\\nonumber &\hspace{-1.1in}
= i^*\Big(D(\omega_1)E_1+\sum_{i=1}^{p}
t^1\cdots t^i(D(\omega_{i+1})E_{i+1}-D(\omega_i)E_{i}) +{\cal
O}(\delta \omega_{p+1})+{\cal O}(\delta E_{p+1})\Big).
\end{align}
The third term in (\ref{Xi}) already contains a $\delta
\omega_{p+1}$ apart from the $D_{(p+1)} e_{(p+1)}$. Now
$D_{(p+1)}e_{(p+1)}$ is proportional to $D_{(p+1)}E_{(p+1)}$
so to first order in $\delta E_p$, this term vanishes if
$D(\omega_i)E_i = 0$ for all $i=1...p$.\footnote{The general
gravitation theory we consider is built as a sum of a desired set of
dimensionally continued topological densities. The coefficients can be taken to be
functions of scalar fields, making the theory dilatonic. The
variation with respect to the connection leads to a non-zero
torsion in this case. The torsion equation is a constraint on the
variation of the scalar fields. Solving it for the connection and
substituting in the Lagrangian one obtains explicitly an action
for the dilatonic fields.}

The only non vanishing term in (\ref{Xi}) is the second which
involves:
\begin{align*}
\frac{\partial }{\partial t^{p+1}}(e_{(p+1)})_{a_1...a_{2n}} = &
-(1-t^1)\cdots(1-t^p)(\delta E_{p+1})^b \wedge
(e_{(p+1)})_{a_1...a_{2n}b}
\\\nonumber
= & -(\delta e_{(p)})_{a_1...a_{2n}}.
\end{align*}
So we arrive at a simple expression for the variation of the action,
once the equation of motion for the connection and continuity of the
metric have been substituted:
\begin{gather*}
\delta {\cal L} (\omega_1,\dots,\omega_{p+1},e) = \int_0^1
dt^1\cdots dt^p\  \widehat{\zeta_p}\
f(\Upsilon_1\cdots\Upsilon_{p}\Omega_{p}^{n-p} \delta e)+(...).
\end{gather*}

Then, variation of an $\omega_i$  will vanish automatically upon
imposing the zero torsion condition and the continuity of the metric
at the intersections\footnote{The equations of motion for $\omega$
are satisfied by the zero torsion condition but are not, for $n \geq
1$ identical to it. There are potentially solutions of non-vanishing
torsion~\cite{Mardones-91}.}. Second, from the variation of the
frame $E^a$ we obtain field equation for gravitation and its
relation to the matter present, by
\begin{gather}
\delta_E S_g +\delta_E S_{\text{matter}}=0.
\end{gather}
The field equations are actually algebraically obtained, on the 
gravity side, using (\ref{varvol}). Note that although intersections
describe physically situation such as collisions there is a non-zero
energy momentum tensor at the intersection when the theory is not
linear in the curvature $2$-form. The dimensionally continued n-th
Euler density produces a non-zero energy tensor down to $d-n$
dimensional intersections. Explicitly, the gravitational equation of
motion for a simplicial intersection $\{1,...,p+1\}$, carrying
localised matter ${\cal L}_{m(1...p+1)}$ is
\begin{gather}\label{eqm}
(-1)^p\sum_n \alpha_n\int_0^1 dt^1\cdots dt^p\  \hat{\zeta_p}\
(\Upsilon_1\cdot\cdot\Upsilon_{p})^{a_1\dots a_{2p}}
(\Omega_{(p)})^{a_{2p+1}...a_{2n}} e_{ba_1...a_{2n}}
 =\frac{\delta}{\delta E^b} {\cal
L}_{m(1...p+1)}.
\end{gather}
We have dropped the wedge notation. The $(-1)^p$ factor comes from permuting
$\delta E^b$ to the left hand side.


\chapter{Homotopy parameters and simplices}\label{tdimensions}

It was shown in the previous chapter that
one can formally rewrite the Euler number in terms of a discontinuous
connection. One will then have boundary and intersection terms in the
integral. This amounts to turning the manifold into a honeycomb-type
lattice. The action, including boundary terms was found to be:
\begin{gather}\label{rewrite}
\int_M {\cal L}(\omega) = \sum_i \int_i {\cal L}(\omega_i)
+\sum_{k \geq 2} \frac{1}{k!}\sum_{i_1...i_k}\int_{\{i_1...i_k\}}
{\cal L}(\omega_{i_1},...,\omega_{i_k}).
\end{gather}
which came from the composition rule
\begin{equation}\label{comprule}
\sum_{s=1}^{p} (-1)^{s-p-1} {\cal L}(\omega_1,..,
\widehat{\omega_s},..,\omega_{p})
={\bm d}{\cal L}(\omega_1,..,\omega_{p})
\end{equation}
${\cal L}(\omega) = f(\Omega^n)$ and explicit formulae for the
intersection terms were found. We will find somewhat simpler
expressions for them in the next section.

I shall re-derive these results by introducing a closed form $\eta$
in a space $W \subset F$. $F$ is a product space of the manifold $M$
and the space of Homotopy parameters appearing in the definition of
the intersection terms. Our composition rule (\ref{comprule}) will
be shown to be equivalent to The condition that $\eta$ be closed.
The results can be presented in a simpler way by introducing a
multi-parameter generalisation of the Cartan Homotopy Operator.

The entire honeycomb is described by a few simple equations. All
sorts of intersections are accommodated in the scheme given by these
equations and by the shape of $W$. For the Euler density, we find an
explicit expression\footnote{This form $\eta$ is already known in
the mathematics literature~\cite{Gabrielov-75}. It is the Secondary
Characteristic form. The application of this approach to the
dimensionally continued Euler form is a new departure.} for $\eta$
and show that it is closed in $F$. The form of the intersection
terms will be clarified greatly.

Then we turn to a dimensionally continued action where the metric
enters into the action. We will show that the gravitational
intersection Lagrangians obey the same composition rule
(\ref{comprule}), further justifying our action (\ref{fundaction}).
This is because the dimensionally continued $\eta$ is still closed
over the domain of integration $W \subset F$.

We can write the action which generates all the intersection terms
as:
\begin{gather}
S = \int_W \eta,
\end{gather}
where $\eta$ is given by (\ref{final?}) for the Euler density and
(\ref{etadc}) for the dimensionally continued Euler density.

\section{A geometrical approach}

We want to describe the situation in the vicinity of an intersection
of co-dimension $p$ between different bulk regions. In this vicinity
there will also be intersections of lower co-dimension.

\begin{figure}
\begin{center}\mbox{\epsfig{file=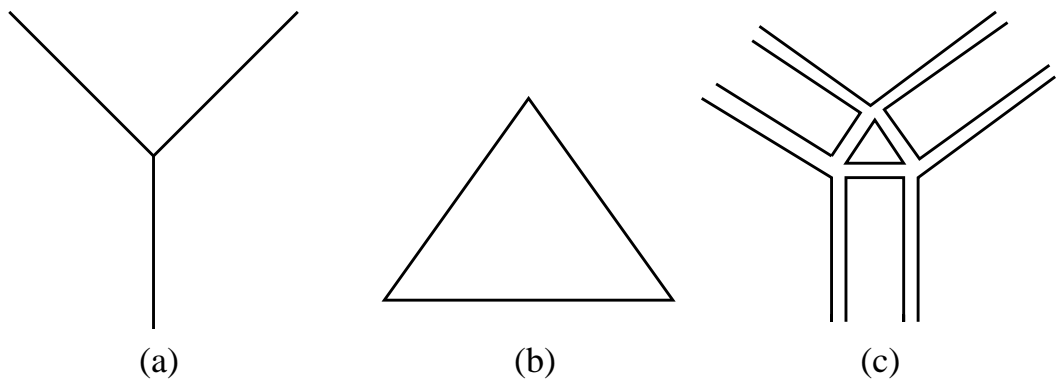, angle=-90, width=14cm}}
\caption{(a) Simplicial intersection; (b) Simplex in t-space; (c) A
projection of $W$.}\label{simp}
\end{center}
\end{figure}

At each intersection, we have a meeting of connections $\omega_i$ in
the different regions. Let us for the moment deal only with
simplicial intersections. We define the simplicial intersection of
codimension $p$ to be a surface of codimension $p$ where $p+1$
regions meet (fig. \ref{simp}a). It was found in chapter
\ref{inters} that the ${\cal L}(\omega_0,...,\omega_{p})$ is an
integral over $p$ different homotopy parameters interpolating
between the connections

Let us interpolate in the most symmetrical way. We introduce a
$p$-dimensional simplex (appendix \ref{simplexsection}) in the space
of some parameter $t$ (fig. \ref{simp}b ). Let us define the
interpolating connection:
\begin{gather}
\omega(t) \equiv \sum_{i=0}^{p} C^i(t)\ \omega_i, \qquad
\sum_{i=0}^{p} C^i(t) =1.
\end{gather}
Sometimes we will use the specific parameterisation denoted by the
Latin index $t^i$. $C^0 = 1-\sum_{j=1}^p t_j$ and $C^i = t^i$,
$i=1,..,p$.
\begin{gather}
\omega(t) = \omega_0 + \sum_{i=1}^{p}t^i \chi_i, \qquad \chi_i
\equiv \omega_i - \omega_0.
\end{gather}
To avoid confusion we shall use a Greek index $t^\alpha$ to denote
the general parameterisation.

Each order of intersection causes us to lose a dimension but gain an
extra connection. each new connection means an extra parameter of
continuous variation. As it were, in integrating, each time we lose a
$dx$ we gain a $dt$. With this in mind, we can think of our action as
living in a d-dimensional space which is a mixture of space-time and t
directions.

So we introduce the Space $F =  M \times S_{N-1}$, with $S_{N-1}$, a
simplex of dimension $N-1$, $N$ being the total number of regions on $M$.
At each of the $N$ points of the simplex there lies a continuous
connection form $\omega_i$ on $M$ with its support on some open set in $M$
containing the region i. Each contribution to our action will live on
some d-dimensional subspace of $F$. The technical reason for
introducing this is that the connection is continuous on $F$ and
integration is well defined. There is also an aesthetic reason. It is
quite a nice feature of the problem that the mathematics will take on
its simplest form when the t-space is a simplex. It provides a
geometrical picture which can simplify many calculations. For example,
the treatment of a non-simplicial intersection becomes easy as we shall
see.

Let us define a $d$-dimensional differential form in this space $F$
(where for convenience the $dx$'s are suppressed).
\begin{gather}\label{zetaform}
\eta \equiv \sum_{l=0}^{n} \frac{1}{l!} dt^{\alpha_1}
\wedge\cdot\cdot\cdot\wedge dt^{\alpha_l}
\wedge\eta_{\alpha_1...\alpha_l}(x,t),
\\\nonumber\eta_{\alpha_1...\alpha_l}\equiv\eta_{\alpha_1...\alpha_l
\mu_{l+1}...\mu_{d}}dx^{\mu_{1}}\wedge\cdot\cdot\cdot\wedge dx^{\mu_d}.
\end{gather}
We can now proceed to integrate this form over different faces of
$S_{N-1}$. A $p$-face (which we call $s_{0...p}$ or just $s$) is a
subsimplex of $S_{N-1}$ which interpolates between a total of $p+1$
different connections. Let us define ${\cal L}_{0...p}$ to be the
integral over the $p$-dimensional face:\footnote{Strictly there
should be a factor of $(-1)^{P(0,...,p)}$ in the middle term to
account for the orientation with respect to $S_{N-1}$. However, we
can choose $s$ to have the positive orientation by assuming the
points $0...p$ are in the appropriate order.}
\begin{gather}\label{defLeta}
{\cal L}_{0...p} \equiv \int_{s_{0...p}} \eta =
\frac{1}{p!}\int_{s_{0...p}} dt^{\alpha_1}\wedge\cdots\wedge
dt^{\alpha_{p}}\wedge\eta_{\alpha_1...\alpha_{p}},
\end{gather}
$\eta$ here being understood to be evaluated at $t=t(s)$ so that the
integral is a function of $x$ only. This integral picks out terms in
$\eta$ which are a volume element on the appropriate face. We would
like, for an appropriate choice of $\eta$, to identify this term
with ${\cal L}(\omega_0,...,\omega_p)$ as previously defined with
respect to the Euler density. We shall see that this can indeed be
done and we shall find a simple form for $\eta$.
\\
\begin{definition}
${\bm d}_{(F)} = {\bm d}_{(M)} + {\bm d}_{(t)}$ is the exterior
derivative on $F$. ${\bm d}_{(t)}$ and ${\bm d}_{(M)}$ are the
exterior derivative restricted to the simplex and $M$ respectively.
\end{definition}
\begin{proposition}\label{propositiona} The appropriate condition on $\eta$ such that
${\cal L}_{0...p}$ obeys the composition rule (\ref{comprule})is that
$\eta$ be a closed form, ${\bm d}_{(F)} \eta = 0$.
\end{proposition}
\begin{proposition}\label{propositionb} The form of $\eta$ corresponding to the Euler
density is
\begin{gather}\label{final?}
\eta = f\Big(\big[{\bm d}_{(t)}\omega(t)+\Omega(t)\big]^{\wedge n}\Big).
\end{gather}
$\Omega(t)$ is the curvature associated with $\omega(t)$. A similar
form has previously been considered in ref.~\cite{Gabrielov-75}.
\end{proposition}
\begin{proposition}\label{propositionc}
The intersection Lagrangian can be recovered by the specific choice:
\begin{gather}\label{expliciteta}
\eta_{1...p} = A_p f\big(\chi_1\wedge...\wedge\chi_p\wedge
\Omega(t)^{\wedge(n-p)}\big),\\
\nonumber A_p =(-1)^{p(p-1)/2}\frac{n!}{(n-p)!}
\\\label{Lagrange}\Rightarrow{\cal L}
(\omega_0,...,\omega_{p})
= A_p \int_{s_{01..p}} d^{p}tf\big(\chi_1\wedge...\wedge\chi_p
\wedge\Omega(t)^{\wedge(n-p)}\big).
\end{gather}
$\chi_i \equiv \omega_i - \omega_0$ and $\omega(t) \equiv \omega_0 +
\sum_\alpha t^i\omega_i$. The integration is over the right angled
simplex $\{t|\sum_i t_i \leq 1\}$. We will see that the asymmetry
between the point $0$ and the points $1,...,m$ is merely an
illusion.
\end{proposition}

To prove Proposition \ref{propositiona}, we will need to use Stokes' Theorem on
the face, $s$.
\begin{gather}\label{tinteg}
\int_s {\bm d}_{(t)} \eta = \int_{\partial s} \eta.
\end{gather}
The boundary of the simplex $s_{0...p}$ is
\begin{gather*}
\partial s_{0...p} = \sum_{i=0}^p (-1)^{i} s_{0...\widehat{i}...p}
\end{gather*}
with the orientation being understood from the order of the indices.

Now let us integrate the form ${\bm d}_{(M)}\eta$ over the face. We will
need to remember that in permuting this exterior derivative past the
dt's we will pick up a $\pm$ factor.
\begin{gather*}
{\bm d}_{(M)}\eta = \sum_l\frac{(-1)^l}{l!}dt^{\alpha_1}
\wedge\cdot\cdot\cdot\wedge dt^{\alpha_l}
\wedge {\bm d}_{(M)}\eta_{\alpha_1...\alpha_l}.
\end{gather*}
Using this information we may integrate over the $p$-face $s$
\begin{gather}\label{xinteg}
\int_s {\bm d}_{(M)}\eta = (-1)^{p} {\bm d}\int_s \eta.
\end{gather}
Combining equations (\ref{tinteg}) and (\ref{xinteg}):
\begin{gather}\label{intdFeta}
\int_{s_{0...p}} {\bm d}_{(F)} \eta = (-1)^{p} {\bm d}\int_{s_{0...p}} \eta
+ \sum_{i=0}^p (-1)^{i} \int_{s_{0...\widehat{i}...p}} \eta.
\end{gather}
If our form $\eta$ is closed in F, ${\bm d}_{(F)}\eta$ must necessarily
vanish term by term in the dt's and dx's. The integral on the right
hand side of (\ref{intdFeta}) must therefore vanish. Recalling the
definition (\ref{defLeta}) we have proved Proposition \ref{propositiona}:
\begin{gather}\label{equiv}
{\bm d}_{(F)}\eta =0\quad \Leftrightarrow\quad
{\bm d}{\cal L}_{0...p}=\sum_{i=0}^{p} (-1)^{p-i-1}
{\cal L}_{0...\widehat{{i}}...p}
\end{gather}
The condition that $\eta$ be closed is indeed equivalent to our
composition formula.
\\\\

The proof of Proposition \ref{propositionb} is in Appendix \ref{proofprop}. Proposition \ref{propositionc} follows from proposition
\ref{propositionb} by expanding the
polynomial but we will show it by more brute force method. First we
note that for $\omega(t)= \omega_0 + \sum_i t^i\chi_i$ we get a useful
formula:
\begin{gather}\label{handy}
\frac{\partial \Omega(t)}{\partial t^{i}} = D(t)\chi_i
\end{gather}
where $D(t)$ is the covariant derivative associated with $\omega(t)$.
Now let us verify explicitly that the right hand side of
(\ref{intdFeta}) vanishes. For convenience, we will drop the wedge
notation.
\begin{align}\label{proof}
\int_{\partial s}\eta =& \sum_{i=1}^p (-1)^{i-1}
\int dt^{1}\cdot\cdot\cdot
dt^{p}\frac{\partial}{\partial t^{i}}
\eta_{1...\widehat{i}...p }
\\ \nonumber
=& \sum_{i=1}^p (-1)^{i-1} (n-p+1)A_{p-1}
\int_s dt^{1}\cdot\cdot\cdot dt^{p}
f\Big(\chi_1\cdot\cdot\cdot\widehat{\chi_i}\cdot\cdot\cdot\chi_p
\frac{\partial \Omega(t)}{\partial t^{i}}\
\Omega(t)^{(n-p)}\Big)\\ \nonumber
=&\sum_{i=1}^p (-1)^{i-1} (n-p+1)A_{p-1}
\int_s dt^{1}\cdot\cdot\cdot
dt^{p}f\Big(\chi_1\cdot\cdot\cdot D(t)\chi_i\cdot\cdot\cdot\chi_p\
\Omega(t)^{n-p}\Big)\\ \nonumber=& (n-p+1)A_{p-1} {\bm d}_M
\int dt^{1}\cdot\cdot\cdot dt^{p}
f\big(\chi_1\cdot\cdot\cdot\chi_p\
\Omega(t)^{n-p}\big)\\ \nonumber
=&(n-p+1)\frac{A_{p-1}}{A_p} {\bm d}_M\int_s \eta
\end{align}
In the first line we have used Stokes' Theorem (\ref{tinteg}). In the
second and last line (\ref{Lagrange}) was used. In the third we made
use of (\ref{handy}). In the fourth the Bianchi identity for
$\Omega(t)$, (\ref{Bianchi}) and the invariance of the polynomial
(\ref{Invariance}). A comparison with equation (\ref{intdFeta}) tells
us that ${\bm d}_F \eta$ does indeed vanish provided
\begin{gather*}
A_p =(n-p+1)(-1)^{p-1} A_{p-1}\quad
\Rightarrow\quad A_p =  (-1)^{p(p-1)/2}\frac{n!}{(n-p)!}
\end{gather*}
There is a consistency check we need to make. We want to equate the
left hand side of (\ref{proof}) with a sum of terms
$\sum_{i=0}^p (-1)^{i-1}{\cal L}(\omega_0,...,\widehat{\omega_i},...,
\omega_p)$ . (\ref{Lagrange}) is not manifestly symmetrical since it
is constructed on a right-angled simplex with $\omega_0$ at the
origin. It follows straightforwardly from (\ref{expliciteta}) for
$i\neq 0$ that
\begin{gather}\label{straightforward}
\int_{s_{0..\hat{i}...p}}\eta = {\cal L}(\omega_0,...,
\widehat{\omega_i},...,\omega_p)
\end{gather}
What about the integral over the opposite face $s_{1...p}$?
\begin{align}\label{opposite}
\int_{s_{1...p}}\eta & = \sum_i(-1)^i \int_{s_{1...p}}
dt^1\cdot\cdot\cdot\widehat{dt^i}\cdot\cdot\cdot dt^p
f(\chi_1\cdot\cdot\cdot\widehat{\chi_i}\cdot\cdot\cdot\chi_p\
\Omega(t)^{n-p+1})
\\\nonumber&\hspace{-.3in}=(-1)^{i}\sum_i \int_{0}^1 dt^1\cdot\cdot\
\int_0^{1-\sum_{j=1}^{p-2}t^j}dt^{p-1}\ f\big(\chi_1\cdot\cdot
\widehat{\chi_i}\cdot\cdot\chi_l\
\Omega(t)^{n-p+1}\big)|_{t^p=1-\sum_{j=1}^{p-1}t^j}
\\\nonumber & \hspace{-.3in}=(-1)^{p-1}\int_{0}^1 dt^1\cdot\cdot
\int_0^{1-\sum_{j}t^j}
dt^{m-1}\ f\big(\bigwedge_{k=1}^{p-1}(\omega_k-\omega_p)\
\Omega(\omega_p+\Sigma_{i}t^i(\omega_i-\omega_p)^{n-p+1}\big)
\\\nonumber&={\cal L}(\omega_{1},...,\omega_{p})
\end{align}
The integral has been made manifestly equivalent to an integral over a
right angled simplex with the origin at $\omega_p$. In the second
line, we have made use of the fact that the multiple integrals are all
over the same face to replace the $\widehat{dt^i}$ with
$\widehat{dt^p}$. In the third line, we have used:
\begin{gather*}
\omega(t^p=1-\Sigma_{j=1}^{p-1}t^j)
=\omega_0+\sum_{i=1}^{p-1}t^i\chi_i+
(1-\sum_{i=1}^{p-1}t^i)
\chi_p=\omega_p +\sum_{i=1}^{p-1}t^i(\chi_i-\chi_p)
\end{gather*}
Also we have made use of the following relation, obtained from the
multilinearity and anti-symmetry of the function f with respect to the
$\chi$'s.
\begin{gather*}
f\big((\chi_1-\chi_p)\cdots(\chi_{p-1}-\chi_p)
\cdot\cdot\cdot) =(-1)^{p-i-1}
\sum_{i=1}^pf(\chi_1\cdots\widehat{\chi_i}
\cdots\chi_p\cdot\cdot\cdot).
\end{gather*}
This makes the anti-symmetry of ${\cal L}$ with respect to the
connections clear. Combining (\ref{expliciteta}),
(\ref{straightforward}), (\ref{opposite}), (\ref{proof}) and
(\ref{equiv}) completes the proof of Proposition \ref{propositionc}.

\section{The implications}

We have introduced a more abstract approach. What have we gained by
this?

\subsection{Dual simplices}

Firstly we see that the simplicial
intersection is related to a simplex in the parameter space. It is a
bit like turning the simplex inside out. As pointed out already, the
connection is smooth on $F = M\times S_N$ as the d-dimensional
Lagrangian density $\eta$ weaves its way through $x$ and $t$ space.

We can check that the form of the intersection action is found in this
chapter is the same as in the previous chapter, (\ref
{genl2}). The difference is that
here the domain of integration is over a simplex (also $\Omega(t)$ is
defined in a seemingly different way).

The measure for the simplex is
\begin{gather*}
\int_0^1 dt^1\, \int_0^{1-t^1} dt^2 \cdots \int_0^{1 -
\sum_{i=1}^{p-1}t^i} dt^p.
\end{gather*}
Let us make the co-ordinate transformation:
\begin{gather*}
t^i\rightarrow s^i = \frac{t^i}{1-\sum_{j=1}^{i-1}t^j}.
\end{gather*}
Now the domain of integration is as in the previous chapter.

We want to write the measure in terms of $s$. So we will need to
find the inverse Jacobian matrix of this co-ordinate transformation.
The Jacobian matrix is in upper diagonal form so the determinant is
just the product of diagonal entries.
\begin{gather*}
\det{\frac{\partial s^i}{\partial t^j}}
=  \prod_{i=1}^p \frac{\partial s^i}{\partial t^i}
=  \prod_{i=1}^p \frac{1}{1-\sum_{j=1}^{i-1}t^j}.
\end{gather*}
The following formula is easy to prove:
\begin{gather}\label{easy}
1-\sum_{j=1}^{i-1}t^j = (1-s^1)\cdot\cdot\cdot(1-s^{i-1})
=\prod_{j=1}^{i-1} (1-s^j).
\end{gather}
The determinant of the inverse matrix is, making use of (\ref{easy}):
\begin{gather*}
\det{\frac{\partial t^i}{\partial s^j}}
=\prod_{i=1}^p \prod_{j=1}^{i-1}(1-s^j)
= (1-s^1)^{p-1}(1-s^2)^{p-2}\cdot\cdot\cdot(1-s^{p-1})
= \prod_{i=1}^{p-1} (1-s^i)^{p-i}.
\end{gather*}
So the measure is, in terms of the $s$ parameters:
\begin{gather*}
\int_0^1\cdots\int_0^1 dt^1\cdots dt^p\,
\prod_{i=1}^{p-1} (1-s^i)^{p-i}
\end{gather*}
This is the same as the measure derived in chapter (\ref{inters}).
The discrepancy in the minus sign factors between
(\ref{expliciteta}) and (\ref{zeta}) is a factor of $(-1)^p$. This
comes from the fact that $\Upsilon_i = -\chi_i$. It just remains to
prove that $\Omega(t) = \Omega_{(p)}$ This follows from
\begin{align*}
\omega_{(p)} = & \omega_0
+ (1-s^1)(\omega_1-\omega_0)+\cdots
+(1-s^1)\cdots(1-s^p)(\omega_p-\omega_{p-1})\nonumber\\
= & t^1\omega_0  + t^2\omega_1+\dots
+t^p\omega_{p-1}
+ (1-\sum_{j=1}^p t^j)\omega_p
=\omega(t)
\end{align*}
$\omega_p$ is at the ``corner" of the simplex not $\omega_0$,
but we have already seem that the integral is independent of the
choice of the corner.

We have confirmed that the two definitions of the intersection forms
${\cal L}(\omega_0,\dots,\omega_p)$ are equivalent. We have proved
the same result by two independent approaches.

\subsection{F-space and Homotopy Operator}\label{F-space}

Secondly, we have a simple expression for the Lagrangian density in
$W\subset F$. It is given by (\ref{final?}) (recall  $W$ is the region
of integration in $F$). From equation (\ref{kindofC}) we notice that
${\bm d}_{(t)}\omega(t) + \Omega$ is just a kind of curvature of $\omega(t)$ on
$F$, call it $\Omega_F(t)$. In other words, the action is very trivial
on this enlarged space.
\begin{gather}
\label{etaisEuler} S= \int_W
\eta,  \qquad   \eta = f(\Omega_F(t)^n)
\end{gather}
and it obeys the same transgression formula as the thing we started
with, only now on $F$. Under continuous variation
$\omega(t)\rightarrow \omega'(t)$,
\begin{gather*}
f(\Omega_F(t)^n) -f({\Omega_F}'(t)^n)
={\bm d}_F T\!P(\omega(t),\omega'(t)).
\end{gather*}

The shape of $W$ is interesting. Every $d-1$ dimensional surface is
thickened in the $t$-direction by a $1$-dimensional simplex; These
meet at a d-2 surface in $M$ which looks like a triangular prism in
$W$, etc. (fig. \ref{simp}(c)).

The proof of proposition \ref{propositiona} can be thought of in terms of a
generalisation of Cartan's Homotopy formula to a higher number of
homotopy parameters. Let the operator $K_s$ be defined by $K_s \eta
\equiv \int_s \eta$ and let $K_{\partial s} \equiv \int_{\partial_s}
\eta$. The equation \ref{intdFeta} can be written as
\begin{gather}\label{compisCartan}
(K_s \cdot {\bm d}_F-(-1)^m {\bm d}_F\cdot K_s)\eta = K_{\partial s}\eta
\end{gather}
This reduces to the usual Cartan Homotopy Formula for the 1-simplex
$m=1$.
\begin{gather*}
(K_{01} {\bm d}_F+{\bm d}_F K_{01})\eta = \eta(1) -\eta(0)
\end{gather*}
In fact, the whole of our analysis can be reduced down to the two
equations (\ref{etaisEuler}) and (\ref{compisCartan}) In words: The
whole intersection Lagrangian is a density in some higher dimensional
simplicial product space over our manifold. The composition rules are
an expression of this higher dimensional Cartan Homotopy operator
acting on this Density.

\subsection{Non-simplicial intersections made simple}\label{simplynonsimp}

Thirdly, we now have a very efficient way to deal with a
non-simplicial intersection in $M$. This is where $k > p$ regions
meet at a co-dimension $p$ surface. We can easily deal with a
non-simplicial intersection by integrating over a simplicial complex
in t-space. More than one face of dimension $p$ are associated with
the same $(d-p)$-surface in $M$.

\begin{figure}
\begin{center}\mbox{\epsfig{file=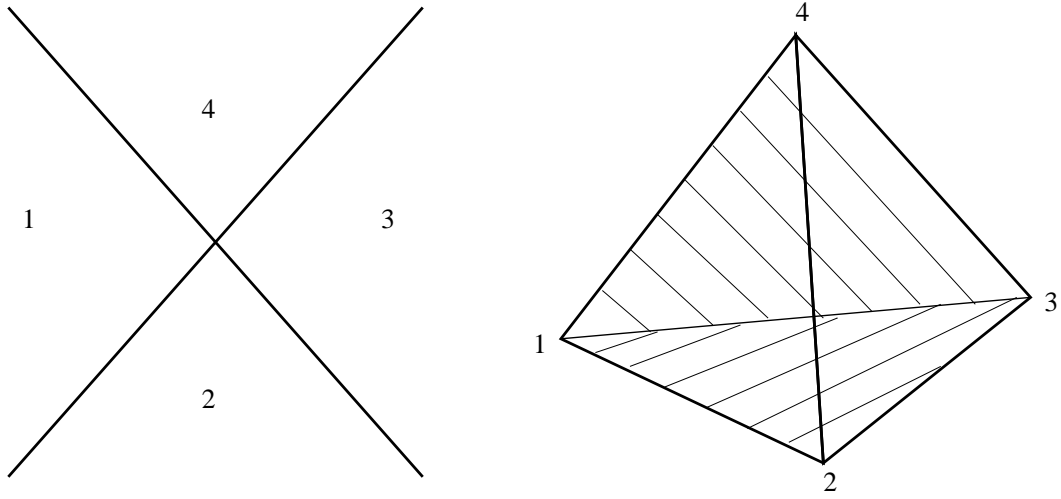, width=14cm}}
\caption{Non-simplicial Intersection. (a) Space-time, (b)t-space.}\label{non-simp}
\end{center}
\end{figure}

Lets consider a simple example. We have $4$ regions, $1,..,4$, meeting at
a co-dimension 2 intersection $I\subset M$ (fig.\ref{non-simp}).
There are four hypersurfaces $\{12\}$, $\{23\}$, $\{34\}$ and
$\{41\}$ meeting at $I$. As described in Lemma \ref{Gravanisinter},
on each hypersurface lives a term ${\cal L}_{ij}={\cal
L}(\omega_i,\omega_j)$. Integrating $\eta$ over the simplicial
complex $s_{123}\cup s_{341}$ will give us the intersection term.
The complex has boundary $s_{12}\cup s_{23}\cup s_{34}\cup s_{41}$.
Applying (\ref{intdFeta}) and ${\bm d}_{(F)}\eta =0$:
\begin{align}
\int_{s_{12}\cup s_{23}\cup s_{34}\cup s_{41}}\eta
=&d\int_{s_{123}\cup s_{234}}\eta
\nonumber\\ \Rightarrow
{\cal L}_{12}+{\cal L}_{23}+{\cal L}_{34}+{\cal L}_{41}
=& d\left({\cal L}_{123}+{\cal L}_{341}\right)\label{4int}
\end{align}
The appropriate term for
the non-simplicial intersection $I$ is
\begin{gather*}
\int_{I\subset M}\int_{s_{123} \cup s_{341}} \eta
\end{gather*}
or equivalently we can integrate over $s_{234}\cup s_{412}$ which has
the same boundary. More symmetrically, integrate over the chain
$c = \frac{1}{2}(s_{123}+s_{234}+s_{341}+s_{412})$.
\begin{gather}
\int_{I\subset M}\int_{c}  \eta.
\end{gather}
So the term which lives on the intersection is $\frac{1}{2}
({\cal L}_{123}+{\cal L}_{234}+{\cal L}_{341}+{\cal L}_{412})$. It is
the degenerate case where two simplicial intersections $\{123\}$ and
$\{341\}$ (or equivalently $\{234\}$ and $\{341\}$) coincide.

\newpage

\section{Dimensionally continued Euler density}\label{dimargument}

So far we have been considering the topological density. This is not
suitable as a Lagrangian. We know that the action yields no equations
of motion. The point is that we can apply what we have learned to the
dimensionally continued densities. The Lovelock Lagrangian is a
combination of such densities:
\begin{gather}
{\cal L} =\sum _{n=0}^{[d/2]} \alpha_n
f(\Omega^{n}\, E^{d-2n}),
\\\nonumber
f(\Omega^{n}\, E^{d-2n})
= \Omega^{a_1a_2}
\wedge\cdot\cdot\cdot\wedge\Omega^{a_{2n-1}a_{2n}}
\wedge E^{a_{2n+1}}\wedge\cdots\wedge
E^{a_d}\epsilon_{a_1...a_{d}}.
\end{gather}
We assume that the connection is a metric compatible (Lorentz)
connection. There are now explicit factors of the vielbein frame
$E^a$ appearing in the action.

We have a manifold $M$, of dimension $d$, with regions, $i$, divided by
walls of matter. The metric on $M$ is continuous but the derivative of
the metric is discontinuous at the surfaces. Once again we rewrite the
Lagrangian in terms of the continuous connections $\omega_i$ and
boundary terms. We interpolate as before:
\begin{gather}
E(t) \equiv \sum_{i=1}^{p+1}C_i(t)\ E_i,\qquad
\omega(t) \equiv \sum_{i=1}^{p+1} C_i(t)\ \omega_i,
\qquad \sum_{i=1}^{p+1}C_i(t) =1.
\end{gather}
The quantities $D(t)E(t)$ and
${\bm d}_{(t)}E(t)$ vanish everywhere on our domain of integration $W $.
In fact, a good way to define $W$ is: $W$ is the region in $F$ where
$E(t,x) = E(x)$.
(of course ${\bm d}_{(M)}E(t,x)$ is a function of $t$
because the derivative of the metric is discontinuous on $M$).
\begin{gather}\label{DtEt=0}
E(x,t) = E(x), \qquad (x,t) \in W,
\end{gather}
\begin{align}\label{ttorsion}
D(t)E(t)^a &= \sum_i C^i {\bm d}_{(M)} E_i^a +
\sum_i C^i {\omega_i}^a_{\ b}\wedge E(t)^b\\\nonumber
&=\sum_i C^i {\omega_i}^a_{\ b}\wedge (E(t)-E_i)^b
= 0, \quad (x,t) \in W.
\end{align}
We have used the zero torsion condition ${\bm d}_{(M)}E_i^a +\omega^a_{\ b}
\wedge  E_i^b =0$.

\begin{figure}
\begin{center}\mbox{\epsfig{file=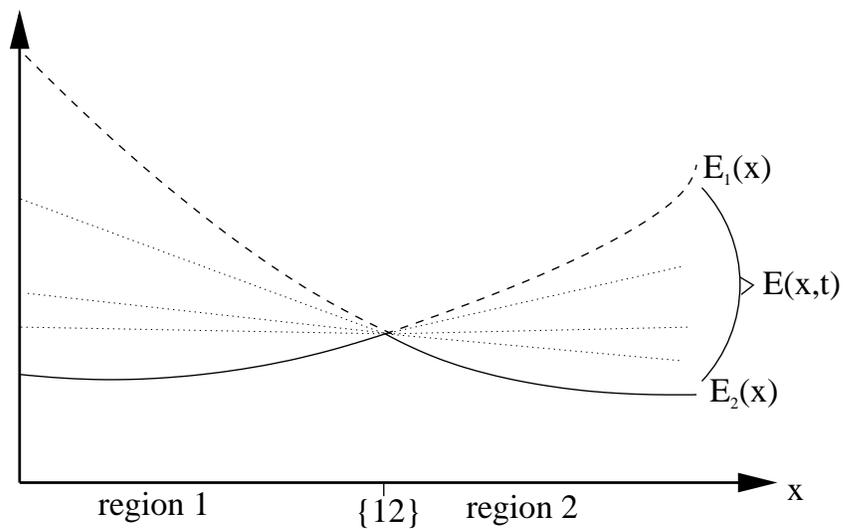, height=7cm}}
\caption{{\small E(x,t)=E(x) at the intersection \{12\}}}\label{Ext}
\end{center}
\end{figure}

\begin{figure}
\begin{center}\mbox{\epsfig{file=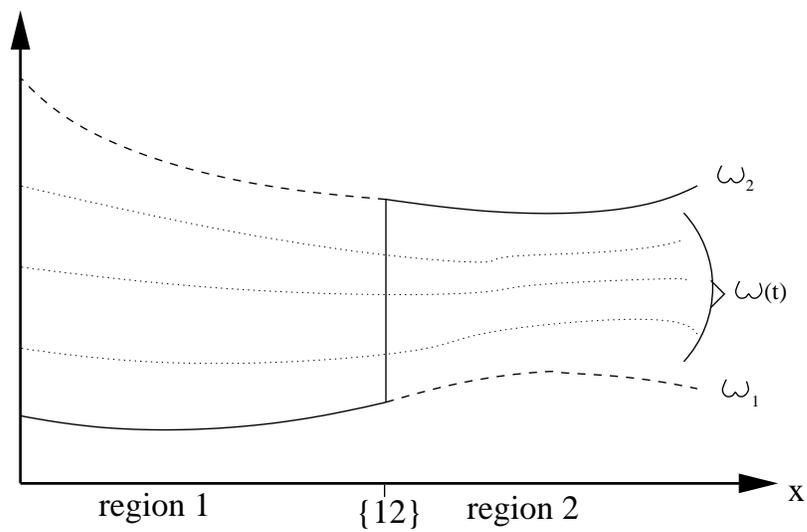, height=7cm}}
\caption{{\small $\omega(t)$ interpolates between the $\omega_i$.}}
\label{omegat}
\end{center}
\end{figure}

The case of a single hypersurface is illustrated in
figures \ref{Ext} and \ref{omegat}. At the intersection,
$E^1=E^2$. At the intersection {\it and only at the intersection},
both are fields living on $W$. $D(t)E(t)=0$, where $E(t)=C(t)E^1+(1-C(t))E^2$.
In region $1$, the above formula does not hold,
we only have $D^1E^1=0$.

\begin{proposition}\label{CompLovelock}
The composition rule, (\ref{comprule}), holds
for the Lovelock action, when evaluated at a point on $M$
where all the regions $1,...,p$ intersect. Define
the form:
\begin{gather}\label{etadc}
\eta_{DC} = f\big(({\bm d}_{(t)}\omega(t)+\Omega(t))^{\wedge n}
\wedge E(t)^{\wedge{(d-2n)}}\big)
\end{gather}
The terms entering into the composition rule
will be terms in the expansion of $\eta_{DC}$
picked out by integrating over the appropriate simplex
$s_{1...p}$ in $F$.
\\\\
We need just to show that $\eta_{DC}$
is closed when restricted to the space $W$. The proof
then follows from Proposition \ref{propositiona}.
We make use of the invariance property of the polynomial
contracted with the epsilon tensor. In $W$:
\begin{align*}
{\bm d}_{(F)} f\big(({\bm d}_{(t)}\omega(t)+\Omega(t))^{\wedge n}
\wedge E(t)^{\wedge{(d-2n)}}\big) & =  D_{F}(t)f
\big(({\bm d}_{(t)}\omega(t)+\Omega(t))^{\wedge n}
\wedge E(t)^{\wedge{(d-2n)}}\big)\\  & =  0
\end{align*}
We have defined the covariant derivative on $F$: $D_F(t) \equiv
{\bm d}_{(t)}+D(t)$. The integral vanishes because $D_F(t)({\bm d}_{(t)}\omega(t)+
\Omega(t))=0$ by (\ref{DFeta}) and $D_F(t)E(t) =0$ as explained above.
\end{proposition}

For the dimensionally continued case, $\eta_{DC}$ and ${\cal L}$ are
no longer Euler densities. It was therefore not obvious that our
composition formula should survive. It does survive though because
$\eta_{DC}$ is still a closed form in  $W \subset F$.

As a consequence of the composition rule, we can apply exactly the same
argument as in Lemma \ref{Gravanisinter} to prove:
\begin{lemma}\label{Finalaction}
The Lovelock action, in the presence of hypersurfaces,
is given by (\ref{fundaction}) which we may now write as
\begin{gather}
S = \int_W \eta_{DC}
\end{gather}
\end{lemma}
Also, as we saw in the previous chapter, the infinitesimal variation
of the action with respect to the connection vanishes, provided we
impose the torsion free condition on the connection and continuity of
the metric. So the equations of motion just come from the explicit
variation with respect to the vielbein.

\section{The interpolating curvature}

We find the explicit form of the interpolating curvature:\footnote{
We can see from(\ref{intercurv}) that even when the bulk regions
themselves are flat, the interpolating curvature may not vanish
\begin{gather}
\Omega_i = 0,\ \forall i \nonumber\\\Rightarrow
\Omega(t) = \frac{1}{2}\sum_{i,j}C^iC^j
[\omega_i-\omega_j,\omega_i-\omega_j] .
\end{gather} }
\begin{align}\label{intercurv}
\Omega(t) =& \sum_i C^i d\omega_i +\frac{1}{2}
\sum_{i,j} C^iC^j[\omega_i,\omega_j]\nonumber\\\nonumber
=&\sum_{i,j} C^iC^j \big( d\omega_i +
\frac{1}{2} [\omega_i,\omega_j]\big)\\\nonumber
=& \frac{1}{2}\sum_{i,j}C^iC^j \Big(\Omega_i+\Omega_j
-\frac{1}{2}[\omega_i,\omega_i]-\frac{1}{2}[\omega_j,\omega_j]
+[\omega_i,\omega_j]\Big)\nonumber\\
= & \sum_{i,j}C^iC^j A_{ij},\nonumber\\
A_{ij} = & \frac{1}{2}\Big(\Omega_i+\Omega_j
-\frac{1}{2}[\omega_i-\omega_j,\omega_i-\omega_j]\Big).
\end{align}
Making use of Appendix \ref{secondtensors},
\begin{align}
\Omega(t) \label{intercurve2}
\sim & \frac{1}{2}\sum_{i\neq j} C^iC^j
\bigg\{ \sum_{k\neq i} \frac{1}{p} \left(\Omega_{(ik)}
-{\frak n}\!\cdot\!{\frak n}_{(ik)} K_{(i)}\wedge K_{(i)}\right)\bigg.
\\\nonumber
&\qquad\qquad\quad +\sum_{k\neq j} \frac{1}{p} \left(\Omega_{(kj)}
-{\frak n}\!\cdot\!{\frak n}_{(kj)} K_{(j)}\wedge K_{(j)}\right)
\\\nonumber
&\qquad\qquad\quad \bigg. +{\frak n}\!\cdot\!{\frak n}_{(ij)}
\left(K_{(i)}
\wedge K_{(i)} + K_{(j)}\wedge K_{(j)}
-2K_{(i)}\wedge K_{(j)}\right)\bigg\}
\end{align}
where the symbol $\sim$ means excluding terms with normal Lorentz indices.

\section{Intersecting hypersurfaces in more general theories via
closed forms}

The existence of the closed form $\eta$ is very important for the
description of intersecting membranes. The question arises:
\emph{are there any other forms which can be generalised to closed
forms in $W$ space?}

We have dealt with Lovelock Lagrangian because it includes the
Einstein-Hilbert Lagrangian of General Relativity, and because of
the many nice mathematical and physical features discussed in
Chapters 1 and 2. The relationship between the Lovelock Lagrangian
and the Euler Characteristic Class was important in the construction
of the closed form. In looking for more generalised forms, the first
thing to consider are other Characteristic Classes. A general
Characteristic form,
\begin{gather*}
P(\Omega^n)
\end{gather*}
can be generalised to the Secondary Characteristic
form~\cite{Gabrielov-75}
\begin{gather}
P([{\bm d}_t \omega(t) + \Omega(t)]^n).
\end{gather}
For example, in four dimensions, there is the Pontryagin form: (the
Hirtzebruch signature)
\begin{gather}\label{Hirzebruch}
\tau \propto \Omega^a_{\ b}\wedge\Omega^b_{\ a}.
\end{gather}
In this case, the invariant tensor which contracts the indices is
$\eta_{a_2a_3}\eta_{a_4a_1}$ rather than $\epsilon_{a_1\dots a_4}$.
The more familiar tensor form of the Hirtzebruch signature is:
\begin{gather*}
\propto \sqrt{g} R^{\mu\nu}_{\ \ \alpha\beta} R^{\rho\sigma}_{\ \
\gamma\delta}\epsilon^{\alpha\beta\gamma\delta}
g_{\mu\rho}g_{\nu\sigma}.
\end{gather*}

The Secondary Characteristic form associated with the Hirtzebruch
signature is:
\begin{gather}
\overline{\tau} = [{\bm d}_t \omega(t) + \Omega(t)]^a_{\ b}\wedge
[{\bm d}_t \omega(t) + \Omega(t)]^b_{\ a}
\end{gather}
which is clearly a closed form in $F$, $({\bm d}_{(t)} + {\bm
d}_{(M)})\overline{\tau} =0$.

In higher dimensions, there are other Pontryagin forms. for example,
in eight dimensions, there are two such forms:
\begin{gather*}
\Omega^{a}_{\ b}\wedge\Omega^b_{\ c}\wedge\Omega^c_{\ d} \wedge
\Omega^d_{\ a},
\\ (\Omega^{a}_{\ b}\wedge\Omega^b_{\ a})\wedge(\Omega^c_{\ d} \wedge \Omega^d_{\ c}).
\end{gather*}

These forms characterise the topology of $M$ (or of the fiber bundle
if $\Omega$ is the curvature of a \emph{gauge} connection). To
describe a theory with local degrees of freedom, we need to
generalise them. There are various ways of doing this, but all of
them involve a non-zero torsion. The forms that one can make
are~\cite{Mardones-91}:
\begin{align}
 {\bm R}_A &:= \Omega^{a_1}_{\ a_2}\wedge\cdots\wedge\Omega^{a_A}_{\ a_1}, \\
 {\bm V}_A &:= E_{a_1} \wedge
 \Omega^{a_1}_{\ a_2}\wedge\cdots\wedge\Omega^{a_A}_{\ b}\wedge E^b,\\
 {\bm T}_A &:= T_{a_1} \wedge
 \Omega^{a_1}_{\ a_2}\wedge\cdots\wedge\Omega^{a_A}_{\ b}\wedge T^b,\\
 {\bm K}_A &:= T_{a_1} \wedge
 \Omega^{a_1}_{\ a_2}\wedge\cdots\wedge\Omega^{a_A}_{\ b}\wedge E^b.
\end{align}
There are various ways of forming $d$-forms out of products of
these.
\begin{gather}
\zeta = {\bm R}_{A_1} \wedge \cdots \wedge {\bm R}_{A_r} \wedge {\bm
T}_{B_1} \wedge \cdots \wedge {\bm T}_{B_t} \wedge{\bm
V}_{C_1}\wedge\cdots \wedge {\bm V}_{C_v} \wedge  {\bm K}_{D_1}
\wedge \cdots \wedge {\bm K}_{D_k},
\end{gather}
where \begin{gather} 2(\sum_{i=1}^r A_i + \sum_{i=1}^tB_i
+\sum_{i=1}^v C_i + \sum_{i=1}^k D_i)+4t+2v+3k = d,
\\
A_i,\ B_i\ \text{even},\quad C_i\ \text{odd},\quad D_i \neq D_j \
\text{if}\ i\neq j.
\end{gather}
Remembering the Bianchi Identity,
\begin{gather*}
DT^a = \Omega^a_{\ b} \wedge E^b,
\end{gather*}
we see that all of these terms vanish if the torsion is constrained
to be zero.

We can generalise as before:
\begin{align*}
 E &\to {E}(t),\\
 \Omega &\to {\bm d}_t {\omega}(t) + {\Omega}(t),
 \\ T &\to {\bm d}_t {E}(t) +{D}(t){E}(t),
 \\\zeta(E,\omega) &\to \overline{\zeta}(E(t),\omega(t)).
\end{align*}

Now we hit a problem. If the torsion is not constrained to be zero,
then $D(t) E(t)$ will not vanish. Neither will $D(t) T(t)$ be
expected to vanish. The only obvious way to make ${\bm
d}_{(F)}\overline{\zeta}$ vanish  is to start with a locally exact
form in $M$. Alternatively, we can impose a more complicated
constraint on the torsion.

Some-times the only constraint needed is to impose the equation of
motion involving the torsion. For example, a $d$-form made from the
tensors $\eta, \epsilon, \Omega, E$. Let
\begin{gather}
\zeta = \Omega^{a_1\dots a_{2n}} \wedge \Xi(E)_{a_1\dots a_{2n}}
\end{gather}
where $\Xi$ is a $(d-2n)$-form and $n >0$.
\begin{proposition}
The form $\overline{\zeta}(E(t),\omega(t))$ is closed when the field
equation from the variation of $\omega(t)$ is satisfied (on shell).
\end{proposition}
{\bf Proof:} The Euler variation w.r.t. $\omega(t)$ gives the field
equation:
\begin{gather}
 \sum_{i=2}^{n}[{\bm d}_t {\omega}(t) + {\Omega}(t)]^{a_3\dots
 a_{2k}} \wedge D\Xi(E(t))_{a_3\dots \widehat{a_{2i-1}}\widehat{a_{2i}}ab \dots a_{2k}}=0.
\end{gather}
By the Bianchi identity, the exterior derivative of ${\cal L}$
vanishes on shell:
\begin{gather}\label{on-shell}
 {\bm d} \overline{\zeta} =  [{\bm d}_t {\omega}(t) + {\Omega}(t)]^{a_1\dots a_{2k}}D\Xi(E(t))_{a_1\dots a_{2k}}
 = 0\quad (\text{on shell}).
\end{gather}

More generally, when $T$ appears explicitly in $\Xi$, this on-shell
condition is not sufficient. Things are a bit more complicated. This
is the subject of further investigations.

It is important to notice that (\ref{on-shell}) applies to the
Lovelock Lagrangian. If we allow torsion in the Lovelock theory, the
thin shell description still applies on shell.
\\

If we are interested in a theory of metric and derivatives of the
metric (i.e. the same field content as GR), we will set the torsion
to zero. This approach seems to lead just to topological terms and
the dimensionally continued Euler densities. At the moment, it seems
that the Lovelock lagrangian is special in this sense. This would be
in accordance with the intuition which comes from Lovelock's
theorem: that the Lovelock Lagrangian is the only one which gives
field equations quasi-linear in second derivatives of the metric.
This quasi-linearity is obviously important in defining thin
hypersurfaces, as discussed in chapter \ref{junction}. Also the
closed form approach is a way of defining thin hypersurfaces. This
consideration leads to the interesting conjecture:
\begin{conjecture}
If the torsion is constrained to vanish and $\eta(E(t),\omega(t))$
is a closed form in $W$, such that $\eta(E,\omega) = {\cal
L}(E,\omega)$ on $M$ then ${\cal L}$ must be either a locally exact
form or a sum of dimensionally continued Euler densities.
\end{conjecture}

The quasi-linearity of the quadratic Lovelock theory is also
important in establishing that it is ghost free about a flat
background. We might also conjecture: ${\bm d}_{(F)} \eta =0
\Rightarrow$ ${\cal L}$ \emph{gives a theory which is ghost-free
about a flat background}(?) The converse is definitely not true. A
generic Lagrangian cubic in the curvature
will be free of ghosts even if there is no corresponding closed
form.

\chapter{An explicit example}\label{example}

As an example, we will look at a co-dimension $2$ intersection of
$N$ hypersurfaces. This may represent an interesting physical
application, related to brane physics. We shall further assume that
the hypersurfaces and intersection are not null. We then find
explicitly the Junction condition for the intersection with the
$n=2$ dimensionally continued Euler density (Gauss-Bonnet) term. We
also express the energy conservation in the form of relations among
the energy tensors involved in this case.

Intersections and collisions have been extensively studied for the
case of 's theory~\cite{Dray-85,Dray-86,Langlois-02,Berezin-02}. In
particular, Langlois et al~\cite{Langlois-02} treated colliding
shells in AdS black hole backgrounds. They reduced the problem to
adding rapidities just as in special relativity. However, there were
some outstanding questions. An important assumption was that upon
traversing an infinitesimal loop round the intersection, there is no
overall Lorentz boost. This is basically the same as saying there is
no deficit angle. Why no possibility of a deficit angle at the
intersection? Intuitively, this would be associated only with matter
localised on the intersection. We shall see that for Einstein's
theory this is true. For the general Lovelock theory things will be
different. It is this difference which is the main physical result
of this work.

\section{The $N$ hypersurface intersection}

\begin{figure}
\begin{center}\mbox{\epsfig{file=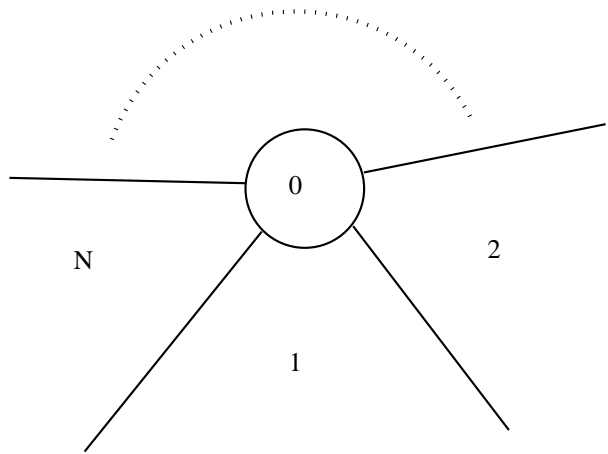, width=8cm}}
\caption{{\small The non-simplical intersection viewed as the
limit $\{0\}\to I$, where I is a co-dimension 2 surface.}}\label{oregion}
\end{center}
\end{figure}

We have a non-simplicial intersection, of co-dimension $2$. One way
to deal with this is to use the method mentioned in chapter
\ref{tdimensions}. Another way is to have a central cylindrical
region and then shrink this region to zero\footnote{These methods
are equivalent. We can see this by the following: Up to total
derivatives, we can eliminate $\omega_I$ from (\ref{Nfundint}).
Adding trivially a set of terms ${\cal
L}(\omega_i\omega_1\omega)+{\cal L}(\omega_1\omega_i\omega)=0, \
i=3..N$ and using the composition rule we have
\begin{equation*}
{\cal L}(\omega_1\omega_2\omega_3)+{\cal
L}(\omega_1\omega_3\omega_4)+..+{\cal
L}(\omega_1\omega_{N-1}\omega_N)
\end{equation*}
plus an exact form containing $\omega_I$. For N=4, this agrees with
(\ref{4int}).}. We divide the space-time into $N+1$ regions formed
by $N$ surfaces intersecting a cylinder in the middle. Taking the
cross section of the system, we see a circle with $N$ outgoing
lines, without further intersections (fig. \ref{oregion}).

Call the connection inside the circle $\omega_I$, associated with
the metric $\gamma$. There are $N$ co-dimension 2 simplicial
intersections. We calculate the contributions at the intersections,
and then take the limit whereby they all coincide i.e. we are going
to take the limit of the circle to zero size. $\gamma$ is chosen to
be the induced metric of the resulting co-dimension $2$
intersection, $I$.

The action terms due to the walls are:
\begin{gather*}
\int_{12} {\cal L}(\omega_1,\omega_2,e)+\int_{23} {\cal
L}(\omega_2,\omega_3,e)+..+\int_{N1} {\cal L}(\omega_N,\omega_1,e)
\end{gather*}
and, from the $k=3$ terms in (\ref{fundaction}), we have the
intersection contributions to the action:
\begin{gather} \label{Nfundint}
\int_I {\cal L}(\omega_1,\omega_2,\omega_I,e)
+{\cal L}(\omega_2,\omega_3,\omega_I,e)+
\cdots+{\cal L}(\omega_N,\omega_1,\omega_I,e).
\end{gather}

In order to calculate the equation of motion explicitly in terms of
intrinsic and extrinsic curvature tensors we should introduce the
connection $\omega_{ij}$ at the common boundaries. This is the
connection associated with the induced metric $h$ on the wall
$\{ij\}$.

We are interested in the dimensionally continued theory. By the argument
in Section \ref{dimargument}, the composition rule has validity on the
hypersurfaces and intersection. Using the composition rule:
\begin{gather} \label{bdycompN}
{\cal L}(\omega_i,\omega_j,e)={\cal L}(\omega_i,\omega_{ij},e)+{\cal
L}(\omega_{ij},\omega_j,e)- d{\cal L}(\omega_i,\omega_j,\omega_{ij},e)
\end{gather}
we obtain some more contributions at the intersection.
Combining the term from the total derivative in (\ref{bdycompN})
with (\ref{Nfundint}) we get terms:
\begin{gather}\label{newI}
{\cal L}(\omega_i,\omega_j,\omega_I,e) - {\cal
L}(\omega_i,\omega_j,\omega_{ij},e).
\end{gather}

Using the composition rule again:
\begin{gather*}
{\cal L}(\omega_i,\omega_j,\omega_I,e)-{\cal L}(\omega_i,\omega_j,\omega_{ij},e)
={\cal L}(\omega_i,\omega_{ij},\omega_I,e)
-{\cal L}(\omega_j,\omega_{ij},\omega_I,e)
\end{gather*}
(plus an exact form which we can neglect). Substituting this into
(\ref{newI})we obtain finally
\begin{gather} \label{Ld-2}
{\cal L}_{d-2}=(12)+ (23)+...+(N1) \ ; \quad (ij)\equiv {\cal
L}(\omega_i,\omega_{ij},\omega_I,e)-{\cal
L}(\omega_j,\omega_{ij},\omega_I,e).
\end{gather}
Now we can express everything in terms of the bulk region
connections and two types of second fundamental form:
$\theta_{i|ij}$ of the surface $\{ij\}$ induced by the region $i$;
and the $\tilde{\theta}_{ij}$ of the intersection induced by
$\{ij\}$.\footnote{Note that $\tilde{\theta}_{ij}$ is defined as the
Second fundamental form of $I$ with respect to the wall $\{ij\}$
{\it not} the bulk.}
\begin{gather*}
\theta_{i|ij}=\omega_i-\omega_{ij} \ , \qquad
\tilde{\theta}_{ij}=\omega_{ij}-\omega_I.
\end{gather*}
 Their definition in terms of
the associated normal vectors and extrinsic curvature tensors are
given in the appendix.

In order to write the simplest non trivial equation of motion for
the common  intersection of $N$ $(d-1)$-dimensional surfaces we consider
the $n=2$ dimensionally continued Euler density. Applying
(\ref{genl}) or (\ref{genl2}) we find easily
\begin{gather}\label{Iterm}
{\cal L}(\omega_i\omega_{ij}\omega,e) =
f(\theta_{i|ij},\tilde{\theta}_{ij},e).
\end{gather}

We assume a gravity Lagrangian of the form
\begin{gather}
{\cal L}_G= \beta_0{\cal L}^{(0)}+\beta_1{\cal L}^{(1)}
+\beta_2 {\cal L}^{(2)}
\end{gather}
where ${\cal L}^{(n)}$ is the (dimensionally continued) n-th Euler
density and $\beta_2$ is constant of dimensions (length)$^2$, the
coupling of the Gauss-Bonnet term. This theory is sometimes called
the Einstein-Gauss-bonnet theory.

Varying the frame $E^a$ we obtain the equations of motion.
We express the second fundamental form $\theta^{ab}$ in terms of
the usual extrinsic curvature tensor $K_{ab}$ by equation (\ref{thetainK})
\begin{equation*}
\theta^{ab}= 2 {\frak n}\!\cdot\!{\frak n} \ {\frak n}^{[a}  K^{b]}_c E^c
\end{equation*}
where ${\frak n}^{\mu}_{ij}$ is the normal vector of a
$(d-1)$-dimensional surface,$\{ij\}$, with orientation induced by
the bulk region $i$. The $i,j$ indices have been suppressed for the
sake of sanity. $\tilde{\theta}^{ab}$  is defined similarly for
$v^{\mu}$, the normal vector of the intersection embedded in a given
$(d-1)$-dimensional hypersurface. We define the corresponding
extrinsic curvature tensor:
\begin{gather*}
C_{\mu\nu}=\gamma^{\rho}_{\mu}
h^{\sigma}_{\nu} \nabla_{\rho} v_{\sigma}
\end{gather*}
with $\gamma_{\mu\nu}=
h_{\mu\nu}-\epsilon(v) v_{\mu} v_{\nu}$.
We have
\begin{gather*}
\tilde{\theta}^{ab}= 2 v\cdot v \ v^{[a}  C^{b]}_c E^c.
\end{gather*}
Plugging this into (\ref{Iterm}) and (\ref{juncminushalf}), we get the
junction condition for the intersection:
\begin{equation} \label{intereqmotion}
\sum 2\Delta\left\{ ({\cal K}\bar C)^{ab}+(\bar{{\cal K}}C)^{ab}- \frac{1}{2}
\gamma^{ab}
{\rm Tr}({\cal K} \bar C+\bar{{\cal K}}C) \right\} = \tilde{T}^{ab}_{(d-2)}
\end{equation}
where ${\cal K}^{ab} = \gamma^a_c\gamma^b_d K^{cd}$ is the
projection of $K^{ab}$ onto the tangent space of $I$. The sum in
(\ref{intereqmotion}) is over all terms in (\ref{Ld-2}), one for
each hypersurface. We use the notation $\bar{\cal K}_{ab}={\cal
K}_{ab}-\gamma_{ab} {\cal K}$, where ${\cal K}=\gamma^{ab} {\cal
K}_{ab}$, and compact matrix multiplication, for example $(\bar{\cal
K}C)^{ab}=\bar{\cal K}^a_cC^{cb}$. $\Delta \{\cdots\}_{ij} \equiv
\{\cdots\}_{ij} -\{\cdots\}_{ji}$ where $\{ij\}$ corresponds to the
positive orientation. $T^{ab}_{d-2}$ is the energy momentum tensor
which is localised at the intersection.

Notice that it was not {\it necessary} to introduce the intrinsic
curvature of the intersection $C$. It is always possible to write
everything in terms of the intrinsic and extrinsic curvatures of the
walls only. This is discussed briefly in the Outlook section.

\section{Energy conservation at the intersection}

Let us see the implications of these results for the question of energy
conservation. We recall that the local expression
of the energy-momentum tensor
conservation is related to the diffeomorphism invariance of the
action, under which the metric tensor changes as $\delta
g_{ab}=2 \nabla_{(a} \xi_{b)}$ where $\xi^a=\delta x^a(x)$ are
infinitesimal coordinate transformations. Note that $2 \nabla_{(a}
\xi_{b)}=\delta g_{ab}$ has to be continuous.
In the previous chapters we have seen that the variation of the
hypersurface action gives terms on the intersection through application of
Stokes theorem. For non-null hypersurfaces, we can use Gauss' law.

Let us first consider the case of an intersection whose action term
is zero. Let $N$ regions intersect, labelled by $i$, at a common
intersection $I$. We write the energy exchange relations in the
system as
\begin{gather} \label{diffmatter}
\delta_{\xi} S_{\text{matter}}=\sum_i \int_i {\bm e}\,  T^{ab}_{d}
\nabla_a \xi_{b}+ \frac{1}{2} \sum_{i,j=i\pm1} \int_{ij} \tilde{{\bm
e}}\, T^{ab}_{d-1} \nabla_a \xi_{b} =0
\end{gather}
where the normal vectors obey ${\frak n}_{ij}=-{\frak n}_{ji}$,
$j=i\pm1$. Then by $\xi_b=\xi_{||b}+({\frak n}\!\cdot\!{\frak n})
{\frak n}_b \ {\frak n}^c \xi_c$ with $\xi_{||b}=h^c_b \xi_c$ we
obtain
\begin{align} \label{conserv}
& -\sum_i \int_i  {\bm e}\, \nabla_a T^{ab}_{(d)} \xi_b+
\\ &  \nonumber
+\sum_{ij} \int_{ij} \tilde{{\bm e}}\, \left\{({\frak
n}\!\cdot\!{\frak n}) {\frak n}_a T^{ab}_{(d)} h^c_b \
\xi_c-\frac{1}{2} D_a  T^{ab}_{(d-1)} \xi_b\right\} + \\\nonumber  &
+\sum_{ij} \int_{ij} \tilde{{\bm e}}\, \left\{{\frak n}_a
T^{ab}_{(d)} {\frak n}_b \ {\frak n}^c \xi_c + \frac{1}{2}
T^{ab}_{(d-1)} K(n)_{ab} \ ({\frak n}\!\cdot\!{\frak n}) {\frak n}^c
\xi_c\right\}  + \\\nonumber &+ \int_I \tilde{{\bm e}}_{(d-2)}\,
\sum_{ij} \frac{1}{2}(v\!\cdot\! v) v_a T^{ab}_{(d-1)} \xi_b =0
\end{align}
where ${\frak n}^a, \ K_{ab}$ are assumed to carry an index $ij$.
Also $j=i\pm1$; the same for $v^a$ which is the normal on $I$
induced by $ij$ pointing outwards. Along with the known
relations~\cite{Davis-02}, we obtain then the ones related with the
intersection
\begin{gather} \label{energyedge}
\sum   \epsilon(v)  \ v_a T^{ab}_{d-1} \gamma^c_b =0 \ , \quad \sum
v_a T^{ab}_{d-1} v_b \ v^c =0
\end{gather}
where the sum is over all shared boundaries. $\gamma_{ab}$ is the
induced metric at $I$.

Equation (\ref{energyedge}) implies that the total
energy current density at the intersection or
collision is zero. This is valid though when the energy
tensor at the intersection vanishes identically.
On the other hand, as we have learned,
the energy tensor is not zero in general and the energy
conservation has to take into account this lower dimensional
energy tensor existing at the intersection hypersurface. In such a
case there is an additional term in
(\ref{diffmatter}) that can be written as
\begin{gather*}
\frac{1}{2N} \int_I \tilde{{\bm e}}_{(d-2)}\,\sum_{ij} T^{ab}_{d-2}
\nabla_a \xi_b
\end{gather*}
where we sum over the contribution from each side of every shared
boundary for $N$ regions. $T^{ab}_{d-2}$ is the total energy
momentum tensor on $I$. We decompose $\xi_b=\xi_{||b}+ ({\frak
n}\!\cdot\!{\frak n}) {\frak n}_b \ {\frak n}^c \xi_c + (v \cdot v)
v_b \ v^c \xi_c$ where $\xi_{||b}=\gamma_b^c \xi_c$. We then have
\begin{gather}
\int_I \tilde{{\bm e}}_{(d-2)} \left[ {\cal D}_a (T^{ab}_{d-2}
\xi_{||b} ) - {\cal D}_a T^{ab}_{d-2} \xi_{||b} + T^{ab}_{d-2}
\frac{1}{N}\sum_{ij} \left\{ ({\frak n}\!\cdot\!{\frak n}) {\cal
K}_{ab} {\frak n}^c+ (v\!\cdot\! v) C_{ab} v^c\right\}  \xi_c\right]
\end{gather}
(the first term is only useful when the intersection is not smooth
itself). The energy exchange relation are then
\begin{align} \label{energyedge2}
& \sum  (v \cdot v) \ v_a T^{ab}_{d-1} \gamma^c_b = {\cal D}_a
T^{ac}_{d-2},
\\
& \sum    v_a T^{ab}_{d-1} v_b \ v^c +T^{ab}_{d-2} \frac{1}{N}
\sum_{ij}  \left\{ ({\frak n}\!\cdot\!{\frak n}) {\cal K}_{ab}
{\frak n}^c+ (v\!\cdot\! v) C_{ab} v^c\right\}=0    \nonumber
\end{align}
where the first sums are over all hypersurfaces.

\section{Colliding Branes and deficit angles}

For a collision of hypersurfaces,
the intersection surface will be space-like. The $v$
vectors are time-like (velocity) vectors.
We assume that the hypersurface matter is a fluid described by:
\begin{gather}
v^av^bT_{ab} = \rho \quad \gamma^{a}_{c}T_{ab}v^b =0.
\end{gather}
The first of (\ref{energyedge})
is satisfied automatically whilst the second becomes:
\begin{gather}\label{cons}
\sum_{\Lambda} \rho_\Lambda v^a_\Lambda=0
\end{gather}
where the upper case Greek index counts the hypersurfaces. We can
recover the results of Langlois, Maeda and Wands~\cite{Langlois-02}
by first introducing the ortho-normal basis at the intersection. The
basis is taken to line up with the two vectors $v_I$ and $n_I$ of
one of the hypersurfaces.
\begin{gather*}
E_{(0)}=v_\Lambda, \quad E_{(1)} = n_\Lambda.
\end{gather*}
We can write the other $v$ vectors in the following way, motivated
by Special Relativity,
\begin{gather*}
v_\Xi = \gamma_{\Xi |\Lambda} E_{(0)} + \gamma_{\Xi |\Lambda}
\beta_{\Xi |\Lambda} E_{(1)}
\end{gather*}
where the $\beta$ and $\gamma$ have the usual interpretation from
SR: $\beta_{\Xi|\Lambda}$ is the relative speed between the two
hyper-surfaces; from the normalisation of $v$ we see
$\gamma_{\Xi|\Lambda}= (1-\beta_{\Xi|\Lambda}^2)^{-1/2}$. The two
components of equation (\ref{cons}) are:
\begin{align}
\sum_\Xi {\rho_\Xi}\gamma_{\Xi|\Lambda} & = 0,\\
\sum_\Xi {\rho_\Xi}\gamma_{\Xi|\Lambda}\beta_{\Xi|\Lambda} & = 0.
\end{align}
These are the results found in~\cite{Langlois-02}. They are the
conservation of energy and momentum respectively.

The hypersurfaces obey the same rules in terms of the local inertial
frame as do point particle collisions in two dimensions. This is
true for quite general bulk backgrounds. The only essential feature
is the absence of a deficit angle at the collision. This means that
there is a well defined local inertial frame at the collision and
the SR addition of velocities applies.

We have calculated the contribution to the energy-momentum
tensor at the collision due to the junction conditions. Our calculation
implicitly assumed that there was no conical singularity (see definition
\ref{localisedcurvdef}).
There may be some correction to this from a conical singularity.
If we impose some reasonable energy condition such as the weak
energy condition, this space-like matter should vanish- the
two contributions should cancel. The assumption of no conical
deficit is then justified
for the {\it Einstein} theory, because we have seen that there
is no contribution due to the junction conditions.
But this would not be so for the Gauss-Bonnet theory. In that case,
the cancellation would demand that there be a conical singularity at
the collision.
Conversely, if we impose that there be no such singularity,
we must either have space-like matter localised at the collision
or stringent selection rules on how hypersurfaces
can interact through collisions.

\section{A three-way intersection in AdS}

We have seen that there is a possibility to localise matter on an
intersection in the Gauss-Bonnet theory.
But it remains to be seen whether there is an actual solution to
the bulk field equations which gives non-zero energy-momentum.
This chapter is devoted to finding examples.

\subsection{The bulk vacuum solution}

We shall assume the simplest kind of bulk solution- a constant
curvature space-time. A constant curvature space-time satisfies
$R_{\mu\nu\alpha\beta} = \frac{1}{12}R(g_{\mu\alpha} g_{\nu\beta}
-g_{\mu\beta} g_{\nu\alpha})$, $R$ being a
constant~\cite{Hawking-73}. This is also known as a maximally
symmetric space-time. There are three possibilities:
\\
\begin{tabular}{ll}
i) &  de Sitter space ($R>0$),\\
ii) & anti-de Sitter space ($R<0$),\\
iii) & flat space ($R=0$).\\
\end{tabular}

In the Einstein theory, empty space will be one of the above three,
depending on whether the cosmological constant is positive, negative
or zero.
In the Einstein-Gauss-Bonnet theory it is possible that
more than one type of constant curvature space-time will
satisfy the vacuum field equations.
The different possibilities are because the Gauss-Bonnet
is quadratic in the curvature.

Anti-de-Sitter (AdS) space has constant negative curvature.
\begin{gather}\label{AdScurve}
\Omega^{ab} = -\frac{1}{l^2}E^a\wedge E^b.
\end{gather}
The constant $l$ has dimensions of length.
As already mentioned, AdS space is a vacuum
solution of the general Lovelock theory
\begin{gather*}
\sum_{n=0}^{[d/2]}
a_n\Omega^{a_1 b_1}\wedge\cdot\cdot\cdot\Omega^{a^n b^n}
\wedge e_{a_1\cdot\cdot\cdot b_nc} =0
\end{gather*}
provided that the following relation is satisfied:\footnote{ An
interesting choice of coefficients is that of Chamseddine's Gauge
theory of the AdS group. The Gauge symmetry fixes all but two of the
coefficients $l$ and $\kappa$. The three terms in the five
dimensional theory are:
\begin{gather*}
{\cal L} = \frac{\kappa}{l^d}\left(
\frac{1}{d}f(E^{\wedge d} )+ \frac{(d-1)l^2}{2(d-2)}
f(\Omega\wedge E^{\wedge (d-2)})
+\frac{(d-1)(d-3)l^4}{8(d-4)}
f(\Omega^{\wedge 2}\wedge E^{\wedge (d-4)})\right).
\end{gather*}
In this case \ref{AdScurve} is the only constant curvature solution
of the vacuum field equations.}
\begin{gather*}
\sum_{n=0}^{[d/2]}\frac{(-1)^n d!}{(d-2n)!}
\frac{ a_n}{l^{2n}}  =0.
\end{gather*}

We will write the AdS metric in conformally flat form:
\begin{gather}
ds^2 = \frac{1}{((u\cdot x)/l +1)^2} \ \eta_{\mu\nu}dx^\mu
dx^\nu,\qquad (u\cdot x)/l +1 \geq 0,
\end{gather}
where $u\cdot x \equiv \eta_{\mu\nu}u^\mu x^\nu$ and $u$ is a
constant normalised to $\eta_{\mu\nu}u^\mu u^\nu =1$. We will only
be interested in the vicinity of the intersection and will not worry
here about the global details of joining together regions of AdS.

We will work in the moving frame formalism and choose the
ortho-normal frame field:
\begin{gather*}
E^{(\mu)} = \frac{1}{\cal J} dx^\mu,
\qquad {\cal J}(x) \equiv (u\cdot x)/l +1.
\end{gather*}
We use the zero Torsion condition:
\begin{gather}\label{notor}
d E^{(\mu)} = -\omega^{(\mu)}_{\ (\nu)} \wedge E^{(\nu)}.
\end{gather}
Evaluating the left hand side:
\begin{gather}\label{dE}
\frac{\partial}{\partial x^\nu}
\left(\frac{1}{{\cal J}}\right)dx^\nu \wedge dx^\mu
=-\frac{1}{{\cal J}^2}\frac{\eta_{\nu\rho} u^\rho}
{l} dx^\nu\wedge dx^\mu
= \frac{\eta_{\nu\rho} u^\rho}{l}
E^{(\mu)}\wedge E^{(\nu)}.
\end{gather}
Comparing (\ref{dE}) and (\ref{notor}) and making use of
$\omega^{ab} = -\omega^{ba}$ for a  Lorentzian connection,
we compute the connection coefficients:
\begin{gather}\label{foundconnection}
\omega^{(\mu)(\nu)} = \frac{1}{l}(u^\mu E^{(\nu)}
-u^\nu E^{(\mu)}).
\end{gather}

\subsection{Three-way intersection}

\begin{figure}
\begin{center}\mbox{\epsfig{file=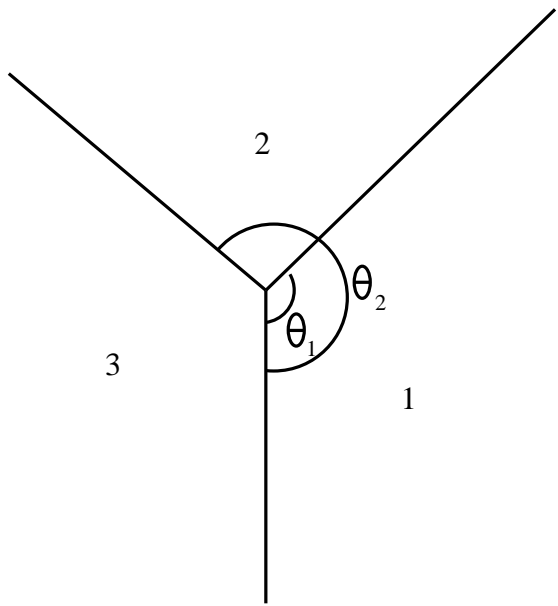, height=8cm}}
\caption{{\small The intersection without conical singularity.}}
\label{nodefecit}
\end{center}
\end{figure}

We will consider the simplest 3-point vertex. There are 3 regions
Region 1: $0 < \theta < \theta_1$, Region 2: $\theta_1 < \theta <
\theta_2$, Region 3: $\theta_2<\theta<\theta_3$, with the
identification $\theta_3 \equiv 0$. One can have a conical
singularity at the intersection with deficit angle $2\pi - \theta_3$
but we will not do so for reasons already mentioned. so we take
$\theta_3 =2\pi$ (fig. \ref{nodefecit}).
Let us work using the co-ordinates $y,z$ for the plane. $x = \rho
\cos{\theta}$, $y=\rho\sin{\theta}$. In each region $i$ let $u_i =
(0,...,0,\cos{\phi_i},\sin{\phi_i})$ such that $u_i\cdot x = \rho
\cos{(\theta -\phi_i)}$. The metric in each region takes the
following form:
\begin{gather}
ds_i^2 = \frac{1}{(\rho\cos(\theta-\phi_i)/l +1)^2} \
\eta_{\mu\nu}dx^\mu dx^\nu.
\end{gather}

The metric should be continuous across the walls:
\begin{align*}
\cos(\theta_1-\phi_1) =& \cos(\theta_1-\phi_{2}),
\\\cos(\theta_2 - \phi_{2})=&
\cos(\theta_{2}-\phi_{3}),
\\ \cos(\phi_{3})=& \cos(\phi_{1}).
\end{align*}
$\cos a =\cos b$ of course has two solutions, $a=\pm b$.
If we are to have any matter on the walls we must choose
the non-smooth minus sign solutions:
\begin{gather*}
\phi_1= -\phi_3=\theta_1-\theta_2,\\
\phi_2=\theta_1 +\theta_2.
\end{gather*}

The contribution to the hypersurface from the
Einstein gravity is
\begin{gather}\label{termE}
-\beta_1\delta E^c\wedge(\omega_2 -\omega_1)^{ab} \wedge {\bm
e}_{abc} = 2(d-2)\frac{\alpha_1}{l} (u_2-u_1)^a \delta E^c\wedge
{\bm e}_{ac}.
\end{gather}
Above, we have used $E^b\wedge {\bm e}_{abc} = -(d-2){\bm e}_{ac}$.

Now, since $(u_2 -u_1)\cdot x =0$ on the hypersurface, $(u_2 -u_1)$
must be normal to the hypersurface.
The norm of this vector with respect to $\eta_{ab}$ is:
\begin{gather*}
|(u_2 -u_1)| = \sqrt{2-2\cos(\phi_{2}-\phi_{1}) }
\end{gather*}
so
\begin{gather*}
(u_2-u_1)^a = \pm\sqrt{2-2\cos(2\theta_2) }\ {\frak n}^a.
\end{gather*}
Similarly $(u_3-u_2)^a = \pm{\frak n}_{\{23\}}^a\sqrt{2-2\cos(2\theta_1) }$
and also a similar relation for the hypersurface $\{31\}$:
$(u_1-u_3)^a = \pm{\frak n}_{\{31\}}^a
\sqrt{2-2\cos(2\theta_1-2\theta_2) }$.
Substituting this into (\ref{termE}), the hypersurface  term
due to the Einstein theory is:
\begin{gather*}
\pm 2(d-2)\frac{\alpha_1}{l}{\frak n}^a\sqrt{2-2\cos(2\theta_2) } \
\delta^b_c \delta E^c\wedge {\bm e}_{ab}.
\end{gather*}
By (\ref{juncminushalf}), the energy momentum on the brane in the
Einstein theory is:
\begin{gather}
(T_{\{12\}})^a_b =\mp (d-2)\frac{\alpha_1}{l}
\sqrt{2-2\cos(2\theta_2) }\
\delta^a_b.
\end{gather}
So there is indeed hypersurface matter. It takes the form of a
$(d-1)$-dimensional cosmological constant. With the Gauss-Bonnet
term this will get modified. We will not calculate this but will
proceed to find the intersection term.

From equation (\ref{eqm}), this term is:
\begin{gather*}
\beta_2 A_2 \int_{s_{123}} d^2t (\omega_2-\omega_1)^{ab}\wedge
(\omega_3-\omega_2)^{cd}\wedge \delta E^f\wedge {\bm e}_{abcdf}.
\end{gather*}
Substituting $A_2 = -2$ and $1/2$ for the volume of the simplex:
\begin{align*}
&-\beta_2  \delta E^f\wedge(\omega_2-\omega_1)^{ab}\wedge
(\omega_3-\omega_2)^{cd} \wedge {\bm e}_{fabcd}\\
&= -4\beta_2 (u_2-u_1)^{a}(u_3-u_2)^{c}
\delta E^f\wedge E^b\wedge E^{d} \wedge {\bm e}_{fabcd}\\
&= 4\beta_2 (u_2-u_1)^{a}(u_3-u_2)^{b}
\delta E^f\wedge E^c\wedge E^{d} \wedge {\bm e}_{fabcd}\\
&= 4(d-4)(d-3)\beta_2 [(u_2-u_1)^{(1)}(u_3-u_2)^{(2)}
-(u_2-u_1)^{(2)}(u_3-u_2)^{(1)}]\delta E^f\! \wedge \!{\bm
e}_{f(1)(2)}.
\end{align*}
The factor in square brackets is
\begin{align*}
&(\cos\phi_2-\cos\phi_1)(\sin\phi_3-\sin\phi_1)
-(\cos\phi_3-\cos\phi_1)(\sin\phi_2-\sin\phi_1)
\\&=  \sin(\phi_2-\phi_3) +\sin(\phi_3-\phi_1)
+\sin(\phi_1-\phi_2)\\
&= \sin(2\theta_1) +\sin(2\theta_2-2\theta_1)
-\sin(2\theta_2).
\end{align*}
since ${\bm e}_{(1)(2)}$ is the natural volume element
on the intersection, we have, for the matter localised there:
\begin{gather}\label{examplelast}
(\tilde{T}_{123})^a_b = 2(d-4)(d-3)\beta_2 [\sin(2\theta_2)
+\sin(2\theta_1-2\theta_2)-\sin(2\theta_1)]\delta^a_b.
\end{gather}
This is a $(d-2)$ dimensional cosmological constant.

This term will vanish if: i) $\theta_1 =\theta_2$, $\theta_1 =0$ or
$\theta_2 =0$ (two of the walls coincide); ii) $\theta_1=\pi$,
$\theta_2=\pi$ or $\theta_1 - \theta_2 =\pi$ (there is a smooth wall
with another branching off). None of these cases are a genuine
$3$-way intersection: either two hypersurfaces coincide or the
matching of the metric is smooth across a hypersurface. Thus we
conclude that for this example, localised matter at the intersection
is inevitable.

There are many other ways to have three walls intersecting in an AdS
bulk. The above is the simplest case of static walls with
cosmological constant type matter. More general solutions have been
found by Lee and Tasinato~\cite{Lee-04}.



 \chapter{Concluding remarks and outlook}

We have found that Lovelock gravity of any
order, $n_{max}$, in any space-time dimension, $d$,
has a well defined description of thin hypersurfaces
of matter. This is obtained from an action principle
which yields junction conditions for each hypersurface
and intersection. The junction conditions relate the
discontinuity in extrinsic curvatures as well as the
intrinsic curvatures to the localised
energy-momentum. For intersections of co-dimension greater
than $n_{max}$ these junction conditions are trivial- there is no localised
matter.

We have proved our results by two different ways of interpolating.
As well as providing a useful check that the numerical factors are
correct, the two approaches have different advantages. The method of
chapter \ref{inters} was used to establish that the action with
surface terms is one and a half order. The method of chapter
\ref{tdimensions} was used to prove the formal equivalence of this
action with the Lovelock action.
\\

The results have been found in terms of the vielbein
and connection $1$-forms. It would be useful to express
them in terms if the extrinsic and intrinsic curvature
tensors. We can give a simple argument:
${\bm e}$ is a $(d-n)$-dimensional volume element. The Lovelock
term of order n is:
\begin{gather}
R^n \wedge {\bm e};
\end{gather}
At a co-dimension one brane:
\begin{gather}
K\wedge (\tilde{R}+K^2)^{n-1} \wedge {\bm e};
\end{gather}
and co-dimension 2:
\begin{gather}
K^2\wedge (\tilde{R}+K^2)^{n-2} \wedge {\bm e};
\end{gather}
co-dimension n:
\begin{gather}
K^n\wedge {\bm e}
\end{gather}
when we run out of $R$'s. This is very schematic: $K$ represents a sum
of extrinsic curvature terms with some coefficients; $\tilde{R}$
denotes a sum over intrinsic curvatures.

More accurately, using the formulae of
\ref{secondtensors}, the Action terms are:
\begin{gather}
S \propto\int_{s_{0..p}} d^pt\,
\int_{\{0...p\}} \tilde{{\bm e}}\,
(\Delta K)^{a_{p+1}...a_{2p}}_{[a_{p+1}...a_{2p}}\,
R(t)^{a_{2p+1}...a_{2n}}_{a_{2p+1}...a_{2n}]}
\end{gather}
where $\tilde{{\bm e}}$ is the volume element on the hypersurface,
$\Omega(t)^{ab}\equiv 1/2 R(t)^{ab}_{\ \ cd}
E^c\wedge E^d$ and
\begin{gather*}
(\Delta K)^{a_{p+1}...a_{2p}} \equiv (K_{\{1|10\}}-K_{\{0|10\}})
^{a_{p+1}}_{c_{p+1}}\cdot\cdot\cdot(K_{\{p|p0\}}-K_{\{0|p0\}})
^{a_{2p}}_{c_{2p}}.
\end{gather*}
The Junction tensors are:
\begin{gather}
{\cal G}^a_b \propto\int_{s_{0..p}} d^pt\,
\delta^{bb_{p+1}...b_{2n}}_{aa_{p+1}...a_{2n}} (\Delta
K)^{a_{p+1}...a_{2p}}_{b_{p+1}...b_{2p}}\,
R(t)^{a_{2p+1}...a_{2n}}_{b_{2p+1}...b_{2n}}.
\end{gather}
Finding the exact numerical factors and performing the
general integral over the simplex are work in progress.
$R(t)$ can be decomposed into terms $\sim \tilde{R} + K^2$
(see equation \ref{intercurve2}).
\\

As mentioned at the end of chapter \ref{junction}, the action may
not, it seems, represent an unambiguous limit of a smooth matter
distribution. It is none-the-less a self-consistent theory of
strictly zero thickness hypersurfaces.
\\

In chapter \ref{tdimensions} we have found a tidy way of expressing
all the terms as
\begin{gather*}
\int_W \eta_{DC},
\end{gather*}
the intersection terms coming from the expansion of a
simple polynomial. This polynomial is given in equation
(\ref{etadc}). The meaning of this is not yet fully clear.

\section{Intersecting or colliding braneworlds}

Intersections can be different types according to the signature
of the induced metric:
\\\\
\begin{centering}
\begin{tabular}{lll}
i)  & positive definite, & space-like (collision).\\
ii)  &  negative definite, & time-like intersection.\\
iii) & indefinite, & can change between time-like, spacelike or null.\\
\end{tabular}\end{centering}
\\\\
I will only discuss the first two cases here since the physical
meaning is more clear.

An intersection of hypersurfaces of a given geometry
which in GR has no localised
matter, will generally have localised matter coming from the higher
order Lovelock terms.
\\\\i)
For a collision we can demand no localised space-like matter and
interpret the constraint on the geometry as conservation of energy.
{\it Even if the coefficients of the higher order terms are small},
this constitutes a qualitative difference between GR and the higher
order Lovelock theories.
\\\\ii)
An important model in String theory is (chiral) matter localised on
time-like intersections of branes. The low energy effective theory
would be expected to include higher curvature terms. With the higher
curvature terms in the Lovelock form, we can have localised matter
classically, not as a black hole but as a kind of defect, matching
several vacuum regions of space-time. It should be pointed out that
in, say, $d=11$ dimensions, only the first five Lovelock terms are
non-zero, $n_{max} =5$. This allows only intersection brane-worlds
of dimension $6$ or greater to be realised in this way. To get
$4$-dimensions, we need more singular branes, such as the
intersecting co-dimension $2$ branes considered by Navarro and
Santiago~\cite{Navarro-04}.


\section{Acceptable singularities}

In Lemma \ref{Gravanisinter} it was assumed that there was no
localised curvature on surfaces of co-dimension $>1$. Indeed, our
action principle will not work without some modification if there is
such a singularity. In four dimensional GR, point particles with
Schwarzschild geometry and thin cosmic strings which produce a
conical geometry are examples of these kind of singularities.

This section is devoted to a discussion of why the
co-dimension $1$ singularities are special.

\subsection{A question of derivatives}

\begin{figure}
\begin{center}\mbox{\epsfig{file=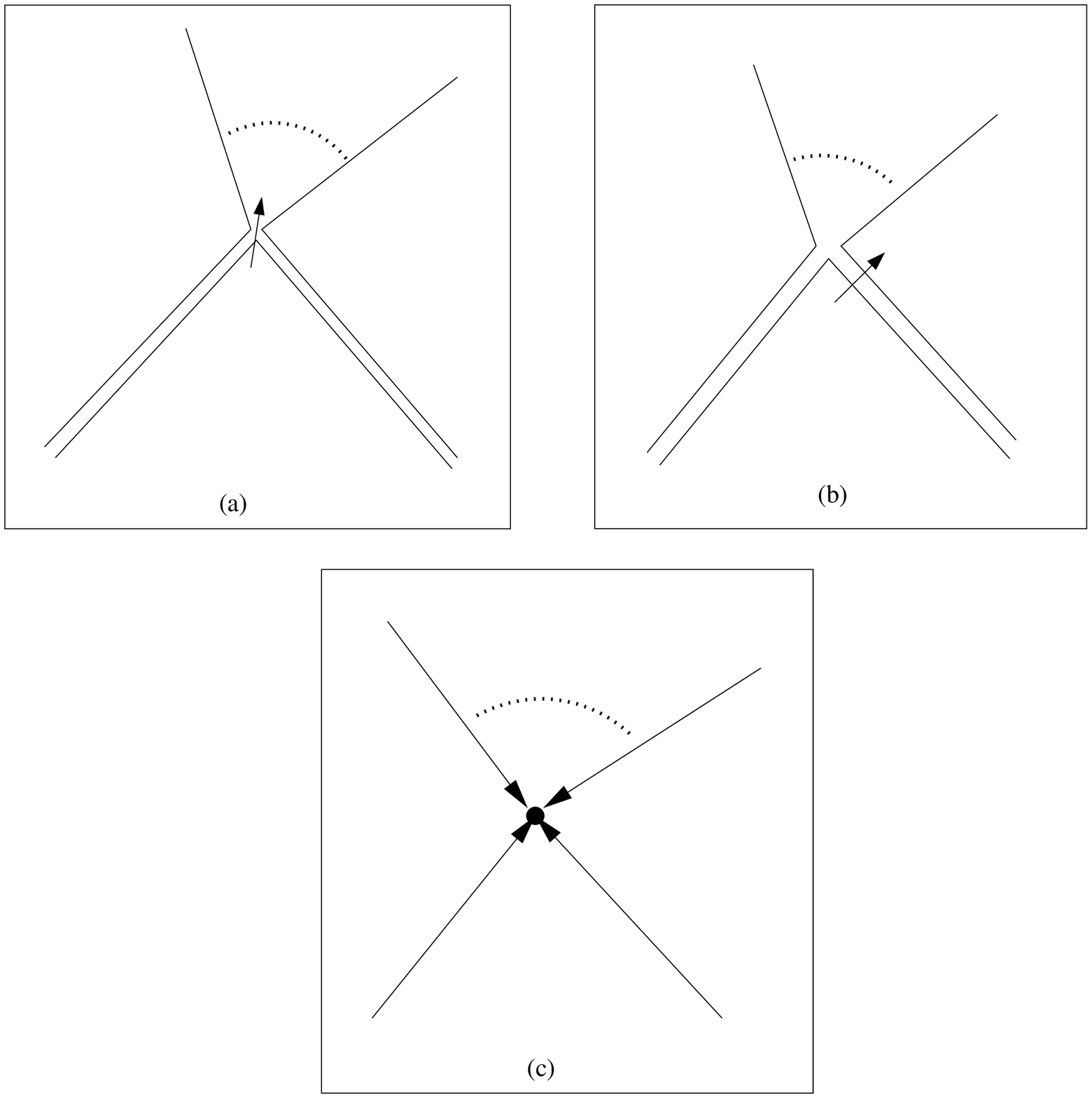, height=15cm}}
\caption{{\small (a) The derivative across the intersection is undefined.
(b) The derivative across the wall can be well defined under integration.
(c) The intersection appears as the shared boundary of the walls.}}
\label{hopeless}
\end{center}
\end{figure}

A crucial feature is the treatment of the derivative of some field
${\bm \sigma}$ which is actually discontinuous. The derivative
itself is undefined but can it be well defined under integration? In
section \ref{Glue} we outlined a way of meaningfully defining $\int
{\bm d} {\bm \sigma}$ across a wall. At the intersection itself
however, the problem seems to be hopeless (see fig.
\ref{hopeless}a). But nonetheless we proceeded to find expressions
for the intersection terms. How has this happened?

There is a sleight of hand involved. From the point of view of the
bulk regions, we treat it as if the intersection has been removed.
However, the co-dimension $2$ intersection reappears, as if by
magic, as part of the boundary of each of the walls (fig.
\ref{hopeless}c). In this way, $\int {\bm d}{\bm \sigma}$ has been
given meaning. There are two ways of interpreting this:
\begin{quote}1) We have cheated;\\
2) There is some underlying mathematical description that is
rigorous.\end{quote} Strong evidence in support of the latter is
found in chapter \ref{tdimensions}. We saw that the expression for
the intersection terms could be written in terms of a singular
mapping from some non-singular theory. Future work would be to make
precise what is meant by ``well defined under integration". This
would involve the mathematics of
distributions~\cite{Choquet-Bruhat-82} or generalised
functions~\cite{Kunzinger-01}.

In throwing out the derivative shown in figure \ref{hopeless},
we see the importance of disallowing deficit angles. Then it seems
such a derivative can not be ignored.


\subsection{Conical singularities and the like}

\begin{figure}
\begin{center}\mbox{\epsfig{file=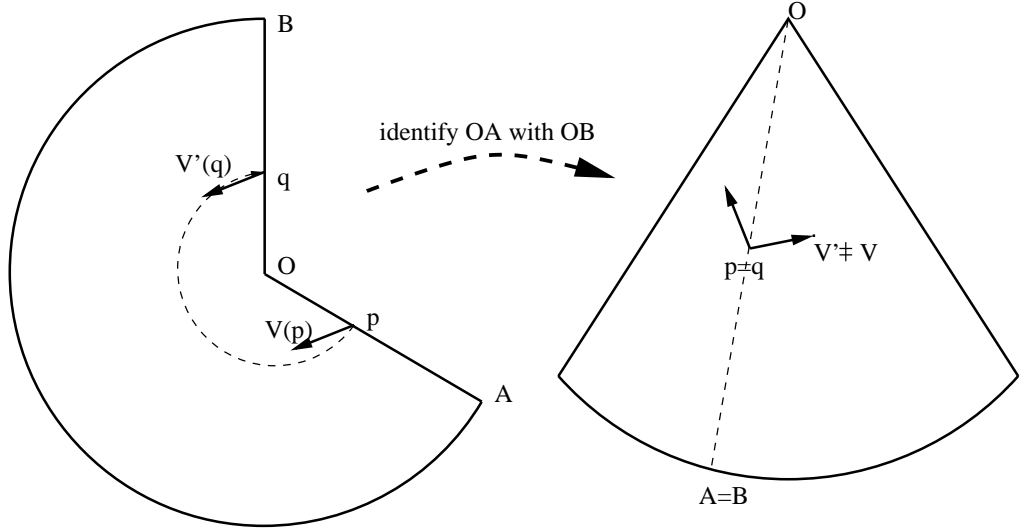, height=7cm}}
\caption{{\small The cone. Parallel transport of a vector around
the singularity gives a different vector. This means that the
relative angle of two vectors at the singularity is ill defined.}}
\label{cone1}
\end{center}
\end{figure}

\begin{figure}
\begin{center}\mbox{\epsfig{file=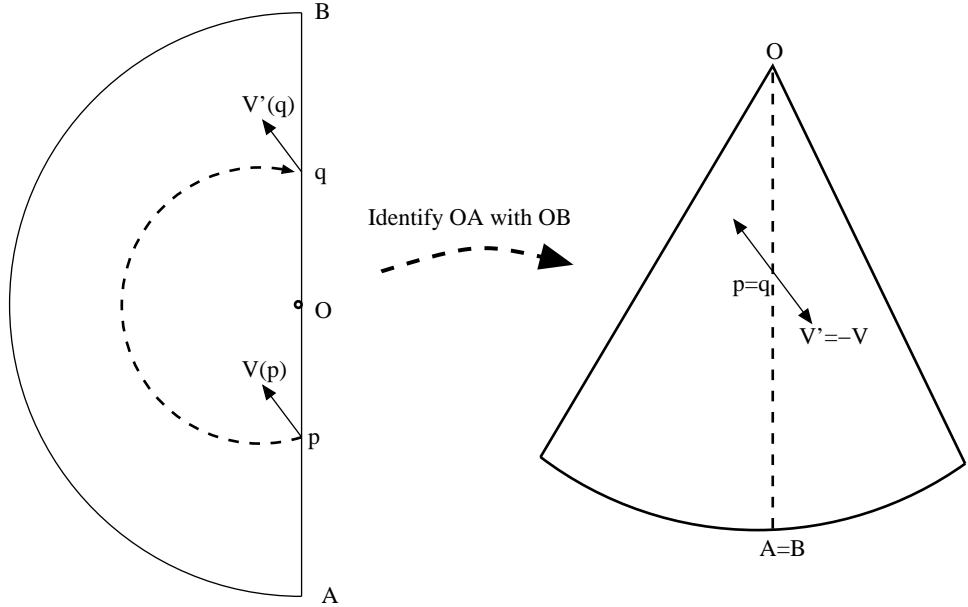, height=8cm}}
\caption{{\small The cone with deficit angle $=\pi$.}}
\label{cone}
\end{center}
\end{figure}

The cone is a topological manifold. One can overlay the tip of the
cone with a co-ordinate region which looks like ${\mathbb R^d}$ but
in these co-ordinates, the metric will be infinite at the tip. As
such there is no sensible way to define an orthonormal frame. The
inner product if two vectors is ill defined. This is obvious if one
represents the cone by a flat space with a wedge taken out (fig.
\ref{cone1}). Everywhere except at the tip, one can draw two arrows
which are perpendicular. At the tip, whether they are perpendicular
or not depends on whether you measure the angle clockwise or
anti-clockwise. \footnote{There is another possibly well defined
situation. If there is a deficit angle of exactly $\pi$ (fig.
\ref{cone}), one can also define unambiguously two orthogonal normal
vectors. Since going round the intersection sends $n \to -n$, $n_1
\cdot n_2 =0$ is well defined.} It is this breakdown of
ortho-normality which we shall regard as an unacceptable
singularity. It is a breakdown of the $d$-dimensional equivalence
principle since, at the singularity, the tangent space does not look
locally like Minkowski space. As well as the cone, or cosmic string,
such things as the Schwarzschild singularity will be considered
unacceptable for the same reason.

Assuming we disallow deficit angles, the localised matter does not
come from localised curvature at the intersections. The ``delta
functions" come from the higher curvature terms (although we don't
really use delta functions- we deal only with Stokes Theorem).
Schematically, for an intersection in the $x-y$ plane:
\begin{gather*}
f(\Omega \wedge E^{\wedge(n-1)})\approx A\delta(x) + B\delta(y),\\
f(\Omega^{\wedge2}\wedge E^{\wedge(n-2)}) \approx C\delta(x,y).
\end{gather*}
If there is a conical singularity, there will be curvature
at the intersection that is not induced by the hypersurfaces:
\begin{gather*}
f(\Omega\wedge E^{\wedge (n-1)}) \sim \delta(x,y)
\end{gather*}
and so our method breaks down.



\subsection{Singular gravity sources}
The theory of General Relativity admits singular sources of gravity
which are hypersufaces. They are particles which, although singular,
produce only a mild form of geometrical singularity. Geodesics
across the hypersurfaces are well defined. There is a well defined
local Lorentz frame at the hypersurface. Hypersurfaces then may be
thought of as fundamental particles compatible with classical
General Relativity.

Generally, where the curvature becomes singular, we expect to be in
the territory of quantum gravity. A completely different notion of
geometry may be necessary there or new types of field theories. So
it may be that this preferred status of hypersurfaces, being
classical, is irrelevant. However, it is conceivable that ``quantum
gravity" is actually a classical theory~\cite{'tHooft-99}
\footnote{For an interesting article on the theological implications
of non-determinism see~\cite{Byl}.}. Even if it is not, the
distinction between hypersurfaces and other singular particles may
still prove to be of value.

There is also a sense in which co-dimension $1$ {\it and} co-dimension
$2$ membranes are special. In both cases, one can have
delta function singularities in the Einstein tensor
without the Weyl tensor diverging near the membrane.
The membrane is not surrounded by a black hole but is a kind of
defect in space-time.

In the Einstein-Gauss-Bonnet theory, it was shown by Bostock et
al~\cite{Bostock-03} that it is possible to have a co-dimension 2
source (Braneworld) where the gravity singularity comes only from
the Gauss-Bonnet term. In this case there is not a deficit angle.
The problem with this is that the Riemann tensor diverges in the
vicinity of the brane. This led them to disregard it since the
quantum effects would be expected to be large in the region of the
brane.

\subsection{Gravity on simplicial manifolds}

There are several approaches to quantum gravity which involve
discretising the geometry of space-time at some small (Planck)
scale. Space-time is divided into cells, with the curvature
concentrated on the edges of the cells. As well as being a simple
model, this is reasonable because it is expected that the structure
of space-time itself be very different on the Planck scale. The
effects of quantum fluctuations on the curvature of a smooth
space-time are so problematic that it has been suggested that the
smooth manifold structure breaks down at small length scales.

One would like to have an action principle for Lovelock gravity on
any triangulated manifold involving boundary terms. The trick is to
take account of the many-valuedness of the connection and the normal
vectors at the singularities, so as to get an unambiguous result.
For example, let us assume that an intersection coincides with the
tip of the cone in fig. \ref{cone1}. Let $E_i^a$ be the vielbein in
one of the bulk regions adjacent to the singularity. Going once
round the singularity will mean $E_i^a \to \Lambda^a_{\ b} E_i^b$,
where $\Lambda$ is some Lorentz transformation. Simply defining
$E^a(t) = \sum_i t^i E_i^a$ is not sufficient. There is a similar
problem with the connection.
 The
fact that one has metric and connection (or vielbein and connection
$1$-form) degrees of freedom seems to be a source of complication.
Chamseddine's formulation of certain Lovelock theories as gauge
theories~\cite{Chamseddine-89} (involving only connection degree of
freedom) may help in this regard. This formulation is valid in any
odd number of dimensions. In four dimensions, there are the
Ashtekkar variables to describe GR in terms of purely affine
connection degrees of freedom, which are important in Loop quantum
gravity~\cite{Perez-04}. A general purely affine formulation of
Lovelock gravity in even dimension is not known at the present.

\section{Geometrical conundrums}
\subsection{Non-simplicial intersections}
The treatment of a simplicial intersection was found to be especially
easy. That being said, there is no great problems with treating
non-simplicial intersections- it is just a matter of summing terms in the
correct way, according to the principles of Section \ref{simplynonsimp}.

Let us denote a co-dimension $p$ (non-simplicial) intersection
by $I^p$.
\begin{conjecture}The intersection lagrangian of $I^p$ is determined
solely by the set of $I^1$'s for which  $\overline{I^1}\cap {I^p}
\neq \emptyset$.
\end{conjecture}
That is to say, in order to know what sum of terms appears in the
intersection, it is sufficient to know which hypersurfaces meet at
that intersection. If this conjecture is true, it would be easy to
write down the action for any kind of intersection- the way of summing terms is
determined
just by the list of hypersurfaces without needing any information about
intermediate co-dimension intersections.

\subsection{Mappings}
In section \ref{dimargument} we have found the action
in the form of an integral over the manifold
we called $W$. $W$ lives in a higher dimensional manifold
$F$ which looks locally like the product $M\times S^N$,
$S^N$ being a simplex. The action is  non-singular in $F$.
It is the projection $W \to M$ which is singular.

We could view $E(t)$ and $\omega(t)$ as fields on $F$.
They are independent fields but, by definition,
the fields, pulled back to $W$ obey:
\begin{gather}
{\bm d}_{(F)}E(t)^a + \omega(t)^a_b E(t)^b = 0,
\end{gather}
by equations (\ref{DtEt=0}) and (\ref{ttorsion}).
This is some kind of zero torsion condition.
What does it mean? It would be interesting to find out.

\subsection{The dual lattice}


The simplicial complex which results from taking the quotient:
\begin{gather*}
\frac{W}{M}
\end{gather*}
tells us which regions are connected to which.
Each bulk region is represented by just a point.
A hypersurface is represented by a line between the
two bulk points. Intersections are represented
by higher dimensional shapes with the bulk points as vertices.
The result is a dual lattice to the original.

For a simplicially valent lattice, these shapes will be simplices.
As  such, $W/M$ is a kind of triangulation of $M$. This is a strange
fact- by suppressing all the $x$-directions we end up with something
of exactly the same topology as $M$.

\appendix



\chapter{Some mathematical preliminaries}

This chapter contains useful background material, presented in a notation
consistent with the main chapters. More details can be found in the relevant
textbooks.

\section{Extrinsic curvature}\label{hyperappendix}
This section is based on Wald~\cite{Wald-84}, but we use the
opposite sign convention for the extrinsic curvature.

We have a hypersurface $\Sigma$. In a surrounding region $O$ we can
have a congruence of geodesic curves. This is a family of geodesics
which pass through each point in $O$ once and only once. The
tangents form a vector field in $O$. We will take $\zeta$ to
coincide with a normal vector on $\Sigma$. Let $\zeta$ be the vector
field
\begin{gather*}
\zeta^a= \frac{d x^a}{d\lambda},
\end{gather*}
where $\lambda$ is the proper time or proper distance along a
geodesic for time-like or space-like curves respectively. The
vectors are normalised
\begin{gather}\label{normalisezeta}
\zeta^a\zeta_a =\mp1
\end{gather}
with the $-/+$ sign for time-like or space-like vectors
respectively.

We define the extrinsic curvature
\begin{gather}\label{defK}
K_{ab} = -\nabla_a \zeta_b.
\end{gather}
It follows from the equation for a geodesic
and from the normalisation (\ref{normalisezeta}) that $K_{ab}$
is orthogonal to $\zeta$:
\begin{gather}\label{Ktangent}
\zeta^a\nabla_a \zeta_b = 0; \\\nonumber \zeta^b\nabla_a\zeta_b
=\frac{1}{2}\nabla_a(\zeta^b\zeta_b) = 0.
\end{gather}

We define the induced metric
\begin{gather*}
h_{ab} = g_{ab}-{\zeta_a\zeta_b}{\zeta\!\cdot\!\zeta}
\end{gather*}
and also the projection operator $\perp$. This is a projection from
the tangent bundle $T(M)$ onto the subspace on
$T(M)|_O$\footnote{This subspace is only well defined in the
neighbourhood $O$.} which is perpendicular to $\zeta$:
\begin{gather*}
\perp^a_b \equiv h^a_b = g^{ac}h_{cb};
\end{gather*}
\begin{gather*}
\zeta_a(\perp^a_bv^b) =  0 , \qquad \forall v \in T(M).
\end{gather*}

From the theorem of Frobenius, since $\zeta$ is hypersurface
orthogonal, we have
\begin{gather*}
\zeta_{[a}\nabla_b\zeta_{c]} = 0.
\end{gather*}
Contracting this with $\zeta^a$ and using (\ref{Ktangent})
we see that the extrinsic curvature is a symmetric tensor:
\begin{gather}\label{Ksymmetric}
\zeta^a\zeta_a K_{[bc]} = \mp K_{[bc]} = 0.
\end{gather}

In the region $O$ one can always write the metric in terms of
Gaussian Normal co-ordinates. One of the co-ordinates, say $w$ is
identified with the proper time of the geodesic congruence. In these
co-ordinates
\begin{gather*}
\zeta^a= \frac{\partial x^a}{\partial w} = (0,..0,1)
\end{gather*}
and the metric takes the form
\begin{gather*}
ds^2 = h_{ij}dx^idx^j \mp (dw)^2
\end{gather*}
where $i,j$ are from $1...d-1$. $h_{ij}$ is the intrinsic metric
tensor on $\Sigma$.
In these co-ordinates, the intrinsic curvature is
\begin{gather}\label{KisGamma}
K_{ij}= -\frac{1}{2}\partial_w h_{ij} =
(\zeta\cdot\zeta)\Gamma^w_{ij}.
\end{gather}

We can also define the extrinsic curvature in terms of
any unit vector, ${\frak n}$, normal to $\Sigma$. This vector field
will agree with $\zeta$ on $\Sigma$. And so the
tangential derivatives will agree on $\Sigma$:
\begin{gather*}
h_{ab} = g_{ab} - \frac{{\frak n}_a{\frak n}_b}{{\frak
n}\!\cdot\!{\frak n}};
\end{gather*}
\begin{gather*}
h_a^c\nabla_c{\frak n}_b = h_a^c\nabla_c\zeta_b.
\end{gather*}
Using (\ref{defK}) and (\ref{Ktangent} we get:
\begin{gather*}
h_a^c\nabla_c{\frak n}_b = -K_{ab}.
\end{gather*}
The above is the most important expression for the
intrinsic curvature for our purposes.

We can also write the extrinsic curvature as a Lie derivative:
\begin{gather*}
K_{ab} = -\frac{1}{2}{\pounds}_{\frak n} h_{ab}.
\end{gather*}

\section{Exterior differential calculus}

An {\it exterior differential form} is basically an anti-symmetrised
tensor. Let ${\bm \psi}$ be an exterior $p$-form. Then:
\begin{gather*}
{\psi}_{a_1...a_p} = \psi_{[a_1...a_p]},
\end{gather*}
where the square brackets $[\ ]$ denote complete anti-symmetry with
respect to the indices. There is an exterior product which preserves
this anti-symmetry, {\it the wedge product}:
\begin{gather*}
({\psi}\wedge{\sigma})_{a_1...a_{p+q}}
= \frac{(p+q)!}{p!\,q!}\,\psi_{[a_1...a_p}
\sigma_{a_{p+1}...a_{p+q}]}
\end{gather*}
where ${\bm \psi}$ and ${\bm \sigma}$ are $p$ and $q$ forms
respectively.

More abstractly, a $p$-form is an anti-symmetric multi-linear map
from the product of co-tangent spaces $T^*_xM\times\cdots\times
T^*_xM$ to ${\mathbb R}$. It can be written in terms of the basis:
\begin{gather*}
{\bm \psi} = {\psi}_{\mu_1...\mu_p}\, dx^{\mu_1}\wedge
\cdots\wedge dx^{\mu_p}.
\end{gather*}
The wedge is the anti-symmetrised product of the basis 1-forms
$dx^\mu\wedge dx^\nu = -dx^\nu\wedge dx^\mu$. One can also write
${\bm \psi}$ in terms of a non-co-ordinate basis of  $T^*_xM$:
\begin{gather*}
{\bm \psi} = {\psi}_{a_1...a_p}\, E^{a_1}\wedge
\cdots\wedge E^{a_p}.
\end{gather*}

The {\it exterior derivative}, ${\bm d}$, takes a $p$-form into a
$(p+1)$-form:
\begin{gather*}
(d \psi)_{\mu_1...\mu_{p+1}} = (p+1)\partial_{[\mu_1}
\psi_{\mu_2...\mu_{p+1}]}.
\end{gather*}
\begin{definition}
A p-form, ${\bm \psi}$ is {\it closed} if $\bm{d \psi}=0$.
A p-form, ${\bm \psi}$ is {\it exact} if there exists a $(p-1)$-form,
${\bm \sigma}$, such that ${\bm \psi}=\bm{d\sigma}$.
\end{definition}

It will be useful to deal also with tensor-valued differential
forms. Let ${\bm \sigma}$ be a type $(p,q+k)$ tensor valued exterior
differential $k$-form. If $p=q=0$ the exterior derivative of ${\bm
\sigma}$ is tensorial. If $p,q \neq 0$ the exterior derivative is
not co-ordinate invariant. The appropriate tensorial derivative
operator is the exterior co-variant derivative:
\begin{gather*}
D{\bm \sigma}^{\mu_1\dots \mu_p}_{\nu_1\dots\nu_q}
= {\bm d} {\bm \sigma}^{\mu_1\dots\mu_p}_{\nu_1\dots\nu_q}
+ \sum_{i=1}^p \omega_\lambda^{\mu_i} \wedge
{\bm \sigma}^{\mu_1\dots\lambda\dots\mu_p}_{\nu_1\dots\nu_q}
-\sum_{j=1}^q \omega^\kappa_{\nu_j}\wedge
{\bm \sigma}^{\mu_1\dots\mu_p}_{\nu_1\dots\kappa\dots\nu_q}.
\end{gather*}
$\omega$ is a connection 1-form.
\begin{gather*}
\omega^\mu_\nu = \Gamma^\mu_{\nu\lambda} dx^\lambda.
\end{gather*}
It will be helpful to write the exterior co-variant derivative
in terms of the tensor notation.
\begin{gather*}
D{\bm \sigma}^{\mu_1\dots \mu_p}_{\nu_1\dots\nu_q}
=\left(\nabla_{\kappa}{\sigma}^{\mu_1\dots
\mu_p}_{\nu_1\dots\nu_{q+k}} +\frac{1}{2}\sum_{i=q+1}^{q+k}
T^\lambda_{\nu_i\kappa} {\sigma}^{\mu_1\dots
\mu_p}_{\nu_1\dots\lambda\dots\nu_{q+k}}\right) dx^\kappa \wedge
dx^{q+1}\wedge\cdots\wedge dx^{q+k}.
\end{gather*}
If the torsion tensor $T^\mu_{\nu\kappa}$ is zero, the
exterior co-variant derivative is:
\begin{gather*}
D{\bm \sigma}^{\mu_1\dots \mu_p}_{\nu_1\dots\nu_q}
=\nabla_{\kappa}{\sigma}^{\mu_1\dots \mu_p}_{\nu_1\dots\nu_{q+k}}\,
dx^\kappa\wedge  dx^{\nu_{q+1}}\wedge\cdots\wedge dx^{\nu_{q+k}} .
\end{gather*}

In terms of some non-co-ordinate basis $E^a$:
\begin{gather*}
D{\bm \sigma}^{a_1\dots a_p}_{b_1\dots b_q} =\nabla_{c}{\bm
\sigma}^{a_1\dots a_p}_{b_1\dots b_{q+k}}\, E^c \wedge
E^{b_{q+1}}\wedge\cdots\wedge E^{b_{q+k}}.
\end{gather*}
\begin{gather*}
\omega^a_b = E^a_\mu D E^\mu_b
\end{gather*}
or
\begin{gather*}
\omega^a_{bc} = E^a_\mu E^\nu_c\nabla_\nu E^\mu_b.
\end{gather*}

I collect here some useful equations. Let ${\bm \psi}$ and ${\bm
\sigma}$ be exterior differential $p$ and $q$ forms respectively.
\begin{align}
{\bm \psi}\wedge{\bm \sigma}= & (-1)^{pq}{\bm \sigma}\wedge{\bm \psi}\\
d({\bm \psi}\wedge {\bm \sigma}) = & d{\bm \psi}\wedge {\bm \sigma}
+(-1)^p {\bm \psi}\wedge d{\bm \sigma}\\
D({\bm \psi}\wedge {\bm \sigma}) = & D{\bm \psi}\wedge {\bm \sigma}
+(-1)^p {\bm \psi}\wedge D{\bm \sigma}
\end{align}
\begin{align}
{\bm \psi}^{[ab}\wedge{\bm \sigma}^{cd]}
= & (-1)^{pq}{\bm \sigma}^{[ab}\wedge{\bm \psi}^{cd]}\\
{\bm \psi}^{[ab}\wedge{\bm \sigma}^{c]}
= & (-1)^{pq}{\bm \sigma}^{[ab}\wedge{\bm \psi}^{c]}\\
{\bm \psi}^{[a}\wedge{\bm \sigma}^{b]} = & (-1)^{pq+1}{\bm
\sigma}^{[a}\wedge{\bm \psi}^{b]}
\end{align}

\subsection{Orthonormal frames}

Let $\xi$ be a set of basis vectors $\{\xi_1,...,\xi_d\}$ of the
vector space $\mathbb{R}^d$. A linear frame $u(\xi) =
\{V_1,...,V_d\}$ is an ordered set of basis vectors of the tangent
space $T_p(M)$. $u$ is a linear isomorphism from $\mathbb{R}^d$ to
$T_p(M)$. The set of all $u(\xi)$ at each point $p$ in $M$ forms a
Principal Fiber Bundle with structure group $GL(d,
\mathbb{R})$~\cite{Kobayashi}.

There is a natural Lorentzian metric on $\mathbb{R}^d$, the
Minkowski metric $\eta$. We restrict $\xi$ to be an orthonormal
basis $\eta(\xi_a,\xi_b) =\eta_{ab}$, $\eta_{ab} \equiv {\rm
diag}(-1,1,...,1)$. The orthonormal basis $E =u(\xi)= {E_1,...,E_d}$
is again an ordered set of basis vectors of $T_p(M)$ but $u^{-1}$
also induces a canonical metric on $M$:
\begin{gather}\label{framedef}
g(E_a,E_b) = \eta_{ab}.
\end{gather}
Conversely, given a metric on $M$, we can always find a frame E such
that \ref{framedef} is satisfied. The set of all $E$ at each point
$p$ in $M$ forms a Principal Bundle with structure group
$SO(d-1,1)$, the restriction of $GL(d, \mathbb{R})$ to those
transformations which preserve the Minkowski metric.

The physical picture is this: The tangent space at any point $p$ on
the manifold looks like Minkowski space. $E$ is simply a set of
vectors:
\begin{gather*}
E_a = E^\mu_a \partial_{\mu}
\end{gather*}
such that
\begin{gather*}
g_{\mu\nu} E^\mu_a E^\nu_b = \eta_{ab}.
\end{gather*}
i.e. $E$ describes a local inertial frame. There is a gauge freedom
in the choice of $E$ corresponding to the set of all inertial
frames, related by local Lorentz transformations. If space-time is
curved, $E_a$ will be a non-co-ordinate basis. i.e. there will be no
co-ordinates $x^a$ on $M$ such that $E_a = \partial_a$. This
expresses the fact that there is no global inertial frame.

In this Thesis, I will deal with the co-frames $E^a$ which are the
dual vectors to $E_a$. In a corruption of notation, I have referred
to them throughout as frames. It should be remembered that they are
really co-frames. $E^a$ is also referred to as the vielbein.

\subsection{Integration of forms on oriented manifolds}\label{manifoldint}

{\it The support } of a form on a manifold $M$ is the domain in
which the form is non-zero: $supp\ \omega = \{x\in
M|\omega(x)\neq0\}$.

\begin{definition}{{\it Integration on an oriented manifold.}}
Let $\{U_i, \phi_i\}$ be an open cover of the manifold $M$. We
define a partition of unity $\omega = \sum_i f_i \omega$ such that
$f_i$ is a smooth function non-zero only in region $U_i$ and at
every point $x\in M$, $\sum_i f_i(x) =1$. We define the integral
\begin{gather}\label{manifoldintegration}
\int_M \omega=\sum_i \int_{U_i}f_i \omega
=\sum_i \int_{\phi_i(U_i)\subset {\mathbb R}^d}(\phi_i^{-1})^*(f_i \omega).
\end{gather}
The sum is finite if the support of $\omega$ is compact.
in which case the $U_i$ can be chosen to be a countable basis.
\end{definition}

\begin{definition}{{\it Domain with regular boundary.}}
Let $M$ be an oriented manifold with an atlas $\{U_i,\phi_i\}$. A
subset D is called a domain with boundary if for each point $x \in
D$ there is a chart $(U, \phi)$ at $x$ such that either:
\\
\begin{tabular}{lll}
&(a) & $\phi(U_i\cap {D})$
is an open neighbourhood of $\phi(x)$ in ${\mathbb R}^d$
\\&&($U_i$ is contained within D);
\\or& (b) & $\phi(U_i\cap {D})$
is an open neighbourhood of $\phi(x)$ in the
half space\\&& $H=\{(x^1,...,x^d)\in{\mathbb R}^d\ |\ x^1\leq 0\}$.
\end{tabular}\\
In the case (b), the boundary cuts through $U_i$
and is defined by $x^1 = 0$.
\end{definition}

\begin{definition}{{\it Manifold with regular boundary.}}
For the important case $D = M$, then $M$ is a manifold with
boundary.
\end{definition}

\begin{definition}{{\it The orientation on $\partial D$
induced by $M$}}. Let the orientation on $M$ be
$dx^1\wedge\cdot\cdot\cdot\wedge\ dx^d>0$. Then the orientation
defined on the surface $\partial D$ ($x^i=0 \in \phi(\partial D)$)
is
\begin{gather}\label{inducedorientation}
(-1)^{i-1} dx^1\wedge\cdot\cdot\cdot\widehat{dx^i}
\cdot\cdot\cdot\wedge\ dx^i>0.
\end{gather}
We use an arbitrary co-ordinate $x^i$. This will be useful later in
the case of domains with irregular boundaries. We have defined $x^i
\leq 0$ to be the interior of $D$. If we had defined it the other
way round, the minus sign factor in (\ref{inducedorientation}) would
be $(-1)^i$.

The integral over a section of the boundary is
\begin{gather*}
\int_{\phi(\partial D\cap U)}dx^1\cdot\cdot\cdot\widehat{dx^i}
\cdot\cdot\cdot\ dx^d= (-1)^{i-1}\int_{\phi(D\cap U)}
dx^1\cdot\cdot\cdot
\widehat{dx^i}\cdot\cdot\cdot\ dx^d|_{x^i=0}
\\\nonumber (-1)^{i-1}\int_{a^1}^{b^1}\cdot\cdot\cdot
\int_{a^d}^{b^d}dx^1\cdot\cdot\cdot
\widehat{dx^i}\cdot\cdot\cdot\ dx^d|_{x^i=0}.
\end{gather*}
\end{definition}
\begin{theorem}[Stokes' Theorem on an Oriented Manifold]
Let $D$ be a domain in $M$, $\omega$ a $(d-1)$-form in $M$ such that
$supp\ \omega \cap D$ is compact.
\begin{equation}\label{stokesD}
\int_D d\omega=\int_{\partial D} i^*\omega.
\end{equation}
\end{theorem}

{\bf Proof:}
Using the definition (\ref{manifoldintegration})
of integration on a manifold:
\begin{gather*}
\int_{\partial D} i^*\omega = \sum_i\int_{\partial D}
i^*(f_i\omega);
\end{gather*}
\begin{gather*}
\int_{D} d\omega = \sum_i\int_{D}d(f_i\omega).
\end{gather*}
So in order to prove Stokes' Theorem, one just needs to prove
\begin{gather}\label{stokesonUi}
\int_{\partial D}i^*(f_i\omega)
= \int_{D}d(f_i\omega),\qquad \forall i.
\end{gather}
Let
\begin{gather*}(\phi_i^{-1})^*(f_i\omega) = \sum_j
a_j(x)\ dx^1\wedge
\cdot\cdot\cdot\ \widehat{dx^j}\cdot\cdot\cdot
\wedge\ dx^d.
\end{gather*}
The support of $f_i\omega$ is contained within $U_i$ The functions
$f_i$ are smooth so $f_i \rightarrow 0$ on the boundary of $U_i$. If
$U_i$ does not contain the boundary, $U_i\cap\partial D =\emptyset$,
the right hand side of (\ref{stokesonUi}) can be seen to vanish as
required.
\begin{align*}
\int_{D}d(f_i\omega) =& \sum_j(-1)^{j-1}\int_{\phi_i(U_i)}
\frac{\partial a_j(x)}{\partial x^j}\ dx^1\cdot\cdot\cdot
\ dx^d\\\nonumber
=&\sum_j \int\ dx^1\cdot\cdot\cdot\
\widehat{dx^j}\cdot\cdot\cdot\ dx^{d}\ [a_j(x_j =max)-(x_j=min)]=0.
\end{align*}
If $U_i$ intersects the boundary, $U_i\cap D = \{(x^1...x^d \in {\mathbb R}^d|
x^1\leq 0\}$, then the integral does not vanish as above.
The integrals over $x^2,...x^{d}$ vanish
but the integral over $x^1$, in general not being over the whole
of $supp\ f_i\omega$, may not vanish.
\begin{align}\label{intdfomega}
\int_{D}d(f_i\omega) =& \sum_j(-1)^{j-1}\int_{\phi_i(U_i\cap D)}
\frac{\partial a_j(x)}{\partial x^j}\ dx^1\cdot\cdot\cdot
\ dx^d\\\nonumber
=&\int_{\phi_i(U_i\cap D)} \left.a_1\ dx^2
\cdot\cdot\cdot\ dx^{d}\right|_{x^1=0}
\\\nonumber=&\int_{\phi_i(U_i\cap\partial D)}a_1\ dx^2\cdot\cdot \cdot\ dx^d.
\end{align}
The left hand side of \ref{stokesonUi} manifestly vanishes if
$U_i\cap\partial D=\oslash$. On the other hand, if $U_i\cap\partial D \neq \oslash$
\begin{gather}\label{intboundaryfomega}
\int_{\partial D} i^*(f_i\omega) = \int_{\phi_i(U_i\cap\partial D)}
a_1(x^1=0)\ dx^2\cdot\cdot\cdot\ dx^{d}.
\end{gather}
Comparison of \ref{intdfomega} and \ref{intboundaryfomega}
verifies \ref{stokesonUi} and
completes the proof of Stokes theorem.
\\

\begin{definition}{{\it Domain with regular $p$-corners.}}
~\cite{Schwartz-68} Let $M$ be an oriented manifold with an atlas
$\{U_i,\phi_i\}$. A subset $D$ is called a domain with regular
$p$-corners if for each point $x \in D$ there is a chart $(U, \phi)$
at $x$ such that $\phi(U_i\cap {D})$ is an open neighbourhood of
$\phi(x)$ in either ${\mathbb R}^d$ or an $s$-corner:
$\mathbb{S}=\{(x^1,...,x^d)\in{\mathbb R}^d\ |\ x^1,...,x^s \leq
0\}$ for s=1...p. An $s$-corner is where $s$ regular sections of
boundary meet at $x^{1}=x^2=...=x^s = 0$.
\end{definition}
Some polygons have faces which are not described by the $s$-corner
prescription but can be broken into pieces which do.

We now extend the validity of Stokes' theorem (\ref{stokesD}) to a
domain with regular $p$-corners. For patches which do not intersect
the boundary or which intersect regular sections of the boundary the
same argument as before applies. We now need to argue
(\ref{stokesonUi}) for the case where $\phi_i(U_i\cap D)$ is an
s-corner, $s\geq 2$.
\begin{align*}
\int_{D}d(f_i\omega) & = \sum_j(-1)^{j-1}\int_{\phi_i(U_i\cap D)}
\frac{\partial a_j(x)}{\partial x^j}\ dx^1\cdot\cdot\cdot \
dx^d\\\nonumber & =\sum_{k=1}^s(-1)^{k-1}\int a_k(x^k=0)\
dx^1\cdot\cdot\cdot\ \widehat{dx^k}\cdot\cdot\cdot\ dx^{d}.
\\\nonumber
& = \int_{\partial D}i^*(f_i\omega).
\end{align*}
In the last line (\ref{inducedorientation}) has been used.
\\

On an oriented manifold we define the $d$-dimensional volume
element:
\begin{gather}
{\bm e} = \frac{1}{d!}\epsilon_{a_1...a_d}E^{a_1}\wedge\cdots \wedge
E^{a_d} = E^1\wedge\cdots\wedge E^d =
\sqrt{|\det(g)|}dx^1\wedge\cdots\wedge dx^d.
\end{gather}
We also define a $(d-r)$-dimensional volume element:
\begin{gather}
{\bm e}_{a_1\dots a_r} = \frac{1}{(d-r)!}
\epsilon_{a_1...a_d}E^{a_{r+1}}\wedge\cdots\wedge E^{a_d}
\end{gather}
e.g.
\begin{gather*}
{\bm e}_{(1)\dots (r)} =
\epsilon_{(1)...(d)}E^{(r+1)}\wedge\cdots\wedge E^{(d)}.
\end{gather*}
This is a volume element on a hypersurface
with the choice of adapted frames $\{E^{(1)},...,E^{(r)};$ $E^{(r+1)}
,...,E^{(d)}\}$ where $\{E^{(1)},...,E^{(r)}\}$ forms a basis
of the space of normal vectors and $\{E^{(r+1)}
,...,E^{(d)}\}$ forms a basis of tangent vectors.
${\bm e}_{a_1\dots a_r}$ will also be useful in writing the
dimensionally continued Euler densities.
\\

Useful formulae:
\begin{align}
\delta {\bm e}_{a_1\dots a_r} & = \delta E^b \!\wedge\!
{\bm e}_{a_1\dots a_r b} \label{varvol}\\
\label{Etimese} E^b\!\wedge\! {\bm e}_{a_1...a_r} & = r!{\bm
e}_{[a_1...a_{r-1}}
\delta_{a_r]}^b\\
n^{a_1\dots a_p} E^{b_{p+1}}\!\wedge\!\cdots\!\wedge\!
E^{b_r}\!\wedge\! {\bm e}_{a_1\dots a_r c} & = (-1)^r(r-p+1)!
n^{a_1\dots a_p} \!\wedge\! {\bm e}_{a_1...a_p [c}
\delta^{b_{p+1}\dots b_r}_{a_{p+1}\dots a_r]}
\end{align}
for $n^{a_1\dots a_p}$ a rank $p$ tensor, $p,q,r \geq 0$, $r > p$.
\\

We will define the canonical volume element
on a boundary in such a way that Gauss Law
takes its usual form.
Let ${\bm \sigma} = \sigma^\mu {\bm e}_\mu$.
\begin{gather*}
\bm{d\sigma} = (\nabla_\mu \sigma^\mu){\bm e}.
\end{gather*}
\begin{gather*}
\int_{\partial M} {\bm \sigma} = ({\frak n}\!\cdot\!{\frak n})
\int_{\partial M}\sigma^\mu n_\mu n^{\nu}{\bm e}_\nu.
\end{gather*}
So Gauss' Law takes the usual form:
\begin{gather}
\int_M (\nabla_\mu \sigma^\mu){\bm e}
= \int_{\partial M}\sigma^\mu n_\mu \tilde{{\bm{e}}}
\end{gather}
with the convention:
\begin{gather}\label{bndryelement}
\tilde{{\bm e}}  = ({\frak n}\!\cdot\!{\frak n}){\frak n}^{\nu}{\bm
e}_\nu = ({\frak n}\!\cdot\!{\frak n})\sqrt{h}d^{d-1}x.
\end{gather}

\subsection{Integration over simplices and chains}\label{simplexsection}

\begin{definition}{\it{A standard $p$-simplex, $S_p$}} in
${\mathbb R}^p$ is defined, following von Westenholz
~\cite{von-Westenholz-78}, by
\begin{gather}
s_p=\left\{(x^1,...,x^p)\left|\sum_{i=1}^{p}
x^i\leq1;\quad 0\leq x^i \leq1\quad \forall i\right.\right\}
\end{gather}
i.e. the standard $p$-simplex is the closed convex hull
$(A_0,,...,A_p)$ of $(p+1)$ points, taken in a definite order and
such that the $p$ vectors $(A_i-A_0)$, $i=1...p$, are linearly
independent.
\end{definition}
Examples:
\begin{eqnarray*}
&\text{standard 0-simplex:} &\text{a point},\ s_0 =\{0\}.\\
&\text{standard 1-simplex:} &\text{a unit interval},\ s_1 = [0,1].\\
&\text{standard 2-simplex:} &\text{an oriented triangle with points
at (0,0), (1,0), (0,1).}
\\&\text{standard 3-simplex:} &\text{an oriented tetrahedron.}
\end{eqnarray*}
\begin{figure}
\begin{center}\mbox{\epsfig{file=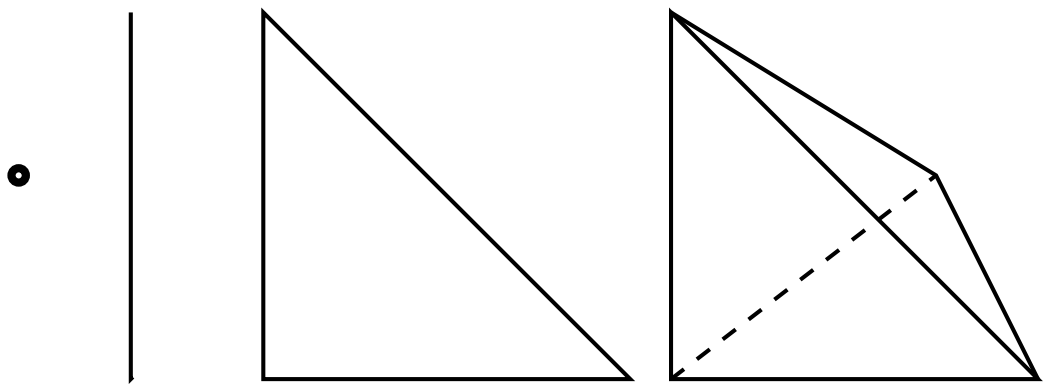, height=5cm}}
\caption{{\small The standard simplices for 0,1,2 and 3 dimensions.}}
\label{simplices}
\end{center}
\end{figure}
{\it The boundary of a standard $p$-simplex} is the oriented sum of
the $(p-1)$-dimensional faces:
\begin{gather}\label{sboundary}
\partial s_p = \sum_{k=0}^p (-1)^k (A_0,...,\widehat{A_k},...,A_p).
\end{gather}

\begin{definition}{{\it A differentiable singular $p$-simplex, $\sigma_p$}}
on a differentiable manifold $M$ is given by a $C^{\infty}$ map of a
standard $p$-simplex into $M$, $f:\ s_p\rightarrow M$. To be
differentiable at the boundaries, it must be possible to extend the
map to $f: U\rightarrow M$, where $U$ is an open subset of ${\mathbb
R}^p$ containing $s_p$. Under such a map, $\sigma_p$ is the image of
$s_p$ in $M$.

It is possible to cover ${\mathbb R}^p$ with $s_p$, i.e. they
provide a triangulation of ${\mathbb R}^p$. The $\sigma_p$ do not in
general provide a triangulation of $M$.
\end{definition}

\begin{definition}\label{chaindef}{{\it A differentiable $p$-chain, $c_p$}}
in $M$ is defined as the formal sum of a set of $p$-simplices
$\{\sigma_p^i\}$ with real coefficients.
\begin{gather}
c_p = \sum_i \lambda_i\ \sigma_p^i, \quad \lambda \in {\mathbb R}.
\end{gather}
\end{definition}

The boundary of a $p$-simplex is
\begin{gather}
\partial \sigma_p = \sum_{j=0}^{p} (-1)^k\ \sigma_p^k
\end{gather}
where $k$ denotes the k-th face of the simplex, the image of
$(A_0,...\widehat{A_k},...,A_p)$ in $M$.

\begin{theorem}[Stokes' Theorem for a simplex] Let $\sigma_d$ be a differentiable
d-simplex with boundary $\partial \sigma_d$ in a $d$-dimensional
differentiable manifold $M$. Let $\omega$ be a differentiable
$(d-1)$-form. Then
\begin{gather}
\int_{\sigma_d} d\omega = \int_{\partial \sigma_d} \omega.
\end{gather}
\end{theorem}
The proof is in the textbooks~\cite{von-Westenholz-78}.

\section{The Euler number}\label{Eulerappend}

\begin{figure}
\begin{center}\mbox{\epsfig{file=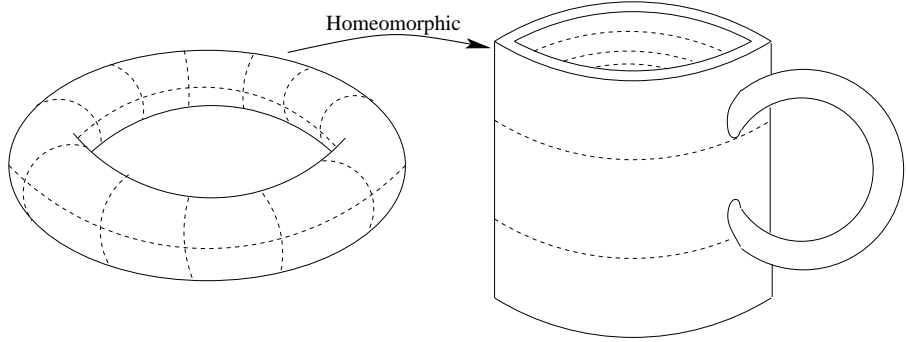, height=4.5cm}}
\caption{{\small The coffee cup and the donut have the same
Euler number, $\chi=0$}}
\label{Eulerfig}
\end{center}
\end{figure}

The Euler number, or Euler-Poincare Characteristic, $\chi(M)$, is a
topological invariant of the manifold. It is preserved under a
homeomorphism, a
global $1$-to-$1$ map from one manifold to another.
For example, the donut and the coffee cup have the same
Euler number (fig. \ref{Eulerfig}).
Mathematical~\cite{Kobayashi} or more physical
treatments~\cite{Nakahara-90} of the Euler number are in many
textbooks. I will give only an outline here. The Euler number is:
\begin{gather}
\chi(M) \equiv \sum_{i=0}^d (-1)^i B_i.
\end{gather}
The $B_i$ are the Betti numbers. $B_p$ is the number of independent
closed $p$-surfaces that are not boundaries of a $(p+1)$-surface.
Also, by de Rham's theorem, $B_p$ is the number of independent
closed $p$-forms modulo exact forms. For a closed manifold $B_p =
B_{4-p}$ so $\chi =0$ for odd dimensional manifolds.

For a simplicial complex, $K$, the Euler number is:
\begin{gather*}
\chi(K) \equiv \sum_r (-1)^r I_r
\end{gather*}
where $I_r$ is the number of $r$-faces in $K$. The equivalence of
this with the Homology definition is the result of the
Euler-Poincare Theorem.

The Poincare-Hopf Theorem provides another interpretation of the
Euler number: Consider any vector field on $M$ which has only
isolated zeros. The Euler number of a compact manifold is equal to
the sum of the indices of the zeros. It is interesting that a
topological quantity is equivalent to an analytic index. The
following theorem also shows it to be equivalent to an integral.

\begin{theorem}[Gauss-Bonnet Theorem]
For a compact manifold, $\chi$ is equal to the integral over $M$ of
the representative, ${\bm \Omega}$, on $M$ of the Euler Class of its
Tangent Bundle~\cite{Choquet-Bruhat-82}.
\end{theorem}

 We will call this representative (give or take a numerical factor)
the Euler density.
The totally antisymmetric
Levi-Civita symbol $\epsilon_{a_1...a_d}$,
with entries $\pm 1$, is a tensor w.r.t. the
$SO(2n)$ or Lorentz Group. For even dimension, $d=2n$,
we can construct the invariant:
\begin{gather}\label{The2nform}
{\bm \Omega} \equiv \frac{(-1)^n}{(4\pi)^{n}n!} \Omega^{a_1a_2}\wedge\cdots\wedge \Omega^{a_{2n-1}a_{2n}}
\epsilon_{a_1...a_{2n}}.
\end{gather}
It is the $2n$-form ${(-1)^n}{(4\pi)^{n}n!}\,{\bm \Omega}$ which we
call the Euler density.

The proof of the Gauss-Bonnet theorem in any even dimension is due
to Chern~\cite{Chern-44}. It involves integrating over a section of
the bundle of ortho-normal frames, known as the sphere bundle. On
this manifold, ${\bm \Omega}$ is a total derivative. There is a
boundary at each singular point, contributing the index at each
point- hence, by the Poincare-Hopf theorem, the proof follows.

\subsection{A useful property of the invariant polynomial}\label{interpappendix}

Let $M$ be a manifold with a Riemannian or Lorentzian metric $g$ and
a Levi-Civita connection. Let $\omega$ be the connection 1-form and
$\Omega$ the curvature form. For ${\rm dim}M=2n$ consider the
integral
\begin{equation*}
\int_M f(\Omega,..\Omega), \qquad  f(\Omega,..,\Omega)=\Omega^{a_1a_2} \wedge ...
\wedge \Omega^{a_{2n-1}a_{2n}} \epsilon_{a_1...a_{2n}}
\end{equation*}
where $\epsilon_{...}$ is the fully anti-symmetric symbol and
$\epsilon_{1..2n}=+1$ and the integral is assumed to exist. The
frame $E$ is ortho-normal in the sense $g(E^a,E^b)=\delta^{ab}$ in
the Riemannian case and
$g(E^a,E^b)=\eta^{ab}=\text{diag}{(-1,1..1)}$ in the Lorentzian
case. When $g$ is Riemannian and $M$ is compact and oriented
$f(\Omega,..\Omega)$ represents the Euler class. The integral  over
$M$, normalised properly, gives the Euler number of $M$, as stated
in the previous section.

Let us now repeat the well known
construction~\cite{Choquet-Bruhat-82,Kobayashi} and show the
following:

\begin{proposition} under a continuous
change of the connection, $\omega \to \omega'$, $f(\Omega,..\Omega)$ changes
by an exact form.
\end{proposition}

Define
\begin{equation*}
\omega_t=t \omega+(1-t)\omega'.
\end{equation*}
Call
\begin{equation*}
\theta=\omega-\omega'
\end{equation*}
and note that
\begin{equation*}
\theta=\frac{d}{dt} \omega_t
\end{equation*}
and for the curvature associated with $\omega_t$
\begin{equation*}
\Omega_t \equiv {\bm d}\omega_t+\omega_t \wedge \omega_t
\end{equation*}
that
\begin{equation}
\frac{d}{dt} \Omega_t=D_t \theta,
\end{equation}
where $D_t$ is the covariant derivative associated with $\omega_t$. Then
\begin{eqnarray} \nonumber
&& f(\Omega,..,\Omega)-f(\Omega',..,\Omega')= \int_0^1 dt
\frac{d}{dt} f(\Omega_t,..,\Omega_t)= n \int_0^1 dt f( d
\Omega_t/dt, \Omega_t, ...\Omega_t)= \\  && =n \int_0^1 dt f(D_t
\theta, \Omega_t, ...\Omega_t)= n \int_0^1 dt  \ {\bm d} f( \theta,
\Omega_t, ...\Omega_t)\label{derivation1},
\end{eqnarray}
where symmetry and multi-linearity of $f$ have been used, as well as
$D_t\Omega_t=0$.

If we define
\begin{equation*}
{\cal L}(\omega)=f(\Omega,..\Omega)
\end{equation*}
and
\begin{equation*}
{\cal L}(\omega,\omega') \equiv -n \int_0^1 dt f(\omega-\omega',
\Omega_t, ...\Omega_t),
\end{equation*}
we can write
\begin{equation} \label{basic}
{\cal L}(\omega) \equiv {\cal L}(\omega')-{\bm d}{\cal
L}(\omega,\omega').
\end{equation}
Now, assume that, for example, $M$ is non-compact and without a
boundary. If ${\cal L}(\omega,\omega')$ vanishes fast enough
asymptotically, then
\begin{equation*}
\int_M {\cal L}(\omega)
\end{equation*}
(assumed to exist) does not depend on $\omega$. It is this property that makes
${\cal L}(\omega)$ so interesting when, with a little modification, it is used as a
Lagrangian for gravity for ${\rm dim}M>2n$,
see section \ref{topapproach}.

\newpage

\section{Honeycombs}

The study of the honeycomb has a long and illustrious history. It
was studied by MacLaurin in the eighteenth century and Lord Kelvin
in the nineteenth. In 1999 Thomas Hales~\cite{Hales-99} proved the
two dimensional honeycomb conjecture: any partition of the plane
into regions of equal area has perimeter which is at least that of
the regular honeycomb tiling.

Back in 36 B.C. Marcus Terentius Varro discussed the problem.
He rejected the theory that bees built hexagonal structures
because they have six feet. He suspected that the bees were
great geometers. The mathematicians of the ancient world were
highly sophisticated in geometry and the three regular tessellating
shapes were well known. It was known that of the three the hexagon
enclosed the most area.
\begin{figure}
\begin{center}\mbox{\epsfig{file=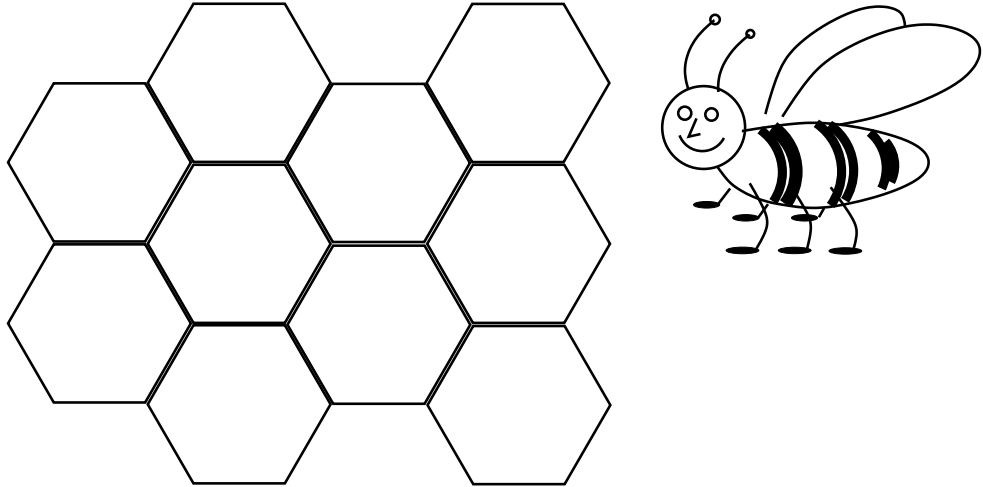, width=13cm}}
\caption{{\small Why do bees make a hexagonal honeycomb?
Is it because they have six feet or do they know something we
don't?}}
\label{bee}
\end{center}
\end{figure}

The three dimensional problem remains unsolved. It has been shown
that the rhombic dodecahedron of the bees is not the solution.
Lord Kelvin's solution, while being better than the bees' effort,
was also proven false. Kelvin proposed a truncated octahedron,
the Voronoi Cell of the body centred cubic packing of spheres.
While this is minimum with respect to small deformations, it has
recently been shown not to be the true minimum. In 1994 Weaire
and Phelan found a counter example based on two different shapes.
Kelvin's proposal is still thought to be correct for identical shapes.

The honeycomb illustrates key points of our study:
\\\\
$*$ The hypersurfaces intersect in such a way as to break up the
bulk space into many regions.
\\\\
$*$ It is simplicially valent. The intersections have the
minimum number (3) of hyper-surfaces meeting there.



\chapter{Useful formulae}

\section{The variational principle for a gravity theory}\label{Boundvar}

For a general gravity theory, we have an action functional built by
contracting products of the Riemann tensor with the metric tensor:
\begin{gather*}
S[g^{ab},\partial_cg^{ab},\partial_c\partial_dg^{ab}].
\end{gather*}
The Euler variation of the action with respect to the metric tensor
is:
\begin{gather}\label{Eulervar}
\delta S= \int \delta g^{ab} \frac{\partial {\cal L}}{\partial g^{ab}}
+\partial_c \delta g^{ab}
\frac{\partial {\cal L}}{\partial g^{ab}_{\ \ ,c}}
+\partial_c\partial_d \delta g^{ab}
\frac{\partial {\cal L}}{\partial g^{ab}_{\ \ ,cd}}
\end{gather}
where the comma denotes a partial derivative. We
partially integrate the second term in (\ref{Eulervar}):
\begin{gather*}
\partial_c \delta g^{ab}
\frac{\partial {\cal L}}{\partial g^{ab}_{\ \ ,c}} =\partial_c
\left(\delta g^{ab} \frac{\partial {\cal L}}{\partial g^{ab}_{\ \
,c}}\right) -\delta g^{ab}\partial_c \frac{\partial {\cal
L}}{\partial g^{ab}_{\ \ ,c}}.
\end{gather*}
Now partially integrate the third term in (\ref{Eulervar}):
\begin{align*}
\partial_c\partial_d \delta g^{ab}
\frac{\partial {\cal L}}{\partial g^{ab}_{\ \ ,cd}} & =
\partial_c\left(\partial_d \delta g^{ab} \frac{\partial {\cal
L}}{\partial g^{ab}_{\ \ ,cd}}\right) -\partial_d \delta
g^{ab}\partial_c \frac{\partial {\cal L}}{\partial g^{ab}_{\ \ ,cd}}
\nonumber \\
& = \partial_c\left(\partial_d \delta g^{ab} \frac{\partial {\cal
L}}{\partial g^{ab}_{\ \ ,cd}}\right) -\partial_d\left( \delta
g^{ab}\partial_c \frac{\partial {\cal L}}{\partial g^{ab}_{\ \
,cd}}\right) + \delta g^{ab}\partial_d\partial_c\left(
\frac{\partial {\cal L}}{\partial g^{ab}_{\ \ ,cd}}\right).
\end{align*}
We get
\begin{gather}
H^{ab}\delta g^{ab} + \partial_c V^c,
\end{gather}
\begin{gather*}
H^{ab} = \frac{\partial {\cal L}}{\partial g^{ab}} -\partial_c
\frac{\partial {\cal L}}{\partial g^{ab}_{\ \ ,c}}
+\partial_d\partial_c\left( \frac{\partial {\cal L}}{\partial
g^{ab}_{\ \ ,cd}}\right),
\end{gather*}
\begin{gather*}
V^c = \delta g^{ab}\frac{\partial {\cal L}}{\partial g^{ab}_{\ \
,c}} -\delta g^{ab}\partial_d\left( \frac{\partial {\cal
L}}{\partial g^{ab}_{\ \ ,cd}}\right) +\partial_d \delta g^{ab}
\frac{\partial {\cal L}}{\partial g^{ab}_{\ \ ,cd}},
\end{gather*}
or after further partial integration:
\begin{gather*}
V^c = \delta g^{ab}\frac{\partial {\cal L}}{\partial g^{ab}_{\ \ ,c}}
-2 \delta g^{ab}\partial_d\left(
\frac{\partial {\cal L}}{\partial g^{ab}_{\ \ ,cd}}\right)
+\partial_d\left( \delta g^{ab}
\frac{\partial {\cal L}}{\partial g^{ab}_{\ \ ,cd}}\right).
\end{gather*}

\section{The variational principle for Lovelock gravity}

The action for General Relativity is
\begin{gather*}
{\cal S} = \int_M R {\bm e} + \int_M {\cal L}_{mat}.
\end{gather*}
The Lagrangian ${\cal L}_{mat}=L_{mat} {\bm e}$ is
due to the (unspecified) matter content such as gauge fields.
The Euler variation w.r.t. $g^{\mu\nu}$ (neglecting boundary terms)
\begin{gather*}
\delta S =\int_M (G_{\mu\nu} - T_{\mu\nu}) \delta g^{\mu\nu}{\bm e},
\qquad
T_{\mu\nu}{\bm e} \equiv
-\frac{\partial {\cal L}_{mat.}}{\partial g^{ab}}
\end{gather*}
leads to the Einstein equations. $T_{\mu\nu}$ is the
Energy-momentum tensor, assuming ${\cal L}_{mat}$ is correctly
normalised.

For the Lovelock theory we have:
\begin{gather*}
{\cal S} = \int_M {\cal L}_{Lovelock}  + \int_M {\cal L}_{mat}.
\end{gather*}
The Euler variation w.r.t. $g^{\mu\nu}$ (neglecting boundary terms)
leads to:
\begin{gather*}
\delta S =\int_M (H_{\mu\nu} - T_{\mu\nu}) \delta g^{\mu\nu}{\bm e},
\end{gather*}
where $H_{\mu\nu}$ is the Lovelock tensor.

Similarly, if we have matter on some intersection $I$, we have a
term in the action:
\begin{gather}
\int_I \tilde{L}_{mat}(\phi, \gamma) \tilde{{\bm e}},
\end{gather}
where $\phi$ represents the matter fields and $\gamma$ the
induced metric on $I$. $\tilde{{\bm e}}$ is the induced volume element
on $I$.
\begin{gather}
\delta \tilde{{\cal L}}_{{\rm mat}}
\equiv-\tilde{T}_{\mu\nu}\delta g^{\mu\nu} \tilde{{\bm e}}
\end{gather}
is the energy-momentum tensor on $I$. According to our formalism
presented in chapter \ref{inters}, $\tilde{T}_{\mu\nu}$ will be
equal to the variation of the appropriate surface term.

These more familiar expressions for the gravitational action
principle are in terms of variation w.r.t. the metric. Throughout
this thesis, I have used the vielbein language so it is useful to be
able to translate between the two. We shall need these identities:
\begin{align}
\delta E^b & =
\delta E^b_\mu E^\mu_c E^c, \label{varyEtog}
\\\nonumber\\
\delta E^b_\mu E^\mu_a & = - E^b_\mu \delta
E^\mu_a,\label{-relation}
\\\nonumber\\
\delta g^{\mu\nu} & =
2\eta^{ab} \delta E_a^\mu E_b^\nu.\label{varygtoE}
\end{align}

When one varies the gravitational part of the action with respect to
the frame, the result is:
\begin{gather*}
\delta {\cal L}_G = \delta E^b {\cal E}_b
={\cal E}^a_b \delta E^b {\bm e}_a
\end{gather*}
(We can always write it in this form by using equation
\ref{Etimese}). Substituting (\ref{varyEtog}):
\begin{align*}
\delta {\cal L}_G & = {\cal E}^a_b E^\mu_c \delta E^b_\mu E^c {\bm
e}_a
\nonumber \\
& = {\cal E}^a_b E^\mu_a \delta E^b_\mu  {\bm e}\label{Gravbit}.
\end{align*}
\\

If we vary the action with respect to $g^{\mu\nu}$, the variation of
the matter part gives the Energy-momentum tensor:
\begin{gather}
\delta {\cal L}_{matter} = -T_{\mu\nu} \delta g^{\mu\nu} {\bm e}.
\end{gather}
Using (\ref{varygtoE}) and (\ref{-relation}):
\begin{align}
\delta {\cal L}_{matter} & = -2T_{\mu\nu}
\eta^{ab} \delta E_a^\mu E_b^\nu {\bm e}\nonumber\\
& = 2T^a_b E^\mu_a \delta E^b_\mu {\bm e}\nonumber\\
& = -2T^a_b \delta E^\mu_a E^b_\mu {\bm e}.\label{Matterbit}
\end{align}

Comparing (\ref{Gravbit}) and (\ref{Matterbit}), the Field
equation is:
\begin{gather}
H^a_b = T^a_b, \qquad H^a_b \equiv -\frac{1}{2} {\cal E}^a_b.
\end{gather}

The variation of the surface terms in the gravity action are of the
form:
\begin{gather*}
\int_I \tilde{{\cal E}}^a_b \delta E^b \tilde{{\bm e}}_a
\end{gather*}
so similarly we get:
\begin{gather}
\tilde{H}^a_b = \tilde{T}^a_b\qquad \tilde{H}^a_b \equiv
-\frac{1}{2} \tilde{{\cal E}}^a_b.
\end{gather}
I shall call $\tilde{H}^a_b$ the {\it Junction Tensor} of $I$.

Of course, we can write this in a co-ordinate basis:
\begin{gather}
H^\mu_\nu = T^\mu_\nu, \qquad H^\mu_\nu \equiv -\frac{1}{2} E_a^\mu
{\cal E}^a_b E^b_\nu.\label{juncminushalf}
\end{gather}

It is important to remember this factor of $-1/2$ when equating the variation
with the Energy momentum tensor. This has been used in equations
(\ref{Lovelocktensor}), (\ref{intereqmotion}) and (\ref{examplelast}).


\section{Second fundamental form and tensors}\label{secondtensors}

An embedded hypersurface, $\Sigma$, has a unique metric connection
$\omega_0$ induced by the bulk metric. Throughout this section
$\iota^*$ is the pull back onto $\Sigma$ acting on differential
forms. The {\it second fundamental form} is:
\begin{gather*}
{\bm I}\!{\bm I} = \iota^*(\omega -\omega_0).
\end{gather*}

The connection transforms non-tensorially but the difference of two
tensors is tensorial:
\begin{gather*}
(\omega -\omega_0)^a_b V^b = (D-D_0)V^a
\end{gather*}
for arbitrary vector $V$. Since the covariant derivative is
tensorial, so the difference of two connections must also be.

There is an important relation between the second fundamental form
and the extrinsic curvature tensor:
\begin{gather}\label{thetainK}
{\bm I}\!{\bm I}^{ab} =  2({\frak n}\!\cdot\! {\frak n}) {\frak
n}^{[a}(\nabla^{b]}{\frak n}_c )E^c=2({\frak n}\!\cdot\! {\frak n})
{\frak n}^{[a}K^{b]}.
\end{gather}
For a proof see e.g. Choquet-Bruhat et.al.~\cite{Choquet-Bruhat-82}.

Concentrate on a hypersurface $\{ij\}$. ${\frak n}_{\{ij\}} =-
{\frak n}_{\{ji\}}$ is the normal vector induced by $i$. I collect
here some formulae~\cite{Myers-87,Muller-Hoissen-90a}
\begin{align}\nonumber
\iota^* (D_{\{i\}} {\bm I}\!{\bm I}_{\{i|ij\}}^{ab}) & = 2 ({\frak
n} \!\cdot\!{\frak n})\, \iota^*\!\left(D_{\{i\}} {\frak n}
^{[a}K_{\{i\}}^{b]} +{\frak
n}^{[a}D_{\{i\}}K^{b]}_{\{i\}}\right)_{\{|ij\}} ,
\\\nonumber\\
[{\bm I}\!{\bm I}_{\{i\}},{\bm I}\!{\bm I}_{\{i\}}]_{\{|ij\}}^{ab} &
= 2 {\bm I}\!{\bm I}^a_c\wedge {\bm I}\!{\bm I}^{cb}\nonumber\\ & =
-2\left(({\frak n}\!\cdot\! {\frak n}) K_{\{i\}}^a\wedge K_{\{i\}}^b
+{\frak n}^a {\frak n}^b K_{\{i\}c}\wedge
K_{\{i\}}^c\right)_{\{|ij\}} \label{KiKj},
\\\nonumber\\\nonumber
\iota^*(D_{\{i\}}{\frak n}_{\{i|ij\}}^a ) & =-K^a_{\{i|ij\}}.
\end{align}
Above, $i|ij$ denotes a quantity on $\{ij\}$ induces by the bulk
region $\{i\}$, in particular $K_{\{i|ij\}} \neq K_{\{j|ij\}}$. Also
$i$ is used as shorthand for $i|ij$  when there is no ambiguity and
${\frak n}$ is shorthand for ${\frak n}_{ij}$.

Now, by $\sim$ I shall mean equality up to terms not explicitly
involving a normal index.
\begin{gather}
\iota^* (D_i {\bm I}\!{\bm I}_{\{i|ij\}}^{ab} ) \sim [{\bm I}\!{\bm
I}_{\{i\}},{\bm I}\!{\bm I}_{\{i\}}]_{\{|ij\}}^{ab} \sim -2\left(
({\frak n} \!\cdot\!{\frak n})K_{\{i\}}^a\wedge K_{\{i\}}^b
\right)_{\{|ij\}}.
\end{gather}
Note that
\begin{gather}
2 K^a \wedge K^b = \left( K^a_c K^b_d -K^a_b K^b_c\right)E^c\wedge
E^d.
\end{gather}
\\

At a hypersurface $\{ij\}$ we can interpolate between the intrinsic
curvature and the curvature of $i$:
\begin{align}
\Omega(t) & =  d\omega_i + td\theta
+\frac{1}{2}[\theta + t\theta,\omega_0 + t\theta]\nonumber\\
& = \Omega_i + tD_{ij} \theta +\frac{1}{2}t^2[\theta,\theta]\\
& = \Omega_i + tD_i\theta + \frac{(t^2-2t)}{2}[\theta,\theta].
\end{align}
Setting $t=1$ and $\iota^* \theta = {\bm I}\!{\bm I}$:
\begin{align}
\Omega_i & = \Omega_{ij} + (D_i\theta-\frac{1}{2}[\theta,\theta])\\
\iota^*\Omega_i & \sim  \Omega_{ij} - {\frak n}\!\cdot\!{\frak n}\,
K_i\wedge K_i.\label{Gauss-Codacci}
\end{align}

\section{Decomposing the bulk Lagrangian}\label{Zappend}

In section \ref{Glue}, we re-wrote $f(\Omega^n E^{d-2n})-f(\Omega_0
E^{d-2n})$ in such a way that the normal derivatives of the
extrinsic curvature were in a total derivative term (equation
\ref{fbreakup}). The other term was (\ref{Zterm}):
\begin{gather*}
{\cal Z} = n\int_0^1 dt\, (1-t)
\theta^{a_1a_2}\Omega(t)^{a_3...a_{2n}} \theta^{a_{2n+1}}_bE^b
e_{a_1...a_{2n+1}}.
\end{gather*}
Now, for convenience, use adapted frames. Let $E^{N}$ be a unit
normal vector and $E^i$ the tangent vectors. We make use of the
following identities (see \ref{thetainK} and \ref{Gauss-Codacci}):
\begin{align}
\theta^{ab} & = 2 N^{[a}K^{b]}_{\,c}E^c + (\text{normal}),
\\
\Omega(t)^{ij}& = \Omega_0^{ij} -{\frak n}\!\cdot\!{\frak n}\,
K^{i}\wedge K^{j} + (\text{normal}).\label{importantOmegat}
\end{align}
There are factors of $\Omega(t)$ appearing in ${\cal Z}$. It is easy
to see that only the tangential components of the tangential form
contribute. First note that if $\Omega(t)$ has a normal index, then
$\theta^{ij} \wedge \theta^{j}_b\wedge E^b$ is proportional to
$E^N\wedge E^N = 0$. Next note that if there are no normal indices
then $\theta^{i_1 i_2} \wedge \cdots \wedge \bm{e}_{i_1\cdots
i_{2n+1}} \propto E^N\wedge E^N =0$. We get:
\begin{multline*}
{\cal Z} = n\int_0^1 dt\, (1-t)\left(2
\theta^{Ni_2}\Omega(t)^{i_3...i_{2n}} \theta^{i_{2n+1}}_bE^b
\bm{e}_{N
i_2...i_{2n+1}}\right.\\\Big.+\theta^{i_1i_2}\Omega(t)^{i_3...i_{2n}}
\theta^{N}_bE^b \bm{e}_{i_1...i_{2n}N}\Big).
\end{multline*}
In the first term $\theta^{Ni_1}$ is proportional to $E^N$. In the
second term, $\theta^{i_{2n+1}}_b\wedge E^b$ will be proportional to
$E^N$. This means that the (normal) form part of $\Omega^{ij}$ does
not contribute to ${\cal Z}$. From (\ref{importantOmegat}), we see
that there are no derivatives of the extrinsic curvature appearing
in ${\cal Z}$. This justifies the statement in section \ref{Glue}
that there are no second normal derivatives of the metric in ${\cal
Z}$.



\section{Proof of $d\eta =0$}\label{proofprop}

In this section, a short proof is given of Proposition
\ref{propositionb}. We start with the definition of $\Omega(t)$ and
also the Jacobi identity~\cite{Chern-74}:
\begin{align}
\Omega(t) & ={\bm d}_{(x)}\omega(t)+\frac{1}{2}[\omega(t),\omega(t)],\\
[[\omega,\omega],\omega] & =0.
\end{align}
From the above one can easily find the following identities.
\begin{align}
{\bm d}_{(x)} [\omega(t),\omega(t)] & = 2[\Omega,\omega(t)],\label{A4}\\
{\bm d}_{(t)} [\omega(t),\omega(t)] & = 2[{\bm d}_{(t)}\omega(t),\omega(t)],\\
\label{kindofC}{\bm d}_{(t)}\omega(t)+\Omega(t) & = {\bm
d}_{(F)}\omega(t)+\frac{1}{2} [\omega(t),\omega(t)].
\end{align}
Also,
from (\ref{A4}), we get the Bianchi identity for $\omega(t)$:
\begin{gather}\label{Bianchi}
D(t)\Omega(t) = 0.
\end{gather}

Let us show that $\omega(t)$ transforms as a connection
\begin{align}
\omega_\alpha & \rightarrow g^{-1}\omega_i g +g^{-1}dg,
\nonumber\\
\omega(t) & \rightarrow \sum_\alpha C^\alpha g^{-1}\omega_\alpha g +
\sum_\alpha C^\alpha g^{-1}dg
\nonumber\\
& = g^{-1}\left(\sum_\alpha C^\alpha \omega_\alpha\right) g +
g^{-1}dg \nonumber\\& =  g^{-1}\omega(t) g +g^{-1}dg.
\end{align}
Hence $\omega(t)$ is a connection and so the invariance
property of $f$ implies, for 2-forms $\psi$:
\begin{gather}\label{Invariance}
\sum_i f(\psi_1\wedge...[\omega(t),\psi_i]...\wedge\psi_n) = 0
\end{gather}
Combining $(\ref{A4}-\ref{kindofC})$:
\begin{gather}\label{DFeta}
{\bm d}_{(F)} \big({\bm d}_{(t)}\omega(t)+\Omega(t)\big)= [{\bm d}_{(t)}\omega(t)+\Omega(t),\omega(t)]
\end{gather}
and so our proposition \ref{final?} follows by the invariance property
of the Polynomial (\ref{Invariance}).
\begin{gather*}
{\bm d} f\Big(\big[{\bm d}_{(t)}\omega(t)+\Omega(t)\big]^{\wedge n}\Big)= 0
\end{gather*}
\\

Let us expand the polynomial.
 \begin{align}\label{expand}
 \big[{\bm d}_{(t)}\omega(t)+\Omega(t)\big]^{\wedge n}
 &=\sum_{l=0}^{n}\ ^nC_l
 \left(\sum_{\alpha=1}^p{\bm d}_{(t)}\omega(t)\right)^{\wedge l}
 \wedge\Omega(t)^{\wedge (n-l)}
 \\\nonumber & \hspace{-1.3in}=\sum_{l=0}^n(-1)^{l(l-1)/2}
 \ ^nC_l \ dt^{\alpha_1}\wedge
 \cdot\cdot\cdot\wedge dt^{\alpha_l}\wedge
 {\bm d}_{(t)}\omega(t)_{\alpha_1}\wedge\cdots\wedge
 {\bm d}_{(t)}\omega(t)_{\alpha_l}
 \wedge\Omega(t)^{\wedge (n-l)}.
 \end{align}
The first term in the expansion evaluated at the 0-simplex $\{i\}$
is just the Euler density in the interior of the region $i$. This,
combined with the recursion relation (\ref{equiv}) completes the
proof by induction of Proposition \ref{propositionb}. As a
consistency check, we can see that the terms in this expansion
reproduce the form of $(\ref{Lagrange})$.

\chapter{Glossary}

{\small

\begin{tabular}{ll}
          &  {\large {\bf Indices}}\\
$\mu,\nu,..$ & Lower case Greek indices for d-dimensional space-time
indices.\\\\
$a,b..$   & Lower case Latin indices from the beginning of the
alphabet
 for local\\ & Lorentz indices.\\
$i,j,..$  & Lower case Latin indices from the middle of the alphabet
label the bulk\\ & regions.\\
$\{ij\},\{ijk\},...$ & The hypersurface or intersection where the bulk regions
                      $i,j,\dots$ meet.\\\\
$\Lambda,\Xi$ & Upper case Greek indices label hypersurfaces.\\&\\\\
          & {\large {\bf Symbols}}\\
$C^k$     &   $k$ times continuously differentiable \\
        &   {\it or} coefficient of $\omega_k$ in
            expansion of $\omega(t)$.\\

$d$       &   The dimension of a manifold.\\
        &\\
${\bm d}$ &   Exterior derivative.\\
        &\\
$E$       &   Co-frame field.\\
        &\\
$f$       &   Invariant $d$-form, $f(\psi) = \psi^{a_1...a_d}
            \epsilon_{a_1...a_d}$.\\
        &\\
$f_i$     &   Partition of unity.\\
        &\\
$I^p$     &   Non-simplicial intersection of co-dimension $p$.\\
        &\\
$s, s_{i_0...i_p}$ & Simplex (of dimension $p$).\\
        &\\
$t,t^\alpha$       & homotopy parameter(s).\\
        &\\
${\bm \eta}$, ${\bm \sigma}$
        &   (bold Greek letters)
            Exterior differential forms.\\
        &\\
$\omega$  & Connection 1-form.\\
          &\\
$\Omega$  & Curvature 2-form.
         \end{tabular}

}

\end{document}